\documentclass[12pt]{article}

\usepackage{times}

\usepackage{graphicx}% Include figure files

\newcounter{lastnote}
\newenvironment{scilastnote}{%
\setcounter{lastnote}{\value{enumiv}}%
\addtocounter{lastnote}{+1}%
\begin{list}%
{\arabic{lastnote}.}
{\setlength{\leftmargin}{.22in}}
{\setlength{\labelsep}{.5em}}}
{\end{list}}

\title{The dynamics of correlated novelties}

\author
{F. Tria$^{1}$,  V. Loreto$^{2,1}$, V.D.P. Servedio$^{2}$ and  S.H. Strogatz$^{3,^\ast}$\\
  \\
  \normalsize{$^{1}$ Institute for Scientific Interchange (ISI), Via Alassio 11C, 10126 Torino, Italy}\\
  \normalsize{$^{2}$ Sapienza University of Rome, Physics Dept., Piazzale Aldo Moro 5, 00185 Roma, Italy}\\
 \normalsize{$^{3}$ Cornell University, Dept. of Mathematics, 310 Malott Hall,
  Ithaca, NY 14853, USA} \\
\\
  \normalsize{$^\ast$ E-mail: strogatz@cornell.edu} }

\date{}

\begin{document} 


\baselineskip24pt


\maketitle


\begin{abstract} 
  One new thing often leads to another. Such correlated novelties are a familiar part of daily life. They are also thought to be fundamental to the evolution of biological systems, human society, and technology. By opening new possibilities, one novelty can pave the way for others in a process that Kauffman has called ``expanding the adjacent possible''. The dynamics of correlated novelties, however, have yet to be quantified empirically or modeled mathematically. Here we propose a simple mathematical model that mimics the process of exploring a physical, biological or conceptual space that enlarges whenever a novelty occurs. The model, a generalization of Polya's urn, predicts statistical laws for the rate at which novelties happen (analogous to Heaps' law) and for the probability distribution on the space explored (analogous to Zipf's law), as well as signatures of the hypothesized process by which one novelty sets the stage for another. We test these predictions on four data sets of human activity: the edit events of Wikipedia pages, the emergence of tags in annotation systems, the sequence of words in texts, and listening to new songs in online music catalogues. By quantifying the dynamics of correlated novelties, our results provide a starting point for a deeper understanding of the ever-expanding adjacent possible and its role in biological, linguistic, cultural, and technological evolution.
\end{abstract}

%\section{Introduction} 

Our daily lives are spiced with little novelties. We hear a new song, taste a new food, learn a new word. Occasionally one of these first-time experiences sparks another, thus correlating an earlier novelty with a later one. Discovering a song that we like, for example, may prompt us to search for other music by the same artist or in the same style. Likewise, stumbling across a web page that we find intriguing may tempt us to explore some of its links.

The notion that one new thing sometimes triggers another is, of course, commonsensical. But it has never been documented quantitatively, to the best of our knowledge. In the world before the Internet, our encounters with mundane novelties, and the possible correlations between them, rarely left a trace. Now, however, with the availability of extensive longitudinal records of human activity online~\cite{lazer_2009}, it has become possible to test whether everyday novelties crop up by chance alone, or whether one truly does pave the way for another.

The larger significance of these ideas has to do with their connection to Kauffman's theoretical concept of the ``adjacent possible''~\cite{kauffman_1996}, which he originally discussed in his investigations of molecular and biological evolution, and which has also been applied to the study of innovation and technological evolution~\cite{johnson_2010_book}. Loosely speaking, the adjacent possible consists of all those things (depending on the context, these could be ideas, molecules, genomes, technological products, etc.) that are one step away from what actually exists, and hence can arise from incremental modifications and recombinations of existing material.  Whenever something new is created in this way, part of the formerly adjacent possible becomes actualized, and is therefore bounded in turn by a fresh adjacent possible. In this sense, every time a novelty occurs, the adjacent possible expands~\cite{kauffman_2000}. This is Kauffman's vision of how one new thing can ultimately lead to another.  Unfortunately, it has not been clear how to extract testable predictions from it.

Our suggestion is that everyday novelties and their correlations allow one to test Kauffman's ideas quantitatively in a straightforward, down-to-earth setting. The intuition here is that novelties, like pre-biotic molecules and technological products, naturally form networks of meaningful associations. Just as a molecule in the primordial soup is conceptually adjacent to others that are one elementary reaction step away from it, a web page is conceptually adjacent to others on related topics. So when a novelty of any kind occurs, it does not occur alone. It comes with an entourage of surrounding possibilities, a cloud of other potentially new ideas or experiences that are thematically adjacent to it and hence can be triggered by it.

We begin by analyzing four data sets, each consisting of a sequence of elements ordered in time: (1) {\em Texts}: Here the elements are words. A novelty in this setting is defined to occur whenever a word appears for the first time in the text; (2) {\em Online music catalogues}: The elements are songs. A novelty occurs whenever a user listens either to a song or to an artist that she has not listened to before; (3) {\em Wikipedia}: The elements are individual wikipages. A novelty corresponds to the first edit action of a given wikipage by a given contributor (the edit can be the first ever, or other contributors may have edited the page previously but that particular contributor has not); (4) {\em Social annotation systems}: In the so-called tagging sites, the elements are tags (descriptive words assigned to photographs, files, bookmarks, or other pieces of information). A novelty corresponds either to the introduction of a brand new tag, or to its adoption by a given user. Further details on the data sets used are reported in the Supplementary Materials.

The rate at which novelties occur can be quantified by focusing on the growth of the number $D(N)$ of distinct elements (words, songs, wikipages, tags) in a temporally ordered sequence of data of length $N$. Figure 1 (A-D) shows a sublinear power-law growth of $D(N)$ in all four data sets, each with its own exponent $\beta<1$. This sublinear growth is the signature of Heaps' law~\cite{heaps_1978}. It implies that the rate at which novelties occur decreases over time as $t^{\beta-1}$.

A second statistical signature is given by the frequency of occurrence of the different elements inside each sequence of data. We look in particular at the frequency-rank distribution. In all cases (figure 1, F-I) the tail of the frequency-rank plot also follows an approximate power law (Zipf's law)~\cite{zipf_1949}. Moreover, its exponent $\alpha$ is compatible with the measured exponent $\beta$ of Heaps' law for the same data set, via the well-known relation $\beta=1/\alpha$ ~\cite{footnote_zipf,serrano_2009,lu_2010}.

Next we examine the four data sets for evidence of correlations between novelties. To do so we need to introduce the notion of semantics, defined here as meaningful thematic relationships between elements. We can then consider semantic groups as groups of elements related by common properties. The actual definition of semantic groups depends on the data we are studying, and can be straightforward in some cases and ambiguous in others. For instance, in the Wikipedia database, we can regard different pages as belonging to the same semantic group if they were created for the first time linked to the same mother page (see Supplementary Materials for further details). In the case of the music database (Last.fm), different semantic groups for the listened songs can be identified with the corresponding song writers. In the case of texts or tags, there is no direct access to semantics, and a slightly different procedure has to be adopted to detect semantically charged triggering events. Also in these cases the triggering of novelties can be observed by looking at the highly non-trivial distribution of words. We refer to the Supplementary Materials for a detailed discussion of these cases.

We now introduce two specific observables: the entropy $S$ of the events associated to a given semantic group, and the distribution of time intervals $f(l)$ between two successive appearances of events belonging to the same semantic group. Roughly speaking, both the entropy $S$ and the distribution of time intervals $f(l)$ measure the extent of clustering among the events associated to a given semantic group, with a larger clustering denoting stronger correlations among their occurrences and thus a stronger triggering effect (see the Supplementary Materials for a complete definition).

All the data sets display the predicted correlations among novelties.
The results for the Wikipedia and Last.fm databases are shown in
figure 2 (A,B,D,E), while we refer to the Supplementary Information
for the texts and tags databases. For comparison, we also reshuffle
all the data sets randomly to assess the level of temporal
correlations that could exist by chance alone. The evidence for
semantic correlations is signaled by a drop of the entropy $S$ with
respect to the reshuffled cases in both the databases considered
(figure 2, A and B). Correspondingly the distribution $f(l)$ features
a markedly larger peak for short time intervals compared to that seen
in the random case (figure 2, D and E), indicating that events
belonging to the same semantic group are clustered in time (figure
2G).

It is interesting to observe that both Wikipedia and Last.fm represent the outcome of a collective activity of many users. A natural question is whether the correlations observed above only emerge at a collective level or are also present at an individual level. We report in the Supplementary Materials the same analysis performed here for single users, showing that in this case each individual reproduces the qualitative features of the whole data set: namely, a significantly higher clustering than that found in the reshuffled data.

Our results so far are consistent with the presence of the hypothesized adjacent possible mechanism. However, since we only have access to the actual events and not to the whole space of possibilities opened up by each novelty, we can only consider indirect measures of the adjacent possible, such as the entropy and the distribution of time intervals discussed above.

To extract sharper predictions from the mechanism of an ever-expanding adjacent possible, it helps to consider a simplified mathematical model based on Polya's urn~\cite{polya_1930,johnson1977urn,mahmoud_polya}. In the classical version of this model~\cite{polya_1930}, balls of various colors are placed in an urn. A ball is withdrawn at random, inspected, and placed back in the urn along with a certain number of new balls of the same color, thereby increasing that color's likelihood of being drawn again in later rounds. The resulting ``rich-get-richer'' dynamics leads to skewed distributions~\cite{Yule_1925,Simon_1955} and have been used to model the emergence of power laws and related heavy-tailed phenomena in fields ranging from genetics and epidemiology to linguistics and computer science~\cite{Mitzenmacher_2003,newman_2005,Simkin_2011}.

This model is particularly suitable to our problem since it considers two spaces evolving in parallel: we can think at the urn as the space of possibilities, while the sequence of balls that are withdrawn is the history that is actually realized.

We generalize the urn model to allow for novelties to occur and to trigger further novelties. Consider an urn $\mathcal{U}$ containing $N_0$ distinct elements, represented by balls of different colors (Figure 3). These elements represent words used in a conversation, songs we've listened to, web pages we've visited, inventions, ideas, or any other human experiences or products of human creativity. A conversation, a text, or a series of inventions is idealized in this framework as a sequence $\mathcal{S}$ of elements generated through successive extractions from the urn. Just as the adjacent possible expands when something novel occurs, the contents of the urn itself are assumed to enlarge whenever a novel (never extracted before) element is withdrawn.

Specifically, the evolution proceeds according to the following scheme. At each time step $t$ we select an element $s_t$ at random from $\mathcal{U}$ and record it in the sequence. We then put the element $s_t$ back into $\mathcal{U}$ along with $\rho$ additional copies of itself. The parameter $\rho$ represents a reinforcement process, i.e., the more likely use of an element in a given context.  For instance, in a conversational or textual setting, a topic related to $s_t$ may require many copies of $s_t$ for further discussion. The key assumption concerns what happens if (and only if) the chosen element $s_t$ happens to be novel (i.e., it is appearing for the first time in the sequence $\mathcal{S}$). In that case we put $\nu+1$ brand new and distinct elements in the urn. These new elements represent the set of new possibilities triggered by the novelty $s_t$. Hence $\nu+1$ is the size of the new adjacent possible made available once we have a novel experience. The growth of the number of elements in the urn, conditioned on the occurrence of a novelty, is the crucial ingredient modeling the expansion of the adjacent possible.

This minimal model simultaneously yields the counterparts of Zipf's law (for the frequency distribution of distinct elements) and Heaps' law (for the sublinear growth of the number of unique elements as a function of the total number of elements). In particular, we find that the balance between reinforcement of old elements and triggering of new elements affects the predictions for Heaps' and Zipf's law. A sublinear growth for $D(N)$ emerges when reinforcement is stronger than triggering, while a linear growth is observed when triggering outweighs reinforcement. More precisely the following asymptotic behaviors are found (see Supplementary Materials for the analytical treatment of the model): (a) $D(N) \sim N^{\frac{\nu}{\rho}}$ if $\nu < \rho$; (b) $D(N) \sim \frac{N}{\log N}$ if $\nu = \rho$; (c) $D(N) \sim N$ if $\nu > \rho$. Correspondingly, the following asymptotic form is obtained for Zipf's law: $f(R) \sim R^{-\frac{\rho}{\nu}}$, where $f(R)$ is the frequency of occurrence of the element of rank $R$ inside the sequence ${\mathcal S}$. Figure 1 also shows the numerical results as observed in our model for the growth of the number of distinct elements $D(N)$ (Fig. 1E) and for the frequency-rank distribution (Fig. 1J), confirming the analytical predictions.

%\section{Introducing semantics}

So far we have shown how our simple urn model with triggering can account simultaneously for the emergence of both Heaps' and Zipf's law. This is a very interesting result {\em per se} because it solves the longstanding problem of explaining the origin of the Heaps' and Zipf's laws through the same basic microscopic mechanism, without the need of hypothesizing one of them to deduce the other. Despite the interest of this result, this is not yet enough to account for the adjacent possible mechanism revealed in real data. In its present form, the model accounts for the opening of new perspectives triggered by a novelty, but does not contain any bias towards the actual realization of these new possibilities.

To account for this, we need to infuse the earlier notion of semantics into our model. We endow each element with a label, representing its semantic group, and we allow for the emergence of dynamical correlations between semantically related elements. The process we now consider starts with an urn $\mathcal{U}$ with $N_0$ distinct elements,  divided into $N_0/(\nu+1)$ groups. The elements in the same group share a common label. To construct the sequence $\mathcal{S}$, we randomly choose the first element. Then at each time step $t$, (i) we give a weight $1$ to: (a) each element in $\mathcal{U}$ with the same label, say $A$, as $s_{t-1}$, (b) to the element that triggered the entry into the urn of the elements with label $A$, and (c) to the elements triggered by $s_{t-1}$. A weight $\eta \leq 1$ is assigned to all the other elements in $\mathcal{U}$. We then choose an element $s_t$ from $\mathcal{U}$ with a probability proportional to its weight and write it in the sequence; (ii) we put the element $s_t$ back in $\mathcal{U}$ along with $\rho$ additional copies of it (figure 3c); (iii) if (and only if) the chosen element $s_t$ is new (i.e., it appears for the first time in the sequence $\mathcal{S}$) we put $\nu+1$ brand new distinct elements into $\mathcal{U}$, all with a common brand new label (figure 3d). Note that for $\eta=1$ this model reduces to the simple urn model with triggering introduced earlier.

This extended model can again reproduce the Heaps' and Zipf's laws (for details, see the Supplementary Materials), and, crucially, it also reproduces the behavior of $S$ and $f(l)$ as measured in real data (figure 2, C and F). Thus, the hypothesized mechanism of a relentlessly expanding adjacent possible is consistent with the dynamics of correlated novelties, at least for the various techno-social systems~\cite{Vespignani_science_2009} studied here.

We speculate that our theoretical framework could be relevant to a much wider class of systems, problems, and issues -- indeed, to any situation where one novelty paves the way for another. One of the most intriguing generalizations would be to the study of innovation in cultural~\cite{innovation_cultural}, technological and biological systems~\cite{jacob_1977,Kauffman_1993}.  A huge literature exists on different aspects of innovation, concerning both its adoption and diffusion ~\cite{bass_1969,valente_1995,rogers_2003,salganik_2006}, as well as the creative processes through which it is generated~\cite{Schumpeter_1934,jacob_1977,Gould_1982,Brian_Arthur_2009}. The deliberately simplified framework we have developed here does not attempt to model explicitly the processes leading to innovations, such as recombination~\cite{Schumpeter_1934,Brian_Arthur_2009}, tinkering~\cite{jacob_1977} or exaptation~\cite{Gould_1982}.  Rather, our focus is entirely on the implications of the new possibilities that a novelty opens up.  In our modeling scheme, processes such as the modification or recombination of existing material take place in a black box; we account for them in an implicit way through the notions of triggering and semantic relations. Building a more fine-grained mathematical model of these creative processes remains an important open problem.  

Another direction worth pursuing concerns the tight connection between innovation and semantic relations. In preliminary work, we have begun investigating this question by mathematically reframing our urn model as a random walk. As we go about our lives, in fact, we silently move along physical, conceptual, biological or technological spaces, mostly retracing well-worn paths, but every so often stepping somewhere new, and in the process, breaking through to a new piece of the space.  This scenario gets instantiated in our mathematical framework. Our urn model with triggering, in fact, both with and without semantics, can be mapped onto the problem of a random walker exploring an evolving graph $\mathcal{G}$. The idea of the construction of a sequence of actions or elements as a path of a random walker in a particular space has been already studied in Ref.~\cite{pnas_cattuto_etal_2009}, where it has been shown that the process of social annotation can be viewed as a collective but uncoordinated exploration of an underlying semantic space. Here we go a step further by considering a random walker as wandering on a growing graph $\mathcal{G}$, whose structure is self-consistently shaped by the innovation process, the semantics being encoded in the graph structure. This picture strengthens the correspondence between the appearance of correlated novelties and the notion of the adjacent possible. Moreover, this framework allows one to relate quantitatively, and in a more natural way, the particular form of the exploration process (modulated by the growing graph topology) and the observed outcomes of observables related to triggering events. We refer to the Supplementary Materials for a detailed discussion of this mapping and results concerning this random-walk framework for the dynamics of correlated novelties.

Two more questions for future study include an exploration of the subtle link between the early adoption of an innovation and its large-scale spreading, and the interplay between individual and collective phenomena where innovation takes place.  The latter question is relevant for instance to elucidate why overly large innovative leaps cannot succeed at the population level. On a related theme, the notion of advance into the adjacent possible sets its own natural limits on innovations, since it implies that innovations too far ahead of their time, i.e. not adjacent to the current reality, cannot take hold. For example, video sharing on the Internet was not possible in the days when connection speeds were 14.4 kbits per second. Quantifying, formalizing, and testing these ideas against real data, however, remains a fascinating challenge.

\bibliography{main}

\begin{thebibliography}{10}
\bibitem{footnote_zipf}
It is important to observe as the frequency-rank plots are far from featuring a
  pure power-law behaviour and the relation $\beta= 1/\alpha$ between the
  exponent $\beta$ of the {H}eaps' law and the $\alpha$ exponent of the
  {Z}ipf's law is to be expected only looking at the tail of the {Z}ipf's plot.
  {T}he frequency-rank plots feature a variety of behaviours depending on the
  specific system. {A} detailed fitting of these curves cannot be obtained
  without taking into account specific features of the observed phenomenon.
  {F}or instance for text corpora the frequency-rank plot features a $1/r$
  trend for intermediate ranks (between $10$ and $10^4$) while a flattening is
  observed for the most frequent words and a larger slope, with an exponent
  compatible with the observed {H}eaps' law, for rare and specialized words.
  {W}ithout modelling the distinction between article, prepositions and nouns
  it would be impossible to recover such a complex frequency-rank plot.
\end{thebibliography}
\bibliographystyle{Science}

\begin{scilastnote}
\item The authors acknowledge support from the EU-STREP project
  EveryAware (Grant Agreement 265432) and the EuroUnderstanding
  Collaborative Research Projects DRUST funded by the European Science
  Foundation. 
  % \item We've included in the template file \texttt{scifile.tex} a
  %   new environment, \texttt{\{scilastnote\}}, that generates a
  %   numbered final citation without a corresponding signal in the
  %   text. This environment can be used to generate a final numbered
  %   reference containing acknowledgments, sources of funding, and
  %   the like, per {\it Science\/} style. Along those lines, we'd
  %   like to thank readers of this document for their attention, and
  %   invite them to address any questions to Stewart Wills, at
  %   swills@aaas.org.
\end{scilastnote}

%%%%%%%%%%%%%%%%%%%%%%%%%%%%%%%%%%%%%%%%%%%%%%%%%%%%%%%%%
%%% FIGURES %%%
%%%%%%%%%%%%%%%%%%%%%%%%%%%%%%%%%%%%%%%%%%%%%%%%%%%%%%%%%

\section*{Figures}

\begin{figure} \begin{center} \includegraphics[width=0.75\columnwidth]{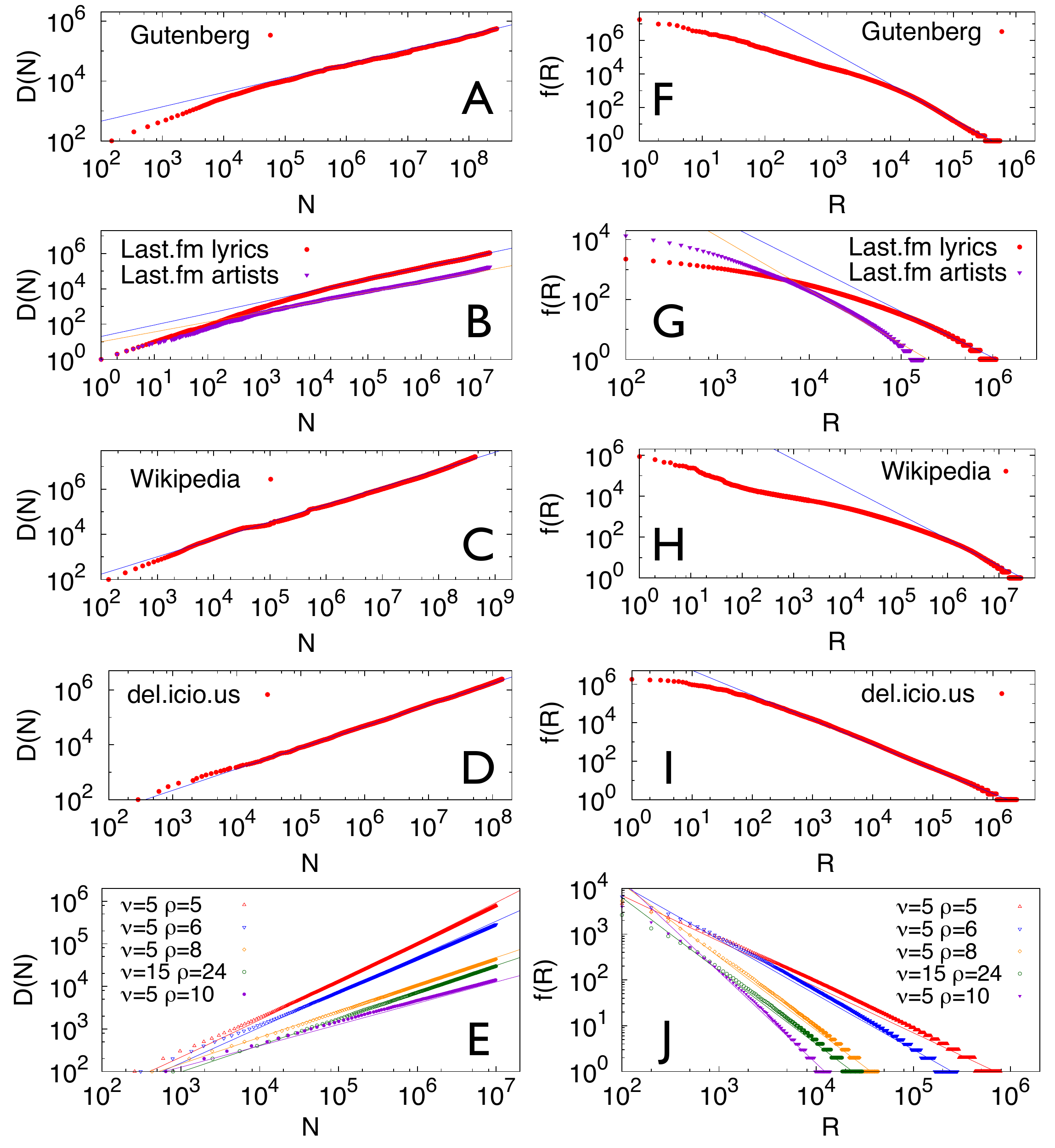} \caption{{\bf Heaps' law (A-E) and Zipf's law (F-L) in real datasets (A-D) and (F-I) and in the urn model with triggering (E,J).} Gutenberg~\cite{gutenberg} {\bf (A,F)}, Last.fm~\cite{last.fm} {\bf (B,G)}, Wikipedia~\cite{wikipedia} {\bf (C,H)}, del.icio.us~\cite{delicious} {\bf (D,I)} datasets, and the urn model with triggering {\bf (E,J)}.  Straight lines in the Heaps' law plots show functions of the form $f(x)=a x^{\beta}$, with the exponent $\beta$ equal respectively to $\beta=0.45$ (Gutenberg), $\beta=0.68$ (Last.fm lyrics), $\beta=0.56$ (Last.fm artist), $\beta=0.77$ (Wikipedia) and $\beta=0.78$ (del.icio.us), and to the ratio $\nu/\rho$ in the urn model with triggering, showing that the exponents for the Heaps' law of the model predicted by the analytic results are confirmed in the simulations. Straight lines in the Zipf's law plots show functions of the form $f(x)=a x^{-\alpha}$, where the exponent $\alpha$ is equal to $\beta^{-1}$ for the different $\beta$'s considered above.  Note that the frequency-rank plots in real data deviate from a pure power-law behaviour and the correspondence between the $\beta$ and $\alpha$ exponents is valid only asymptotically~\cite{footnote_zipf}.}  \end{center} \label{fig:heap_and_zipf_data_model} \end{figure}

\begin{figure} \includegraphics[width=0.98\textwidth]{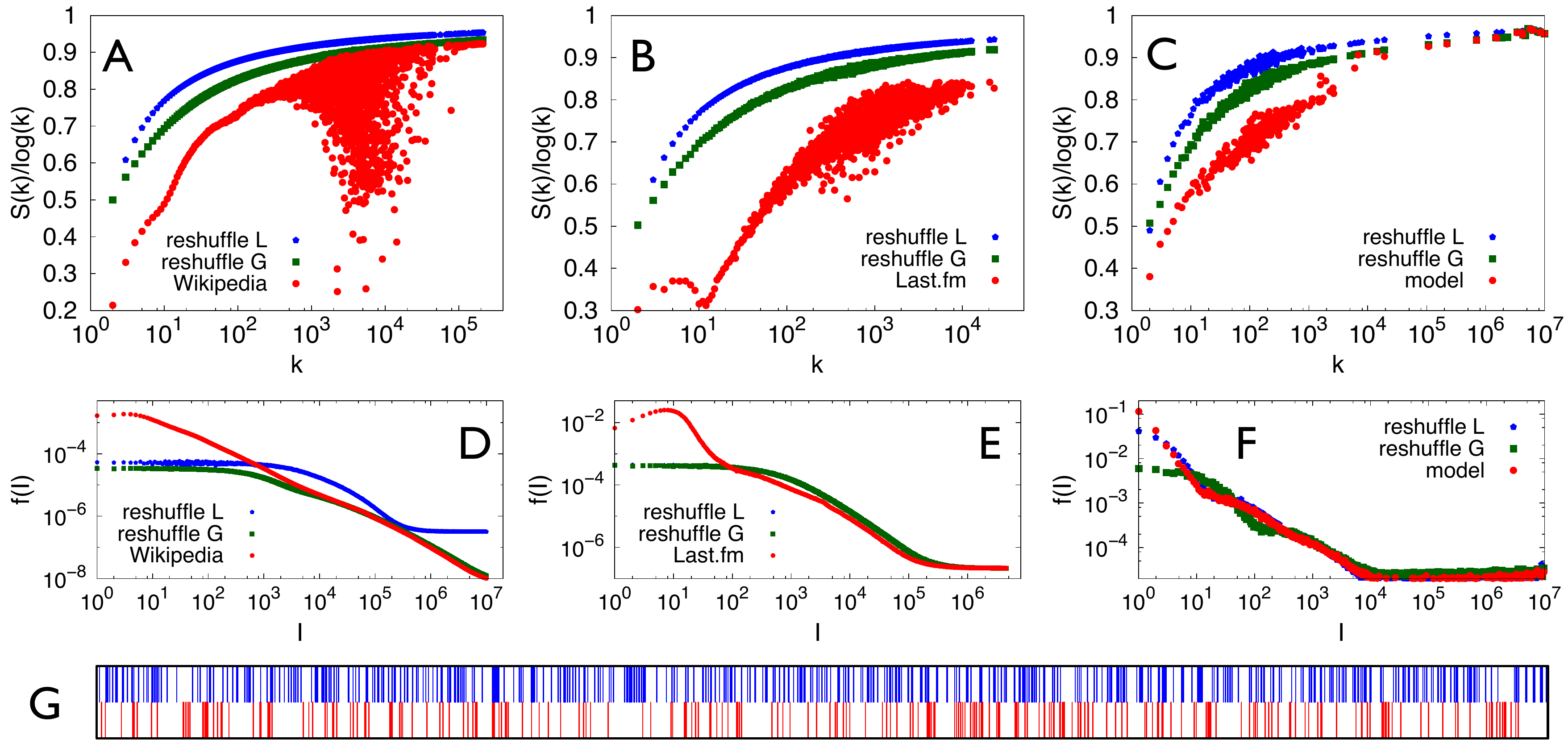}
  \caption{{\bf Entropy and distribution of triggering intervals in
      real data and in the urn model with semantic triggering.} {\bf
      (A,B,C)} Entropy of a sequence associated to a specific label $A$ vs. the
  number of events, $k$, with that label. The entropy is averaged for
  each $k$ over the labels with the same number of occurrences.
  Results are displayed for Wikipedia {\bf (A)}, the Last.fm dataset
  {\bf (B)} and the urn model with semantic triggering {\bf (C)}. For
  the Wikipedia and Last.fm datasets we used the respective sequences
  $S_{unique}$ as described in the Supplementary Materials. The plot for the
  model is an average over $10$ realizations of the process, with
  parameters $\rho=8$, $\nu =10$, $\eta=0.3$ and $N_0=\nu+1$. The
  length of the considered sequences is $N=10^{7}$ and the
  corresponding Heaps' exponent is $\beta= \frac{\nu
    \eta}{\rho}=0.375$ (see Supplementary Materials for the relation
  of the Heaps' and Zipf's exponents with the model parameters). In
  all the cases, results for the actual data are compared with two
  null models, as described in the Supplementary Materials. {\bf (D,E,F)}
  Results for the distribution of triggering intervals (see the
  Supplementary Materials for the definition) for the same data as for the entropy
  measurements. The banner {\bf (G)} shows a $S_{unique}^A$ sequence
  for a particular label $A$ of the Last.fm dataset. The color code is
  red for the actual sequence $\mathcal{S}$ and blue for the local
  reshuffle (see methods section) of $\mathcal{S}$.}
\label{fig:entropia_intervalli}
\end{figure}

\begin{figure}
\includegraphics[width=\textwidth]{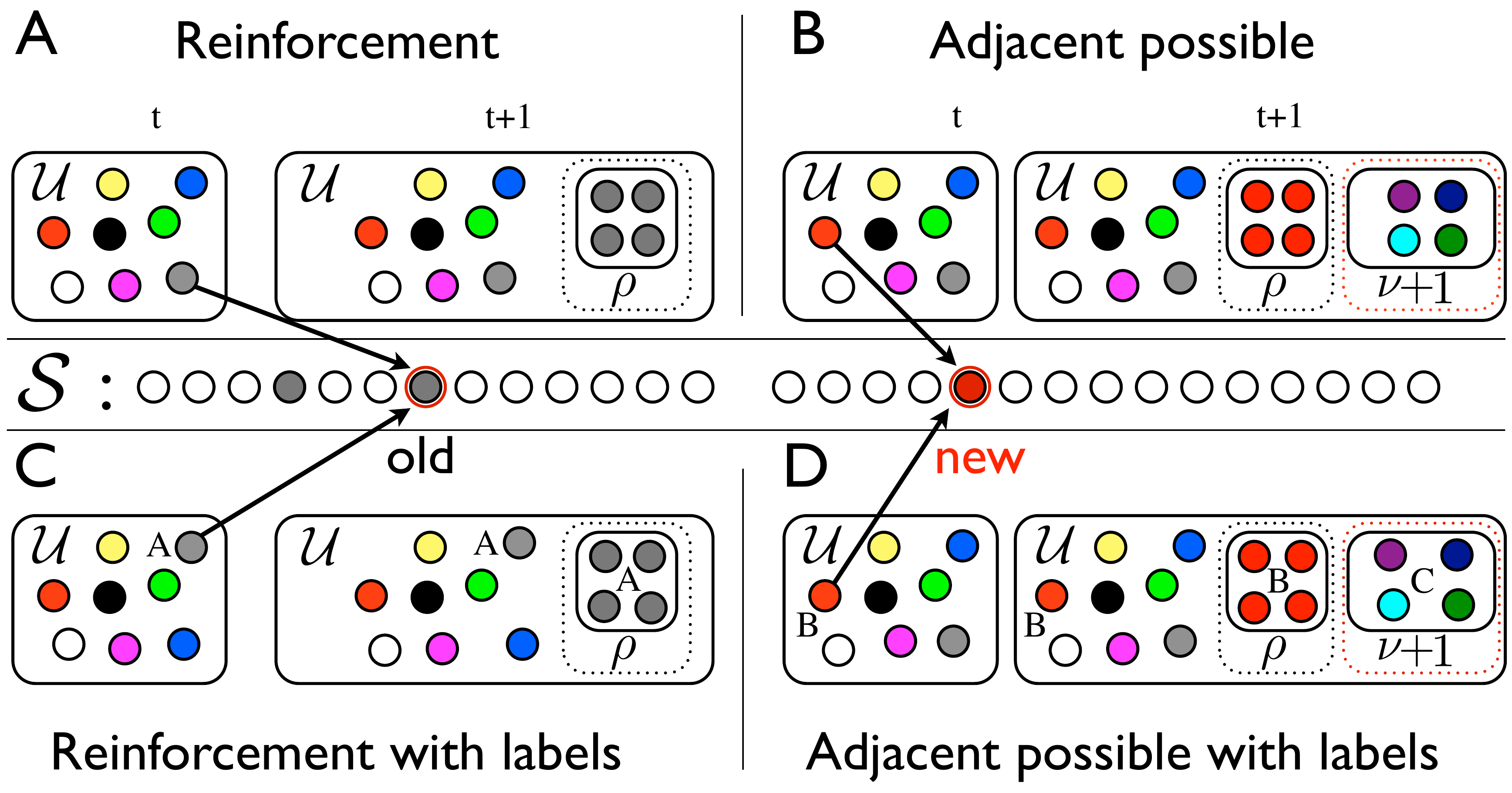}
\caption{{\bf Models.} Simple urn model with triggering {\bf (A,B)}
  and urn model with semantic triggering {\bf (C,D)}. {\bf (A)}
  Generic reinforcement step of the evolution. An element (the gray
  ball) that had previously been drawn from the urn $\mathcal{U}$ is
  drawn again. In this case one adds this element to $\mathcal{S}$
  (depicted at the center of the figure) and, at the same time, puts
  $\rho$ additional gray balls into $\mathcal{U}$. {\bf (B)} Generic
  adjacent possible step of the evolution. Here, upon drawing a new
  ball (red) from $\mathcal{U}$, $\nu+1$ brand new balls are added to
  $\mathcal{U}$ along with the $\rho$ red balls of the reinforcement
  step that takes place at each time step. {\bf (C,D)} Urn model with
  semantic triggering. Same as above except that now each ball has a
  label defining its semantic context. The label is conserved during a
  reinforcement event (e.g., the label $A$ for the gray balls on panel
  c) while it appears as a brand new label, $C$, for the $\nu+1$ balls
  added for an adjacent possible event (panel d).}
\label{fig:rules}
\end{figure}

% Use only LaTeX2e, calling the article.cls class and 12-point type.

% Users of the {thebibliography} environment or BibTeX should use the
% scicite.sty package, downloadable from *Science* at
% www.sciencemag.org/about/authors/prep/TeX_help/ .
% This package should properly format in-text
% reference calls and reference-list numbers.

\newpage

% Include your paper's title here

{\Huge
    Supplementary Material}

\section{Urn model with triggering}

\subsection{Model definition}\label{sec:1}

In the main text we introduced the 
\emph{urn model with triggering}.
Briefly, an ordered
sequence $\mathcal{S}$ was constructed by picking elements (or balls)
from a reservoir (or urn) $\mathcal{U}$ initially containing 
$N_0$ distinct elements. 
Both the reservoir and the sequence
increased their size according to the following procedure. At each time step:
\begin{enumerate}
\item[(i)] an element is randomly extracted from $\mathcal{U}$ with uniform probability and added to
  $\mathcal{S}$;
\item[(ii)] the extracted element is put back into $\mathcal{U}$ together with $\rho$ copies of it;
\item[(iii)] if the extracted element has never been used before in $\mathcal{S}$ (it
  is a new element in this respect), then $\nu+1$ different brand new distinct elements
  are added to $\mathcal{U}$.
\end{enumerate}
Note that the number of elements $N$ of $\mathcal{S}$, i.e.\ the length $|\mathcal{S}|$ of the sequence,
equals the number of times $t$ we repeated the above procedure.  
If we let $D$ denote the
number of distinct elements that appear in $\mathcal{S}$, then the total
number of elements in the reservoir after $t$ steps is $|\mathcal{U}|_t=N_0+(\nu+1) D+\rho t$.\\
In the following, we shall also consider a second and slightly different version of the model, in which the reinforcement
does not act when an element is chosen for the first time. 
Hence, point (ii) of the previous rules will be changed into:
\begin{enumerate}
\item[(ii.a)] 
	the extracted element is put back in $\mathcal{U}$ together with $\rho$ copies of it \emph{only if it is
 not new in the sequence}.
\end{enumerate}
%
%% We now consider two main observables, namely the number of
%% distinct elements $D(t)$ in the sequence ${\mathcal S}$ after $t$
%% iterations of the procedure described above as a function of the time
%% $t$ and the frequency-rank distribution $f(R)$ of the different
%% elements appearing in ${\mathcal S}$.

\subsection{Computation of the asymptotic Heaps' and Zipf's laws}

We discuss here the asymptotic behaviour of both the number of
distinct elements $D(t)$ appearing in the sequence  and the
frequency-rank distribution $f(R)$ of the elements in the
sequence $\mathcal{S}$. 
We will show that both versions of the urn model above predict a
Heaps' law for $D(t)$ and a frequency-rank distribution $f(R)$ with a fat-tail
behavior. Our calculations yield simple formulas for the
Heaps' law exponent and the exponent of
the asymptotic power-law behavior of the frequency-rank distribution 
in terms of  the model parameters $\rho$ and $\nu$.

Strictly speaking, Zipf's law requires an inverse proportionality
between the frequency and rank of the considered quantities~\cite{zipf_1949}.
In the following, however, we shall always refer instead to a generalized version of Zipf's law, in which
the dependence of the frequency on the rank is  power-law-like in the tail of the distribution, i.e.\ at large ranks.

\paragraph{Heaps' law\\}
In the first version of the model, the time dependence  of the
number  $D$  of different elements in the sequence $\mathcal{S}$
 obeys the following differential equation:
\begin{equation}
  \frac{d D}{dt}= \frac{U_D(t)}{U(t)}=\frac{N_0 + \nu D}{N_0+(\nu+1) D+\rho t},\label{eq:D_complete}
\end{equation}
 where $U_D(t)$ is the number of  elements in the reservoir
that at time $t$ have not yet appeared in $\mathcal{S}$, and $U(t)=|\mathcal{U}|_t$ is the
total number of elements in the reservoir at time $t$.
The term $\nu D$ in the numerator of the rightmost expression comes from the fact
that each time a new element is introduced in the sequence,  $U_D(t)$ is
increased by  $\nu$  elements (since $\nu+1$ brand new elements
  are added to $\mathcal{U}$, while the chosen element is no longer new).
Due to the inherently discrete character of $D$ and $t$, Eq.~(\ref{eq:D_complete})
is valid asymptotically for large values of $D$ and $t$.

In the second version of the model,
Eq.~(\ref{eq:D_complete}) has to be modified by replacing the denominator with
\[
	U(t) = N_0+(\nu+1) D+\rho (t-D)= N_0+(\nu+1-\rho) D+\rho t.
\]

To analyze both versions of the model simultaneously, it is convenient to define a
parameter $a \equiv \nu+1$ for the first version and $a \equiv \nu+1-\rho$ for
the second version.

In order to obtain an analytically solvable equation, and 
since we are interested in the behaviour at large times $t \gg N_0$, we
approximate equation (\ref{eq:D_complete}) by 
\begin{equation}
\frac{d D}{dt}= \frac{\nu D}{ a D+ \rho t}.\label{eq:D}
\end{equation}
%
%%%%%%%%%%%%%%%%%%%%%%%%%%
%% This can be solved by defining $z=\frac{D}{t}$, obtaining:
%% 
%% \begin{equation}
%%   z^\prime t + z= \frac{\nu z}{a z+\rho} ,\label{eq:zprime}
%% \end{equation}
%% 
%% \noindent and thus:
%% \begin{equation}
%% \int \frac{a z +\rho}{z(\nu-\rho -a z)}dz = \int \frac{dt}{t},\label{eq:z}
%% \end{equation}
%% 
%% 
%% \noindent Solving the integral we obtain:
%% 
%% \begin{equation}
%%   \frac{\rho}{\nu-\rho}\log{z}  - \frac{\nu}{\nu-\rho}\log{\left( z a+\rho-\nu\right)} =\log{t} \nonumber
%% \end{equation}
%% 
%% \noindent and after some algebra:
%% 
%% 
%% \begin{equation}
%% \frac{z^{\frac{\rho}{\nu}}}{z a+\rho-\nu}=t^{\frac{\nu-\rho}{\nu}} 
%% \end{equation}
%% 
%% \noindent By substituting $z=\frac{D}{t}$ and again after some algebra we obtain:
%% 
%% \begin{equation}
%% D^{\frac{\rho}{\nu}}- a D = (\rho-\nu) t .\label{eq:Dfinal}
%% \end{equation}
%%%%%%%%%%%%%%%%%%%%%%%%%%%%%%%%%
%
By introducing the auxiliary variable $z=\frac{D}{t}$  and performing some
straightforward algebra we obtain the asymptotic
behaviour of $D(t)$ for large $t$:
\begin{enumerate}
\item $\rho > \nu$: $D \sim (\rho-\nu)^{\frac{\nu}{\rho}} t^{\frac{\nu}{\rho}}$;
\item $\rho < \nu$: $D \sim \frac{\nu-\rho}{a}t$;
\item $\rho = \nu$: $D \log D \sim \frac{\nu}{a} t \rightarrow D \sim \frac{\nu}{a} \frac{t}{\log t}$,
\end{enumerate}

%% \noindent where the last estimate cannot be deduced directly from
%% equation~(\ref{eq:Dfinal}), but is deduced substituting $\nu=\rho$ directly
%% in equation~(\ref{eq:zprime}).

% \begin{equation}
% D(t) \sim \begin{cases}
% \left(\frac{\rho-\nu}{\rho+1}\right)^{\frac{\nu}{\rho}} t^{\frac{\nu}{\rho}} &
% \mbox{if}\; \nu< \rho    \\
% & \\
% \frac{\nu}{\nu+1} \frac{t}{\log{t}} & \mbox{if}\; \nu= \rho    \\
% &\\
% \frac{\nu-\rho}{\rho+1} t & \mbox{if}\; \nu>\rho.
% \end{cases}
% \label{eq:D_t}
% \end{equation}

For completeness,
we note that both versions of the model can be regarded as the coarse-grained 
equivalent of a two-color asymmetric Polya urn model~\cite{mahmoud_polya}.
In particular, within that finer framework the substitution matrices (denoted $M_1$ for the first version of the model and $M_2$ for the second) would be:
\[ 
	M_1=\left( \begin{array}{cc}
		\rho & 0 \\
		1+\rho & \nu \end{array} \right)
~~\mbox{and}~~ 
	M_2=\left( \begin{array}{cc}
		\rho & 0 \\
		1 & \nu \end{array} \right).
\] 
In this interpretation, the elements that have already appeared in $\mathcal{S}$
are represented by balls of one color, while those that have not appeared yet correspond to balls of the other color.

\paragraph{Zipf's  law\\} 
Making the same approximations as above, the continuous dynamical equation for the number of
occurrences $n_i$ of an element $i$ in the sequence $\mathcal{S}$ can be written as
\begin{equation}
	\frac{d n_i}{d t} = \frac{n_i \rho  +1}{N_0+aD+\rho t}\cdot
\label{eq:dni_dt}
\end{equation}
Two cases can be distinguished: 
\begin{enumerate}
\item $\nu\leq \rho$, when  
\( \displaystyle \lim_{t  \rightarrow +\infty} D/t =  0\). 
By considering only the leading term for \mbox{$t \rightarrow +\infty$}, one has
\begin{equation}
	\frac{d n_i}{d t} \simeq \frac{n_i}{t}.
\end{equation}
Let $t_i$ denote the time at which the element $i$ occurred for the first time
in the sequence. Then the solution for $n_i(t)$ starting from the initial condition  $n_i(t_i)=1$ is given by
%%$n_i=t/t_i$,
%
\begin{equation}
	n_i =\frac{t}{t_i}. \label{eq:ni}
\end{equation}
Now consider the cumulative distribution $P(n_i\leq n)$.
From Eq.~(\ref{eq:ni}), we can write $P(n_i\leq n)=P(t_i \geq \frac{t}{n})= 1 -P(t_i <\frac{t}{n})$.
This leads to the estimate: 
\begin{equation}
P(t_i<\frac{t}{n})\simeq\frac{D(\frac{t}{n})}{D(t)}=
n^{-\frac{\nu}{\rho}}. \label{eq:cumul1}
\end{equation}

\item $\nu>\rho$, when  $D \simeq \frac{\nu-\rho}{a} t$. Again
  considering $t \gg N_0$, we write:
\begin{equation}
\frac{d n_i}{d t} \simeq \frac{\rho n_i}{(\rho+a
  \frac{\nu-\rho}{a})t} = \frac{\rho n_i}{\nu
  t},
\end{equation}
which yields the solution
\begin{equation}
n_i = \left(\frac{t}{t_i} \right)^{\frac{\rho}{\nu}} .
\end{equation}
Proceeding as in the previous case, we find 
\( 
	P(n_i\leq n)=
	P(t_i \geq	 {t}\,{n^{-\frac{\nu}{\rho}}})= 
	1 -P(t_i< {t}\,{n^{-\frac{\nu}{\rho}}})
\), 
and thus 
\begin{equation}
	P(t_i< {t}\,{n^{-\frac{\nu}{\rho}}})\simeq
	\frac{D({t}\,{n^{-\frac{\nu}{\rho}}})}{D(t)}= 
	n^{-\frac{\nu}{\rho}},\label{eq:cumul2}
\end{equation}
obtaining the same functional expression of the asymptotic power-law
behavior of the frequency-rank distribution as in the previous case.
\end{enumerate}
The probability density function of the occurrences of the elements in the sequence is therefore
\(
	P(n)= \frac{\partial P(n_i<n) }{\partial n}
%	= - \frac{\partial  P(t_i<\frac{t}{n})}{\partial n}
	\sim n^{-\left(1+\frac{\nu}{\rho}\right)}
\), 
which corresponds to a frequency-rank distribution
$f(R) \sim R^{-\frac{\rho}{\nu}} $.

Note that the estimates in equations~(\ref{eq:cumul1})
and~(\ref{eq:cumul2}) have been derived under the assumption that $t/n
\gg1$, i.e.\ in the tail of the frequency-rank distribution. In this
respect, it is important to recognize that Zipf's and Heaps' laws are
not trivially and automatically related, as is sometimes claimed. We
certainly agree that Heaps' law can be derived from Zipf's law by the
following random-sampling argument: if one assumes a strict power-law
behaviour of the frequency-rank distribution $f(R) \sim R^{-\alpha}$
and constructs a sequence by randomly sampling from this Zipf
distribution $f(R)$, one recovers Heaps' law with the functional form
$D(t) \sim t^{\beta}$ with $\beta
=1/\alpha$~\cite{serrano_2009,lu_2010}. But the assumption of random
sampling is strong and sometimes unrealistic. If one relaxes the
hypothesis of random sampling from a power-law distribution, the
relationship between Zipf's and Heaps' law becomes far from trivial.
In our model, and in work by others~\cite{lu_2010}, the relationship
$\beta = 1/\alpha$ holds only asymptotically, i.e.\ only for large
times, with $\alpha$ measured on the tail of the frequency-rank
distribution.

\begin{figure}[tp]
\centerline{\includegraphics[width=\textwidth]{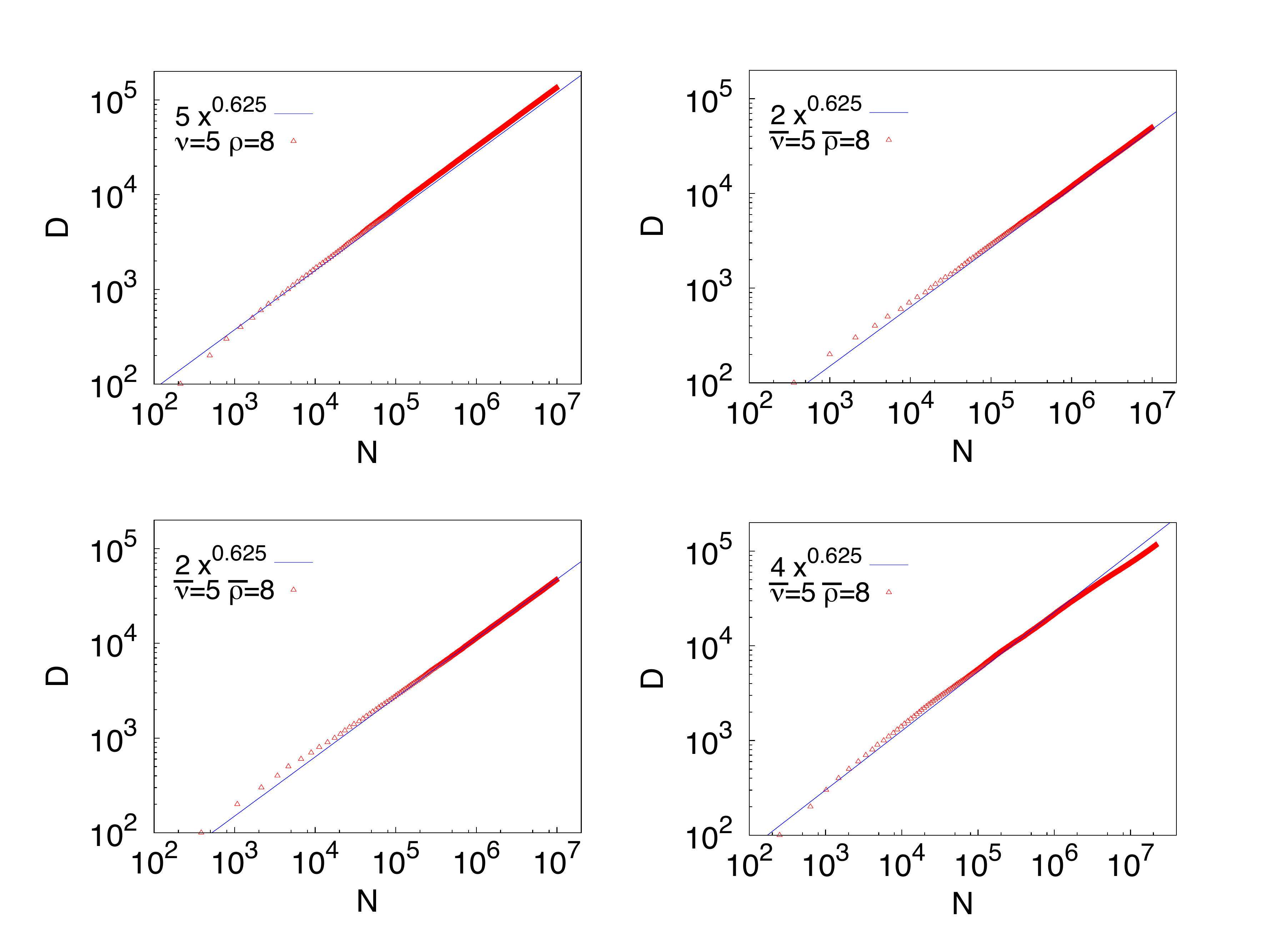}}
\caption{\textbf{Growth of the number of distinct elements (Heaps' law).}
  {Top left}: second version of the model without reinforcement on new words. 
  {Top right}: original model with $\rho$ and $\nu$ sampled from
  uniform distributions.
 % with average respectively $\bar{\rho}$ and  $\bar{\nu}$.  
%  
  {Bottom left}: original model with $\rho$ and $\nu$
  extracted from exponential distributions.
	%  with average respectively  $\bar{\rho}$ and	 $\bar{\nu}$. 
%
  {Bottom right}: original model with
  $\rho$ and $\nu$ extracted from power law distributions.
  % with  exponents $\alpha_{\rho}=\frac{2 \rho -1}{\rho -1}$ and  $\alpha_{\nu}=\frac{2 \nu -1}{\nu -1}$.
  %with average respectively  $\bar{\rho}$ and $\bar{\nu}$.
  All distributions bear the same average values $\bar\rho=8$ and $\bar\nu=5$ 
  (see text for details).
  \label{fig:heaps_normal}
  }
\end{figure}
%
% \begin{figure}
% \centerline{\includegraphics[width=0.6\textwidth]{D_p_6_q_10}\includegraphics[width=0.6\textwidth]{D_p_8_q_10}}
% \centerline{\includegraphics[width=0.6\textwidth]{D_p_9_q_10}\includegraphics[width=0.6\textwidth]{D_p_10_q_10}}
% \centerline{\includegraphics[width=0.6\textwidth]{D_p_11_q_10}\includegraphics[width=0.6\textwidth]{D_p_20_q_10}}
% \caption{{\bf Number of different elements} in $\mathcal{T}$ as a function of
%   the number of steps t (Heap's law). Straight lines are the computed asymptotic
%   expression of D(t) (eq.~\ref{eq:D_t}). $N$ corresponds to $N_0$.}
% \end{figure}

%\paragraph{Zipf's law}
%
\begin{figure}[htp]
\centerline{\includegraphics[width=\textwidth]{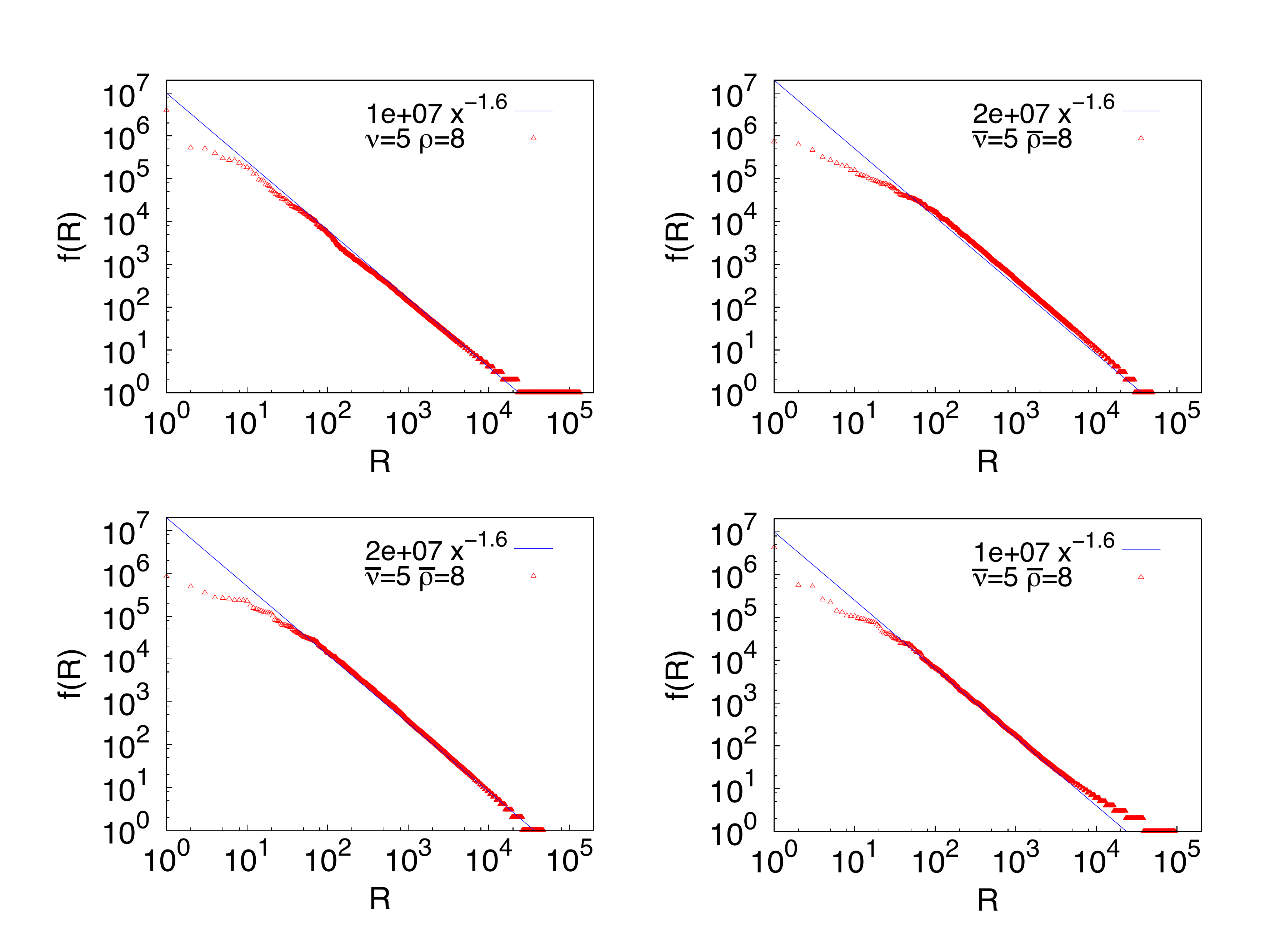}}
\caption{\textbf{Frequency-rank distribution (Zipf's law).}  
  {Top left}: second version of the model without reinforcement on new words. 
  {Top right}: original model with $\rho$ and $\nu$ sampled from
  uniform distributions.
 % with average respectively $\bar{\rho}$ and  $\bar{\nu}$.  \vito{variance?}
%  
  {Bottom left}: original model with $\rho$ and $\nu$
  extracted from exponential distributions.
	%  with average respectively  $\bar{\rho}$ and	 $\bar{\nu}$. 
%
  {Bottom right}: original model with
  $\rho$ and $\nu$ extracted from power law distributions.
%   with  exponents $\alpha_{\rho}=\frac{2 \rho -1}{\rho -1}$ and   $\alpha_{\nu}=\frac{2 \nu -1}{\nu -1}$.
  %with average respectively $\bar{\rho}$ and $\bar{\nu}$.  All
  distributions bear the same average values $\bar\rho=8$ and
  $\bar\nu=5$.  We have checked that the results do not depend on the
  initial condition $N_0$.  This is set in all the simulations to the
  value $N_0=100$.
  \label{fig:zipf_normal}
  }
\end{figure}
%
%% \subsection{Stability of the results for the modified and stochastic versions of the model}
%% We here show that numerical simulations of the modified version of
%% the model, as described at the end of section~\ref{sec:1}, 
%% agree with the analytic predictions. 
%% Moreover, we verify that the same
%% asymptotic behavior is obtained when extracting the parameters $\rho$
%% and $\nu$ from different probability distributions, with the same mean
%% value.
%
% \begin{figure}
% \centerline{\includegraphics[width=0.6\textwidth]{zipf_p6q10}\includegraphics[width=0.6\textwidth]{zipf_p8q10}}
% \centerline{\includegraphics[width=0.6\textwidth]{zipf_p9q10}\includegraphics[width=0.6\textwidth]{zipf_p10q10}}
% \centerline{\includegraphics[width=0.6\textwidth]{zipf_p11q10}\includegraphics[width=0.6\textwidth]{zipf_p20q10}}
% \caption{{\bf Frequency rank plot} of the elements occurrences. The exponent of
%   the straight lines is the computed one ($-q/p$). $N$ corresponds to $N_0$}
% \end{figure}
% 
In the main text we presented numerical results confirming the above analytical predictions for the
first version of our model. Here we report numerical
results for the second version of the model (employing the definition
(ii.a)), summarized in the top-left panels of
Fig.~\ref{fig:heaps_normal} and Fig.~\ref{fig:zipf_normal}. The
robustness of the results with respect to fluctuations of the model
parameters $\nu$ and $\rho$ was checked as follows. At each
time step both $\rho$ and $\nu$ were sampled from a uniform
distribution (top-right), an exponential distribution (bottom-left)
and a fat-tailed distribution with diverging variance, all 
with the same mean values $\bar\rho=8$ and $\bar\nu=5$. For the
uniform distribution, $\rho$ and $\nu$ were sampled from the intervals
$[0,2\bar\rho]$ and $[0,2\bar\nu]$, while for the fat-tailed
distribution, the chosen exponents were $\alpha_{\rho}=\frac{2 \rho
  -1}{\rho -1}$ and $\alpha_{\nu}=\frac{2 \nu -1}{\nu -1}$, which
ensured the desired average values by choosing 1 as the minimum
value.

In the case $\rho<\nu$ we recover the results of the well-known
Yule-Simon model~\cite{Simon_1955}, originally proposed in the context
of linguistics. In this model, new words are added to a text (more
generally a stream) with constant probability $p$ at each time step,
while with complementary probability $(1-p)$, a word that has already
occurred is chosen uniformly from within the text (or stream)
generated so far. This model leads to a Zipf's law with an exponent
$-(1-p)$ compatible with a linear growth in time of the number of
different words. In the framework of our {\em urn model with
  triggering} we recover the same Zipf's exponents as well as the
linear growth of $D(t)$ if $p=1-\frac{\rho}{\nu}$, with
$\rho<\nu$\footnote{We note that if $\nu \gg 1$ when $a=\nu+1$ (first
  version of the model) or $\nu \gg \rho$ and $\nu \gg 1$ when
  $a=\nu+1-\rho$ (second version of the model) our model also
  reproduces the same prefactor of the linear growth of $D(t)$ as in
  the Yule-Simon model. This is evident by setting $a=\nu$ in
  Eq.~(\ref{eq:D}).}. The Yule-Simon model is a paradigmatic example
of a model that generates a fat-tail frequency-rank distribution
$f(R)\sim R^{-\alpha}$ by using a rich-gets-richer mechanism. But it
has the drawback that it does not reproduce both an $f(R)$ obeying a
power-law behavior and a sublinear Heaps' exponent at the same time.
Moreover, the Yule-Simon model cannot reproduce values of $\alpha$
larger than $1$ (which are found empirically in the frequency-rank
distribution of words in certain texts). These problems were at the
basis of the famous Simon-Mandelbrot
dispute~\cite{Mandelbrot_1959,Simon_1960,Mandelbrot_1961a,Simon_1961,Mandelbrot_1961b}.
In our model the introduction of the parameter $\nu$ (describing the
expansion of the adjacent possible) heals these problems by confining
the phenomenology of the Yule-Simon model to the special case
$\rho<\nu$.

\subsection{Heaps' and Zipf's laws for the \emph{urn model with
    semantic triggering}}

We turn now to the counterparts of Heaps' and Zipf's laws for the
\emph{urn model with semantic triggering}. For the sake of
completeness we recall the model's definition. One starts with an urn
$\mathcal{U}$ with $N_0$ distinct elements, divided in $N_0/(\nu+1)$
groups, the elements in the same group sharing a common label. After
choosing the first element at random, the sequence $\mathcal{S}$ is
constructed according to the following scheme:

\begin{itemize}
\item[(i)] a weight $1$ is given to: (a) each element in $\mathcal{U}$
  with the same label, say $A$, as $s_{t-1}$, (b) to the element that
  triggered the enter in the urn of the elements with label A, and (c)
  to the elements triggered by $s_{t-1}$; a weight $\eta \leq 1$ is
  given to any other element in $\mathcal{U}$;
\item[(ii)] an element $s_t$ is chosen from $\mathcal{U}$ with a probability
  proportional to its weight and appended to the sequence; 
\item[(iii)] the element $s_t$ is put back into $\mathcal{U}$ along with
  $\rho$ additional copies of it; 
\item[(iv)] if the chosen element $s_t$ is new (i.e., it appears for the
  {\em first time} in the sequence $\mathcal{S}$) $\nu+1$ brand new
  distinct elements, all with a common brand new label, are added to
  $\mathcal{U}$. These $\nu+1$ new elements are given a weight
  $\eta=1$ at the next time step $t+1$ and each time the same mother
  element $s_t$ is picked.
\end{itemize}
Note that if $\eta=1$ this model corresponds to the simple urn model
with triggering introduced earlier.

\begin{figure}[htp]
\centerline{%
  \includegraphics[width=0.5\columnwidth]{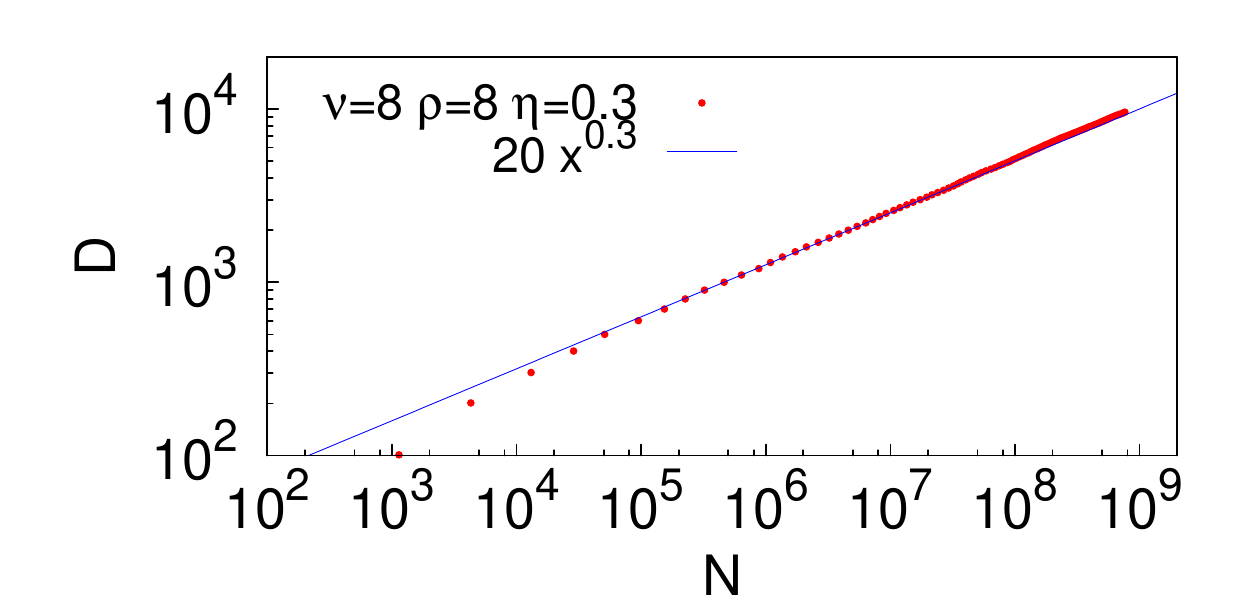}%
  \includegraphics[width=0.5\columnwidth]{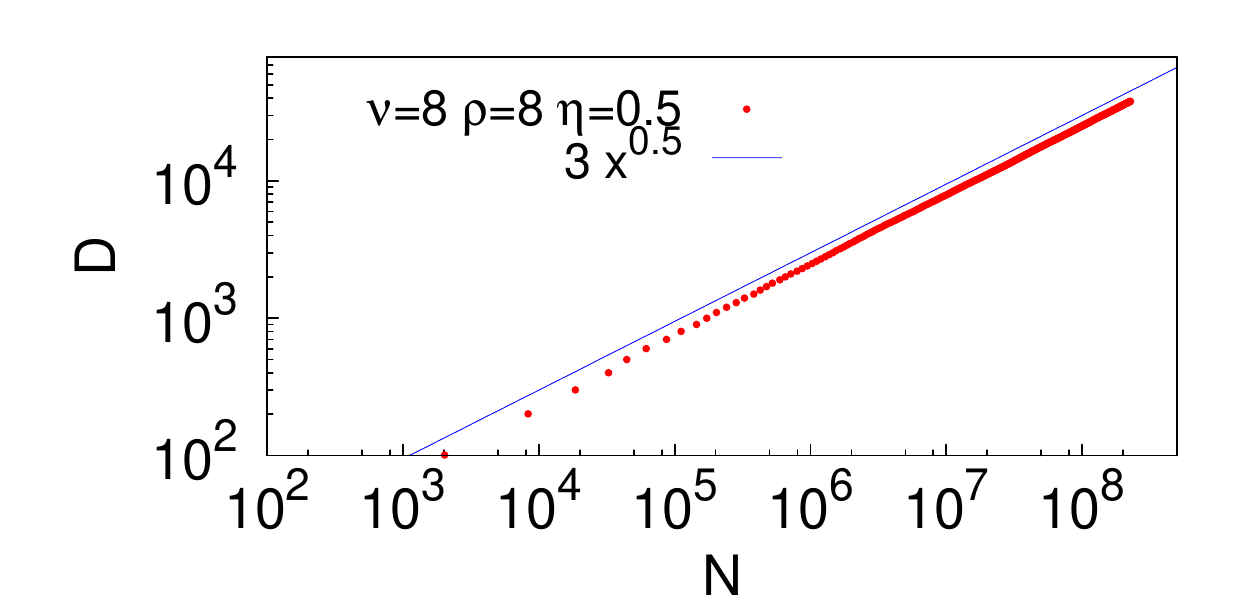}}
\centerline{%
  \includegraphics[width=0.5\columnwidth]{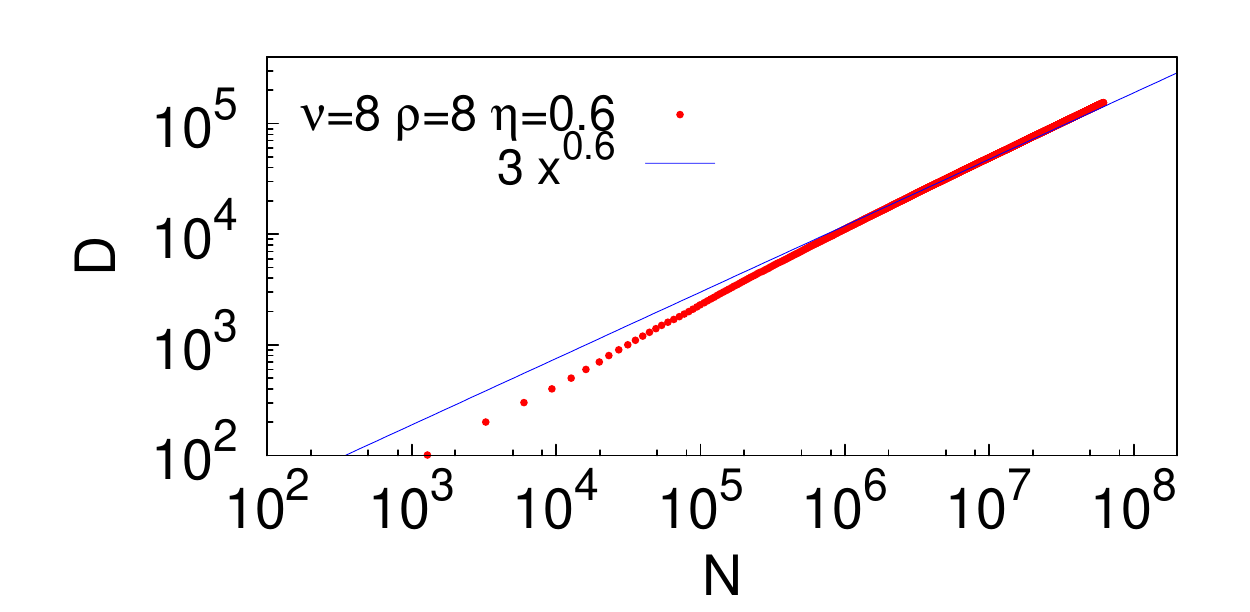}%
  \includegraphics[width=0.5\columnwidth]{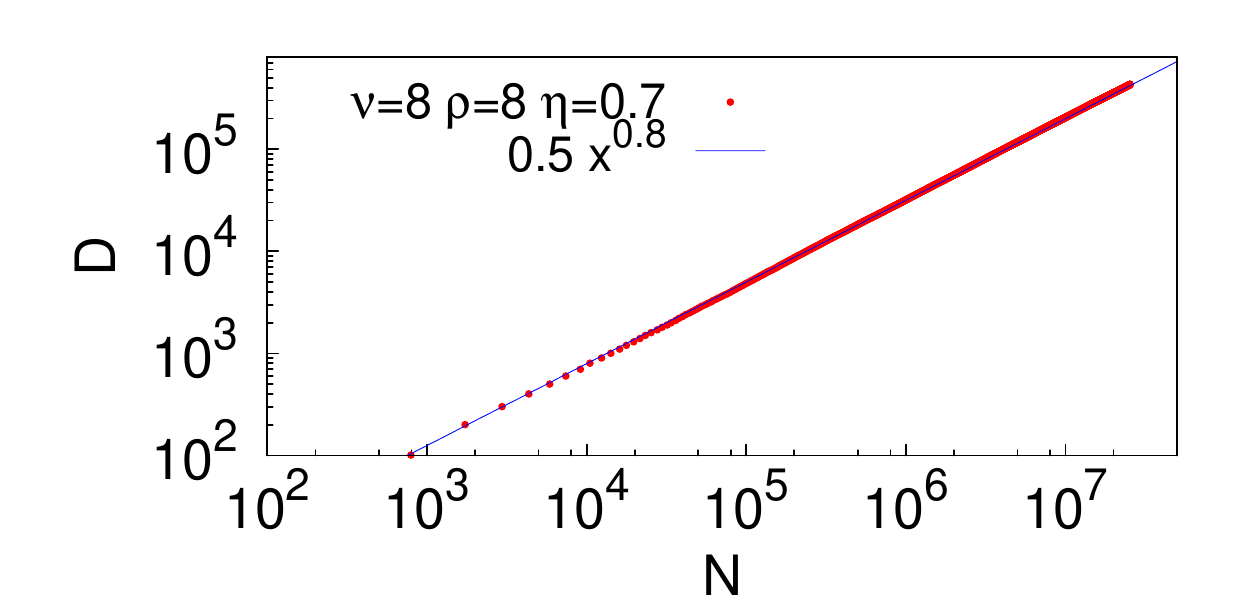}}
\centerline{%
  \includegraphics[width=0.5\columnwidth]{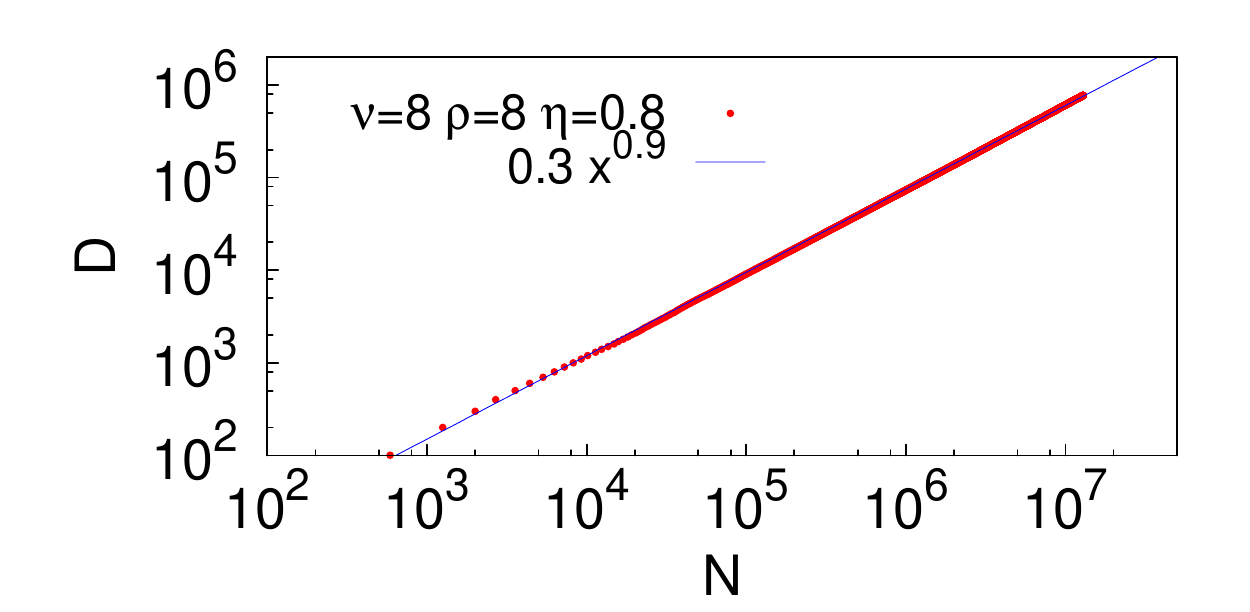}%
  \includegraphics[width=0.5\columnwidth]{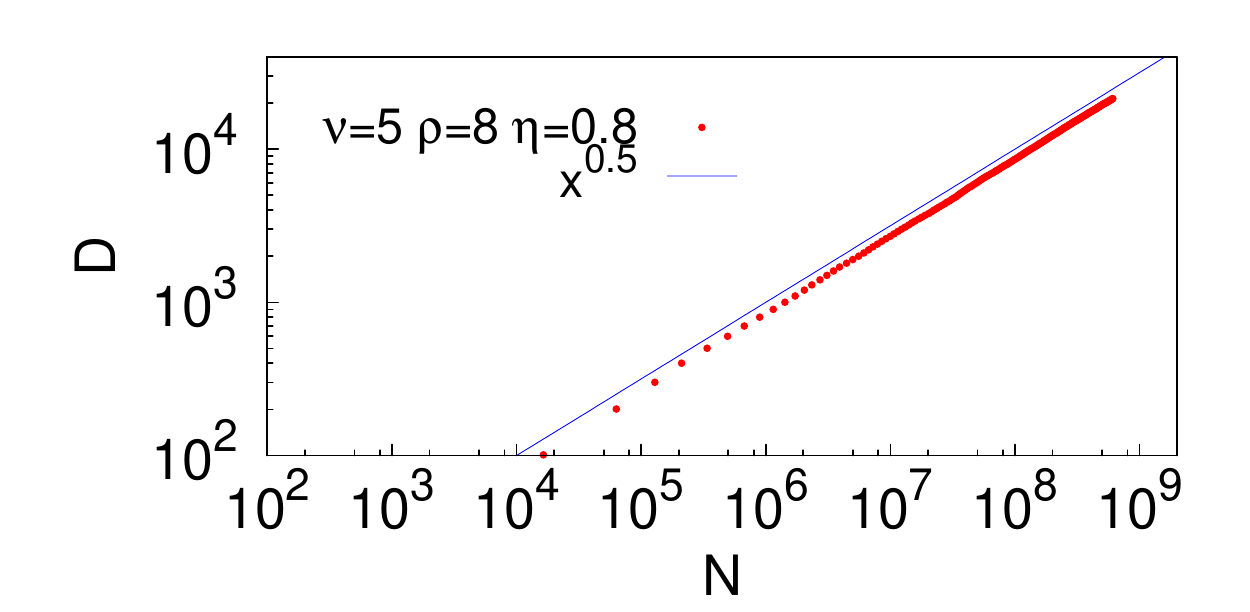}}
\centerline{%
  \includegraphics[width=0.5\columnwidth]{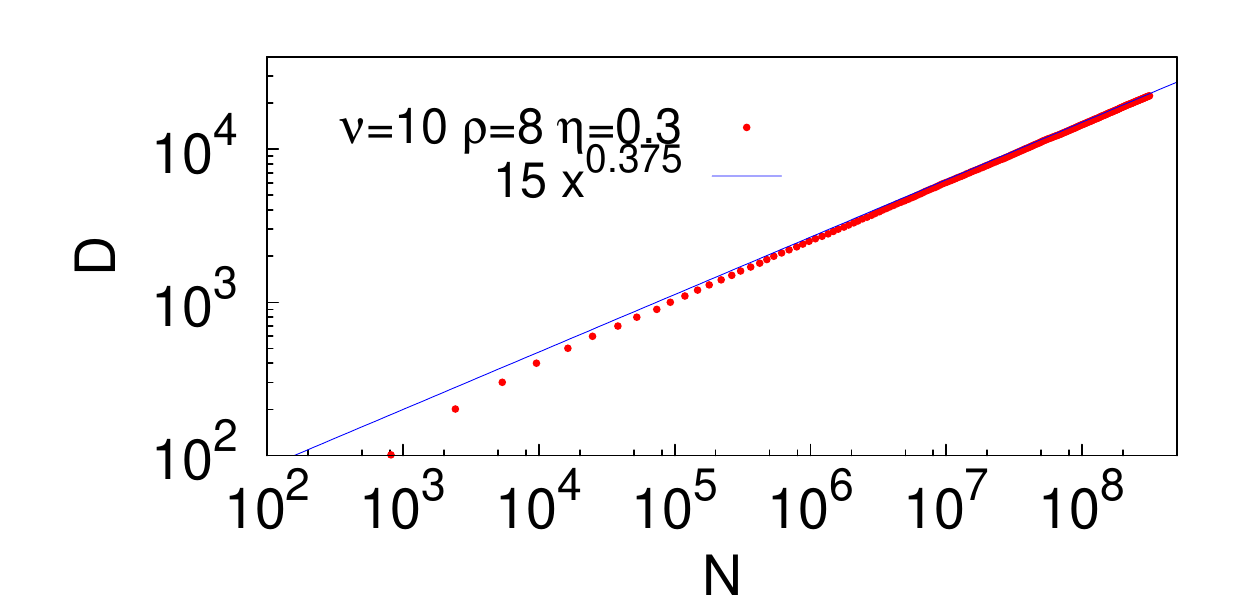}%
  \includegraphics[width=0.5\columnwidth]{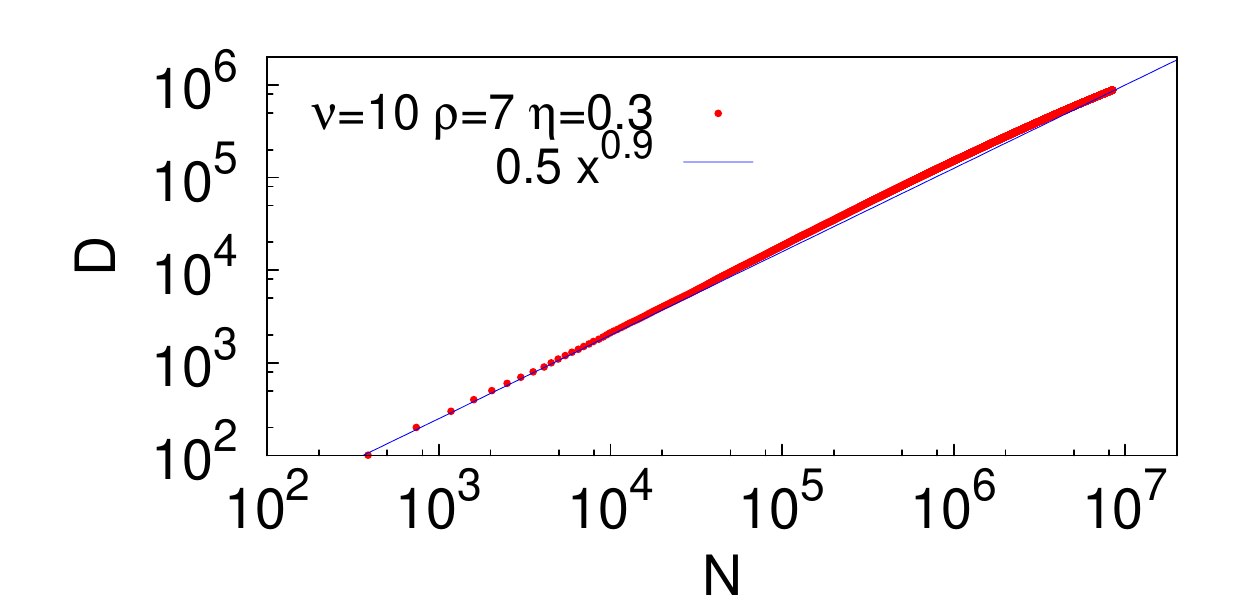}}
\caption{{\bf Growth of the number of distinct elements (Heaps' law).}
  Heaps' law for several values of the parameters of the urn model
  with semantic triggering. Straight lines show functions of the form
  $a x^{\beta}$, where $a$ is a constant. In all the simulations
  $N_0=\nu+1$. The observed exponents are within the theoretical
  bounds $\min(\frac{\nu \eta }{\rho},1) \leq \beta \leq
  \min(\frac{\nu }{\rho},1) $.}\label{fig:Heap_wl}
\end{figure}

\begin{figure}[htp]
\centerline{%
  \includegraphics[width=0.5\columnwidth]{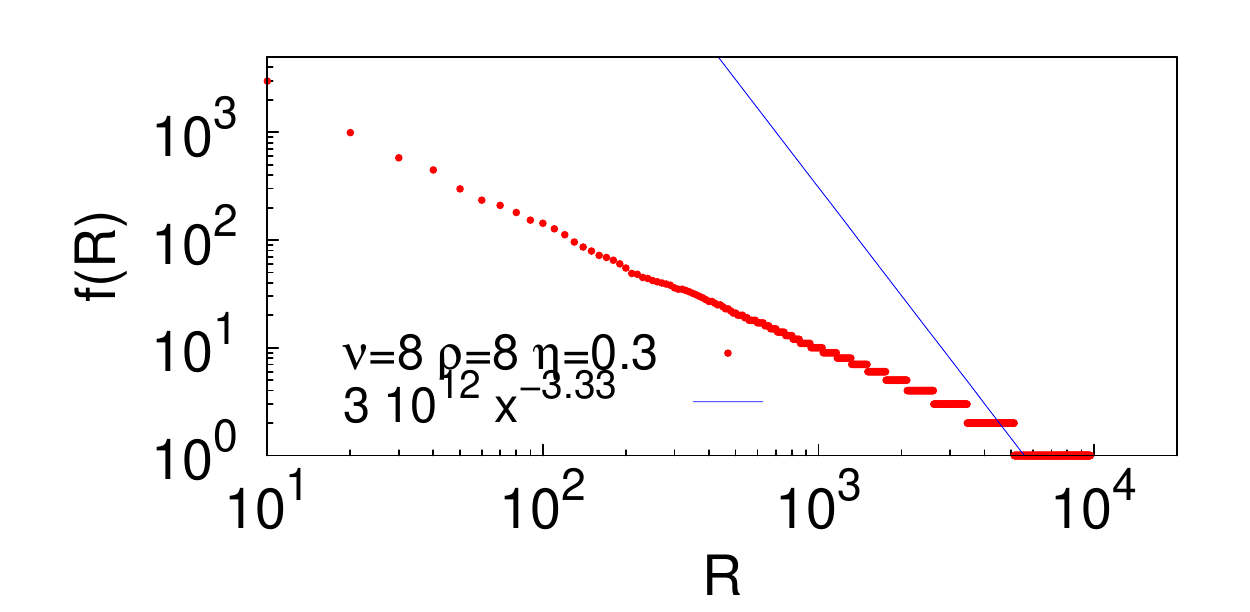}%
  \includegraphics[width=0.5\columnwidth]{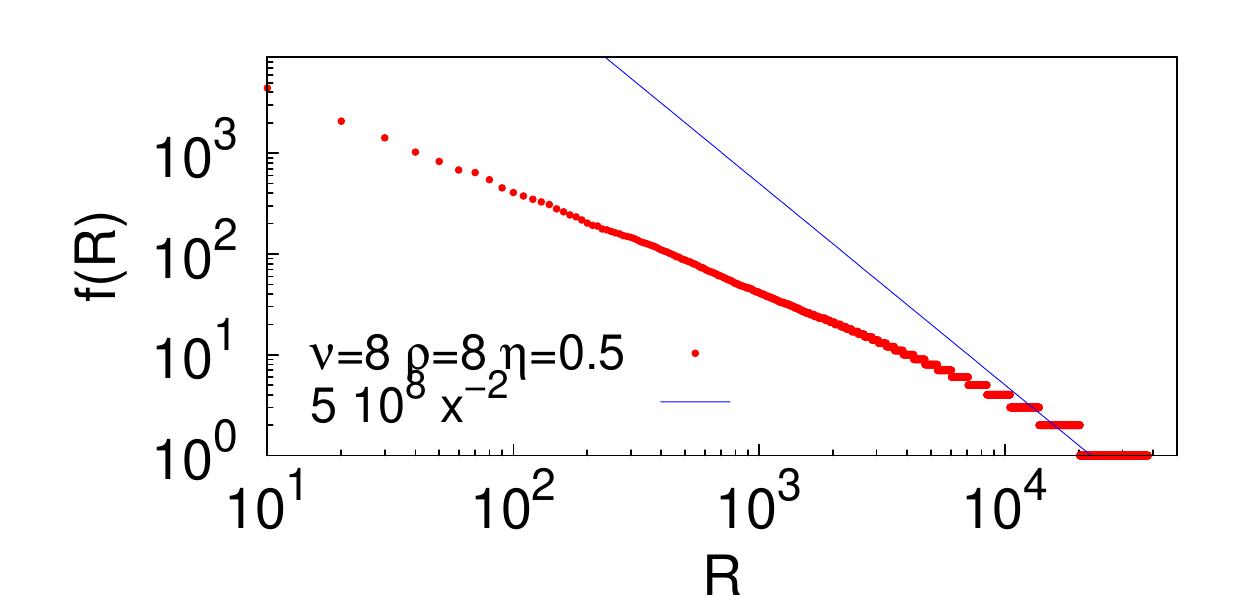}}
\centerline{%
  \includegraphics[width=0.5\columnwidth]{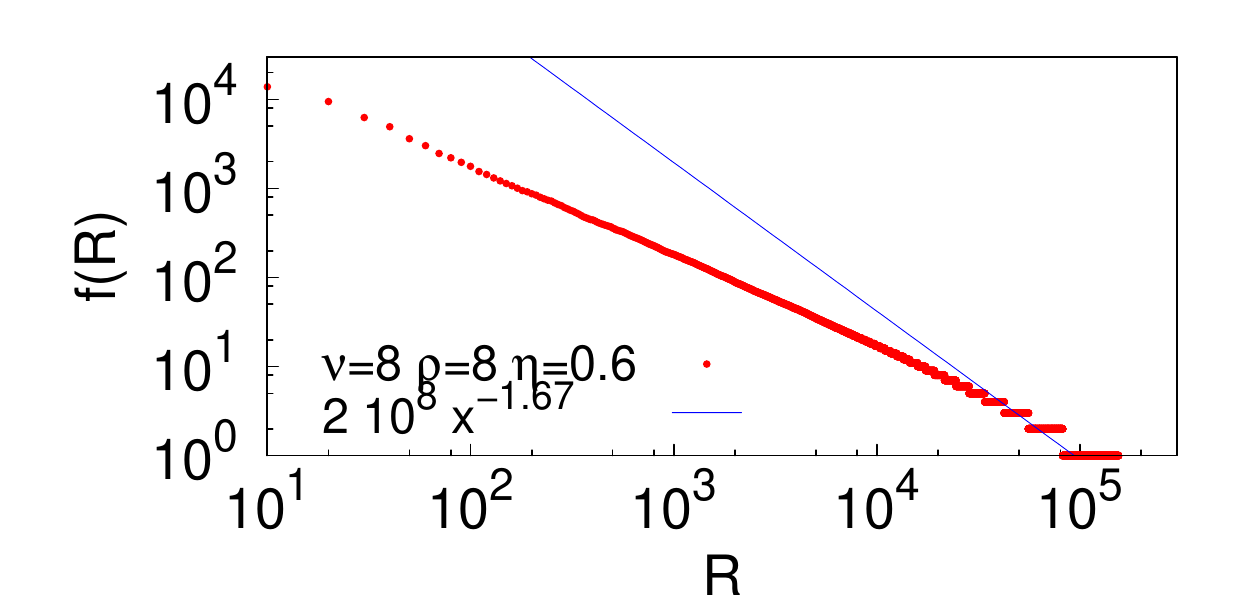}%
  \includegraphics[width=0.5\columnwidth]{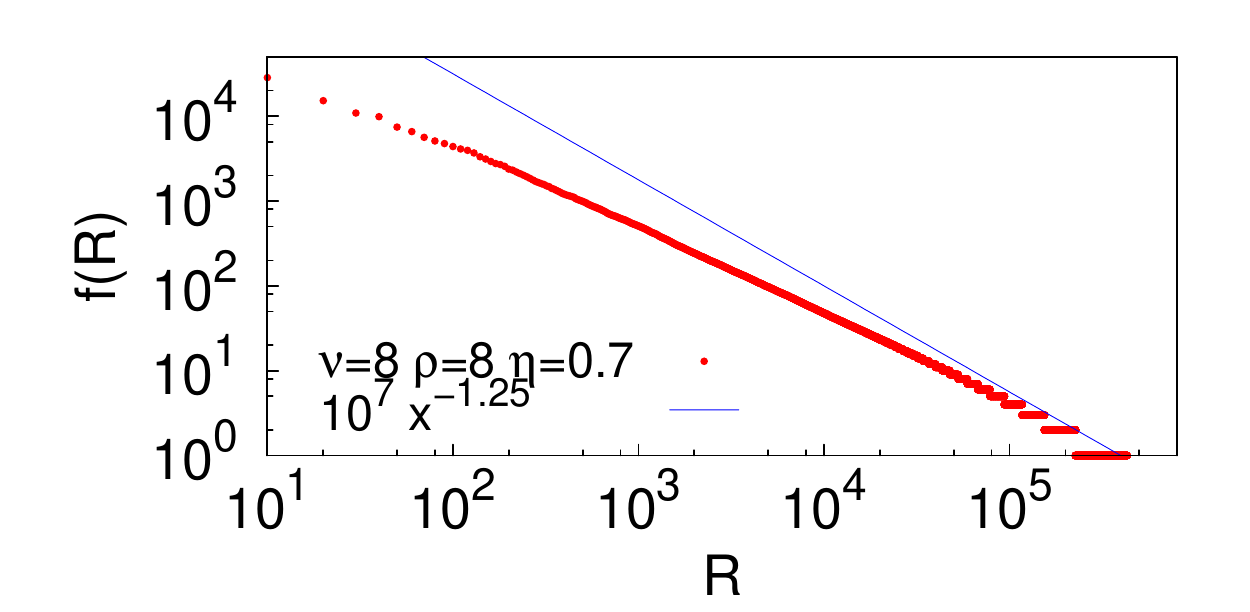}}
\centerline{%
  \includegraphics[width=0.5\columnwidth]{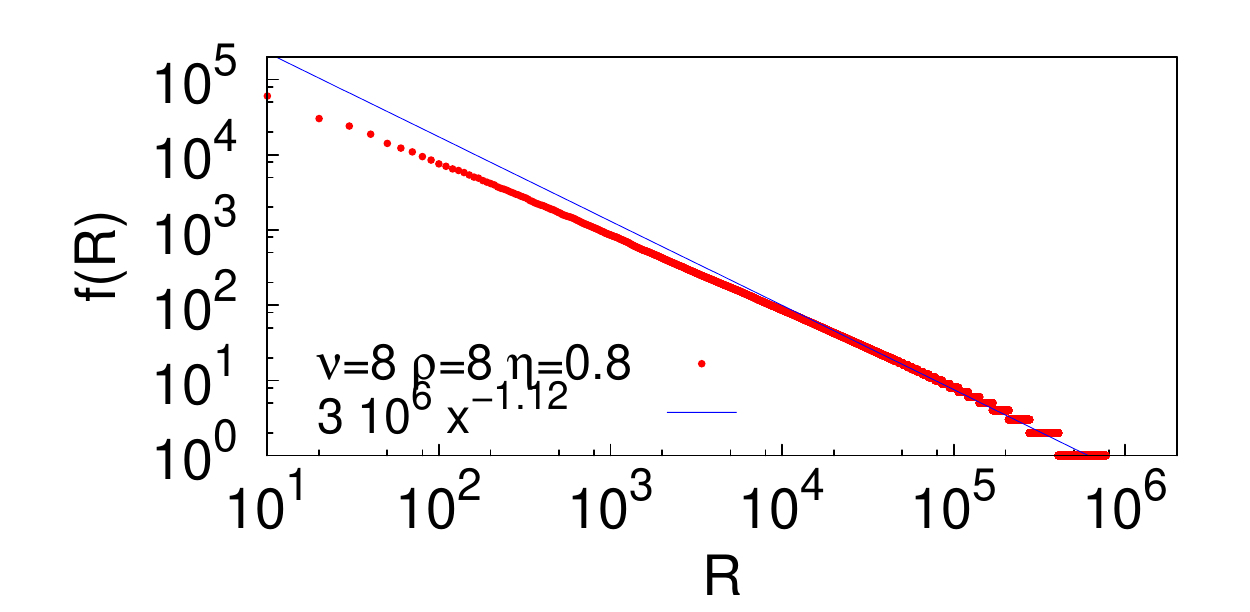}%
  \includegraphics[width=0.5\columnwidth]{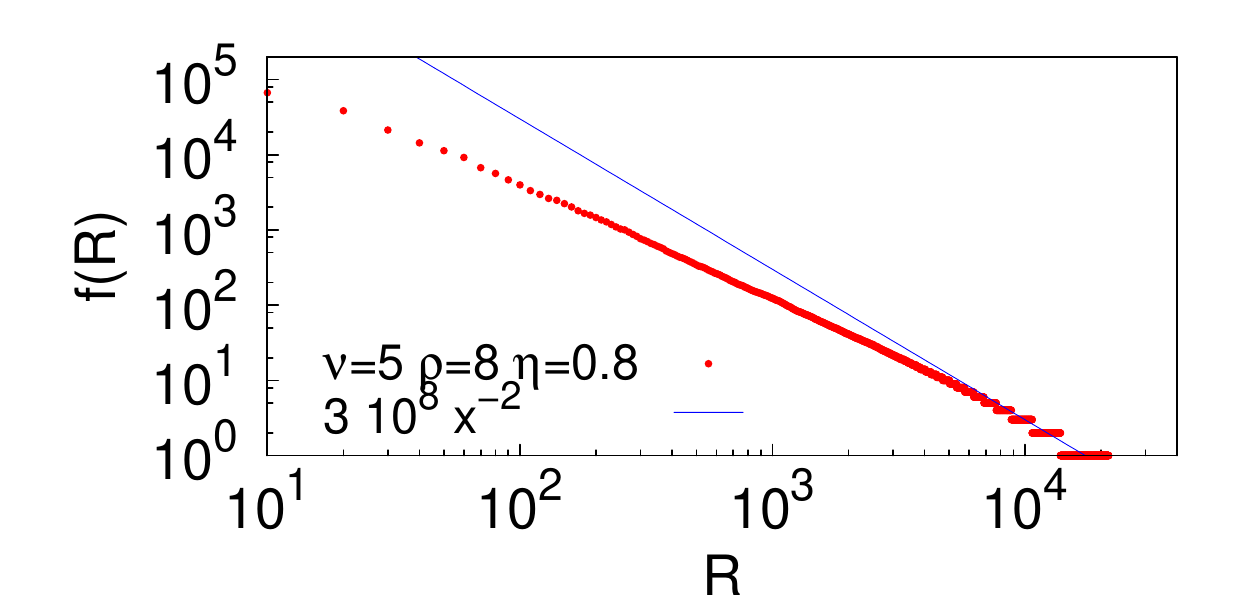}}
\centerline{%
  \includegraphics[width=0.5\columnwidth]{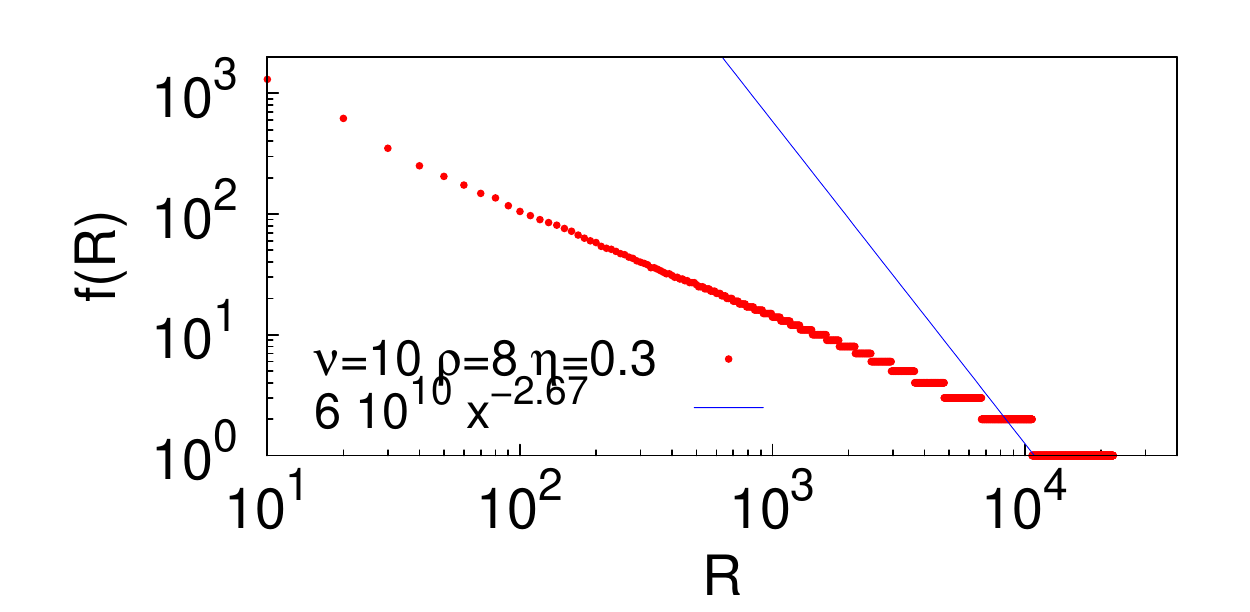}%
  \includegraphics[width=0.5\columnwidth]{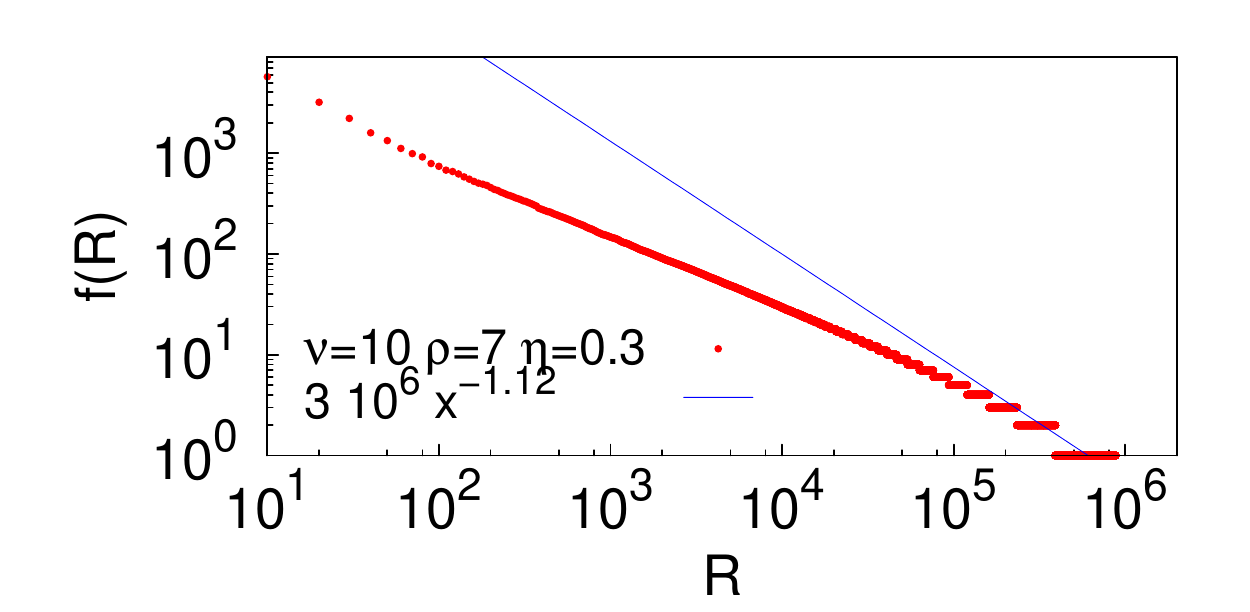}}
\caption{\textbf{Frequency-rank distribution (Zipf's law).} Zipf's law
  for several values of the parameters of the urn model with semantic
  triggering. The exponent $\alpha$ of the tail of the distributions
  is compatible with the exponent $\beta$ of the Heaps' law. It is
  worth remarking how the correspondence gets worst when the exponent
  of the Heaps' law decreases, since in this case one needs extremely
  longer simulations in order to get sufficient statistics on the
  tail. Straight lines show functions of the form $a x^{-1/\beta}$,
  where $a$ is a constant. In all the simulations
  $N_0=\nu+1$.}\label{fig:Zipf_wl}
\end{figure}

Figures~\ref{fig:Heap_wl} and~\ref{fig:Zipf_wl} report numerical
results for the Heaps' and Zipf's laws respectively, for some values
of the parameters of the model $\nu$, $\rho$ and $\eta$. For this
modified model with semantic triggering, the relation between the
exponent $\beta$ of the Heaps' law and the exponent $\alpha = 1/\beta$
of the Zipf's law continues to hold asymptotically, i.e. for large
times, with $\alpha$ measured on the tail of the frequency-rank
distribution. In particular, the time at which the above relation
starts to hold depends on the exponent $\beta$ of the Heaps' law.
Larger times are needed for smaller $\beta$.

We now outline the analysis leading to an estimate for the Heaps' exponent as
a function of the model parameters $\nu$, $\rho$ and $\eta$.  
Observe that if we know the label of the last added element to
the sequence $\mathcal{S}$, say $s$, we can write for the number of
distinct elements $D(t)$ appearing in the sequence ${\mathcal S}$:

\begin{equation}
\frac{dD(t)}{d t}= \frac{N^{s}(t)}{N^{s}(t)+ \eta N^{\bar{s}}(t)} \frac{N_D^{s}(t)}{N^{s}(t)}+ 
\frac{\eta N^{\bar{s}}(t)}{N^{s}(t)+ \eta N^{\bar{s}}(t)} \frac{N_D^{\bar{s}}(t)}{N^{\bar{s}}(t)}=
\frac{N_D^{s}(t)+ \eta N_D^{\bar{s}}(t)}{N^{s}(t)+ \eta N^{\bar{s}}(t)}
\end{equation}

\noindent where $N^{s}(t)$, $N_D^{s}(t)$,   $N^{\bar{s}}(t)$ and
  $N_D^{\bar{s}}(t)$ denote respectively the number of elements with
  label $s$, the number of new (never used in the sequence
  $\mathcal{S}$) elements with label $s$,  the number of elements with
  label different from $s$, and the number of new elements with label
  different from $s$, that are present  in the reservoir $\mathcal{U}$ at time $t$. 

The following relations hold:

\begin{equation}
\nu D(t)=N_D^s(t)+  N_D^{\bar{s}}(t) ~~\mbox{and}~~U(t)=N^s(t)+  N^{\bar{s}}(t),
\end{equation}

\noindent where $U(t)$ is the number of total elements in the
reservoir. It is worth remarking that if $\eta=1$ one recovers Eq.
(\ref{eq:D_complete}).

We now drop the hypothesis of knowing the label of the last
added element, and write a general equation for $D(t)$ of the
form:
\begin{equation}
\frac{dD(t)}{dt}= \sum_k P(k) \frac{N_D^{k}(t)+ \eta
  N_D^{\bar{k}}(t)}{N^{k}(t)+ \eta N^{\bar{k}}(t)}= \sum_k P(k)
\frac{N_D^{k}(t)+ \eta( \nu D(t)- N_D^{k}(t))}{N^{k}(t)+ \eta (U(t) -
  N^{k}(t))} \label{eq:D_label}
\end{equation}
\noindent where the sum is over all the labels $k$ present at time $t$
in the reservoir $\mathcal{U}$ and $P(k)$ is the probability that the last added element 
to the sequence $\mathcal{S}$ at time $t$ had the label $k$.

In order to close the equation (\ref{eq:D_label}), we should estimate
$N^{k}(t)$ and  $N_D^{k}(t)$ for a generic label $k$. 
Let us start by observing that $N_D^{k}(t) \leq \nu +1$, and this term
can be neglected in the large $t$ limit with respect to $D(t)$.

We now leave the more complex problem of estimating $N^{k}(t)$ and we
consider instead the probability $P(n)$ that $N^{k}(t) \equiv n$,
substituting the sum over $k$ in equation (\ref{eq:D_label}) with the
sum over the labels with the same number of occurrences $n$ in the
reservoir. We can thus write (asymptotically):
\begin{equation}
\frac{dD(t)}{dt}= \sum_n P(n) \frac{\eta \nu D(t)}{n (1- \eta) + \eta U(t) }.\label{eq:D_wl}
\end{equation}
We do not explicitly compute $P(n)$, but we consider two opposite limits:

\begin{enumerate}
\item We retain in the sum of equation (\ref{eq:D_wl}) only the terms
  $n \simeq U(t)$. This approximation is sufficiently good when the
  frequency-rank distribution for the elements in $\mathcal{S}$ is
  sufficiently steep, corresponding to a high Zipf's exponent.
  Solving the equation (\ref{eq:D_wl}) within this approximation, we
  obtain the result for the Heaps' exponent $\beta = \min(\frac{\nu
    \eta }{\rho},1)$.

\item When the probability $P(n)$ is large only for $n \ll U(t)$, we
  can neglect in the sum of equation (\ref{eq:D_wl}) the term $n (1-
  \eta)$ with respect to $\eta U(t)$. Solving the
  equation (\ref{eq:D_wl}) within this approximation, we obtain: $\beta
  \simeq \min(\frac{\nu }{\rho},1) $.
\end{enumerate}

Summarizing, we have obtained lower and upper bounds for $\beta$: $
\min(\frac{\nu \eta }{\rho},1) \leq \beta \leq \min(\frac{\nu
}{\rho},1) $, that are satisfied by the simulation results shown in
Figs.~\ref{fig:Heap_wl} and~\ref{fig:Zipf_wl} .

%%%%%%%%%%%%%%%%%%%%%%%%%%%%%%%%%%%%%%%%%%%%%%%%%%
\section{Detecting triggering events\label{sec:triggering}}

As pointed out in the main text, the semantics and the notion of
meaning could trigger non-trivial correlations in the sequence of
words of a text, the sequence of songs listened to, or the sequence of
ideas in a given context. In order to take into account semantic
groups, we introduce suitable labels to be attached to each element of
the sequence. For instance, in the case of music, one can imagine that
when we first discover an artist or a composer that we like, we shall
want to learn more about his or her work. This in turn can stimulate
us to listen to other songs by the same artist. Thus, the label
attached to a song would be, in this case, its corresponding writer.
\begin{figure}[t!]
\centerline{\includegraphics[width=0.9\textwidth]{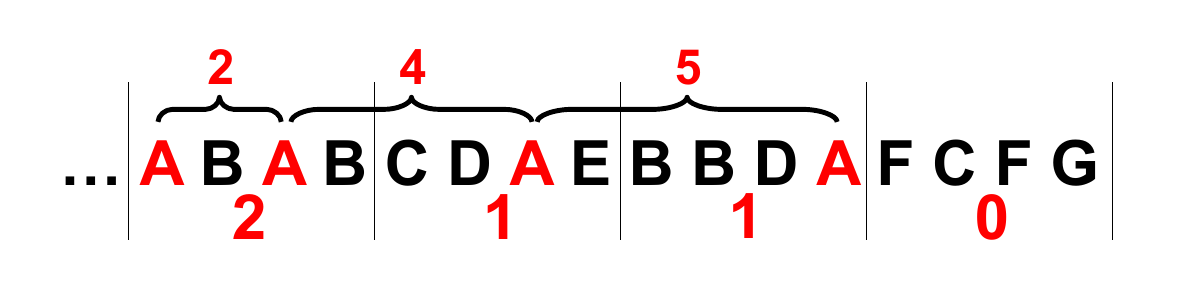}}
\caption{\textbf{Entropy and intervals example.}  Let us indicate with
  the same letters the occurrences, e.g., of lyrics of the same artist
  in the sequence.  Suppose that A has just appeared in the sequence,
  which ends with G.  Thus, A appears 4 times, i.e.,\ $k=4$. We divide
  the subsequence $\mathcal{S}_A \subset \mathcal{S}$ in 4 parts and
  count the occurrences $f_i$ of A in each of them (bottom numbers).
  The normalized entropy of A will be \( S_A(k=4) / \log{4}= (\frac{1}{2}\log 2
  + \frac{1}{4}\log 4 +\frac{1}{4}\log 4+ 0 )/\log 4 = \frac{3}{4} \).
  As a value of $S(k)$ we average all entropies of the elements
  occurring $k$-times in $\mathcal{S}$.  The numbers at the top show
  the length of the inter-times used in the interval distribution
  evaluation.  The \emph{local} reshuffling would shuffle only those
  15 elements occurring after the first occurrence of A, and compute
  the normalized entropy and the time intervals distribution on this
  reduced sequence.
\label{fig:entropy_cartoon}
}
\end{figure}

To detect such non-trivial correlations we define 
the entropy $S_A(k)$ of the sequence of occurrences of
a specific label $A$ in the whole sequence $\mathcal{S}$, as a
function of the number $k$ of occurrences of $A$. To this end we
identify the sub-sequence $\mathcal{S}^{A}$ of $\mathcal{S}$ starting
at the first occurrence of $A$. We divide $\mathcal{S}^{A}$ in $k$
equal intervals and call $f_i$ the number of occurrences of the label
$A$ in the $i$-th interval (see Fig.~\ref{fig:entropy_cartoon}). 
The entropy of $A$ is defined as 
\begin{equation}
S_A(k) = -\sum_{i=1}^{k} \frac{f_i}{k} \log \frac{f_i}{k}.
\label{eq:entropy}
\end{equation}

\noindent In case the occurrences of A were equally distributed among
these intervals, i.e.,\ $f_i=1 \; \forall i=1\ldots k$, $S_A(k)$ would
get its maximum value $\log{ k}$. On the contrary, if all the
occurrences of A were in the first chunk, i.e.,\ $f_1=k$ and
$f_{i\neq1}=0$, the entropy would get its minimum value $S_A(k)=0$.
Each $S_A(k)$ is normalized with the factor $S_A^{max}(k)=\ln{k}$, the
theoretical entropy for a uniform distribution of the $k$ occurrences.
The entropy $S(k)$ is calculated by averaging the entropies relative
to those elements occurring $k$-times in the sequence.

Moreover, we also analyse the distribution of triggering time intervals
$P(l)$. For each label, say $A$, we consider the time intervals
between successive occurrences of $A$. We then find the
distribution of time intervals related to all the labels appearing in
the sequence $\mathcal{S}$ (see also Fig.~\ref{fig:entropy_cartoon}).

In the Wikipedia and Last.fm datasets we can go a step further since
they contain the contribution of many users. In this case we can focus
on a sub-sequence $\mathcal{S}_{unique}$ of $\mathcal{S}$ that
neglects the multiple occurrence of the same element by the same
users, e.g. in Last.fm multiple listening of the same song by the same
users (a specific song can be present anyway several times in the
sub-sequence since that song can be listened for the first time by
different users). We can thus identify for each label, say A, the
sub-sequence $\mathcal{S}_{unique}^{A}$ and correspondingly define the
entropy and the time intervals distribution as described above (see
the following Sections for a detailed discussion of this analysis both
for Last.fm and Wikipedia).

\subsection{Reshuffling methods}

In order to ground the results obtained, both for the entropy and the
distribution of triggering intervals, we consider two suitably defined
ways of removing correlation in a sequence. Firstly, we just globally
reshuffle the entire sequence $\mathcal{S}$. In this way semantic
correlations are disrupted but statistical correlations related to the
non stationarity of the model, responsible for instance for Heap's and
Zipf's law, are still there. Secondly, for each label, we reshuffle
the sequence $\mathcal{S}^{A}$ locally, i.e., from the first
appearance of $A$ onwards. This latter procedure removes altogether
any correlations between the appearance of elements.
\begin{figure}[t!]
\centerline{%
  \includegraphics[width=0.5\columnwidth]{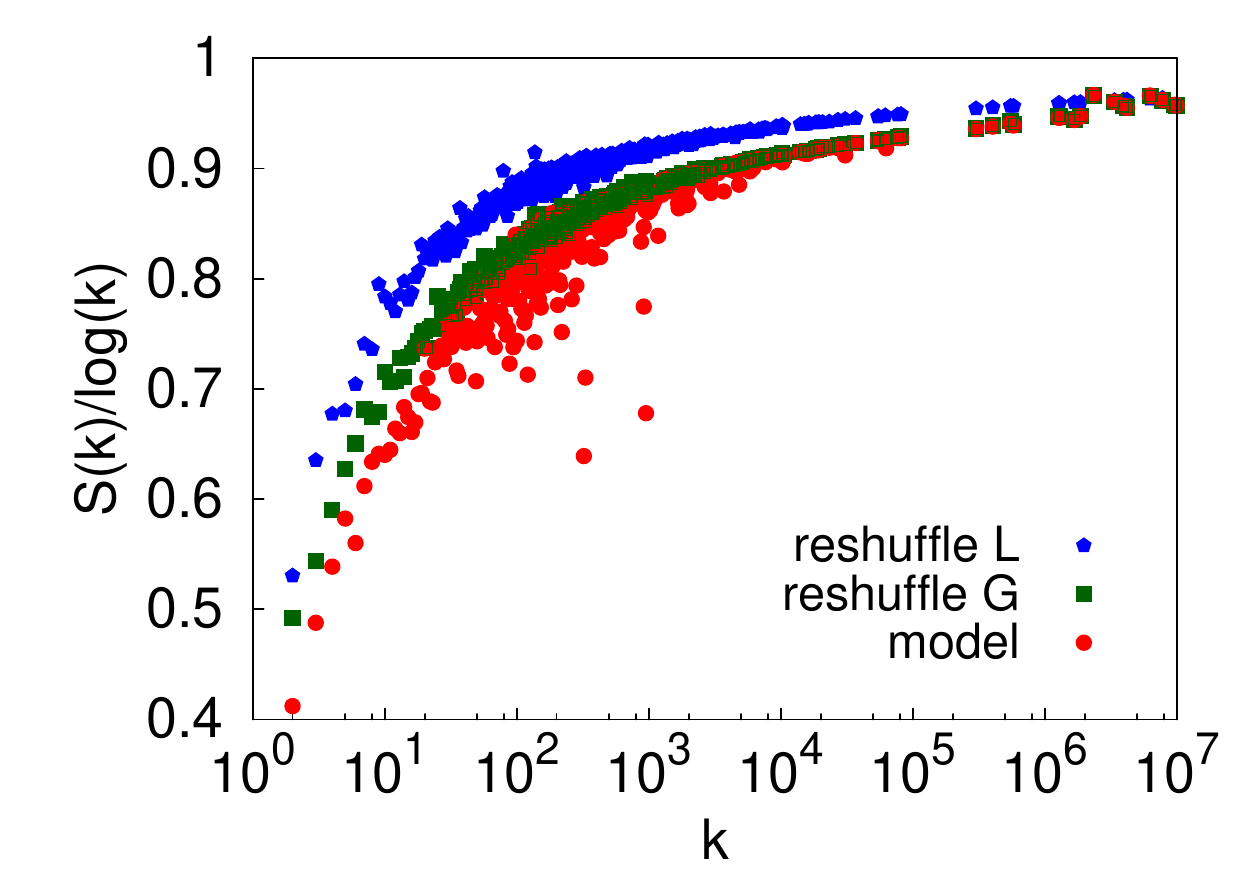}%
  \includegraphics[width=0.5\columnwidth]{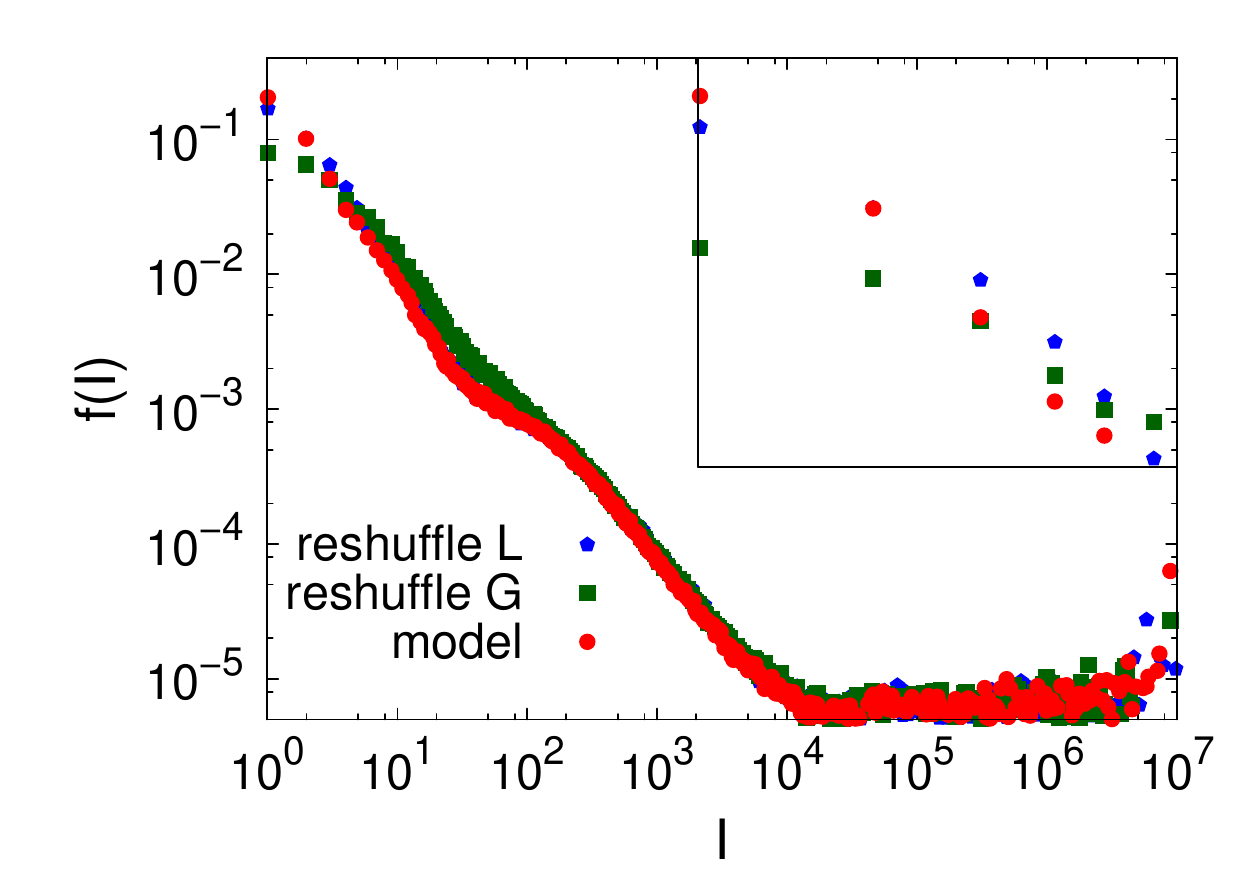}}
\caption{\textbf{Entropy (left) and intervals (right) distribution in
    the random walk model mapping the urn model with semantic
    triggering}. Left: Entropy of a sequence associated to a specific
  label A vs.\ the number of events, k, with that label. The entropy
  is averaged for each $k$ over the labels with the same number of
  occurrences. The plot shows an average over 10 realizations of the
  process with parameters values: $\nu=10$, $\rho=7$, $\eta=0.2$, and
  $N_0=\nu+1$, corresponding to a Heaps' exponent of $\beta=0.29$ (see
  figure~\ref{fig:RW_heaps_and_Zipf}). In each realization the
  sequence $\mathcal{S}$ has length $N=10^7$. Right: Results for the
  time intervals distribution for the same data as for the entropy.
  The color code is red for the actual sequence, green for the global
  reshuffle of the sequence $\mathcal{S}$, and blue for the local
  reshuffle (see text). In the inset a zoom of the first intervals'
  lengths is shown.}\label{fig:RW_triggering}
\end{figure}

\section{The random walk model for the dynamics of novelties}
\begin{figure}[htp]
\centerline{%
  \includegraphics[width=.93\textwidth]{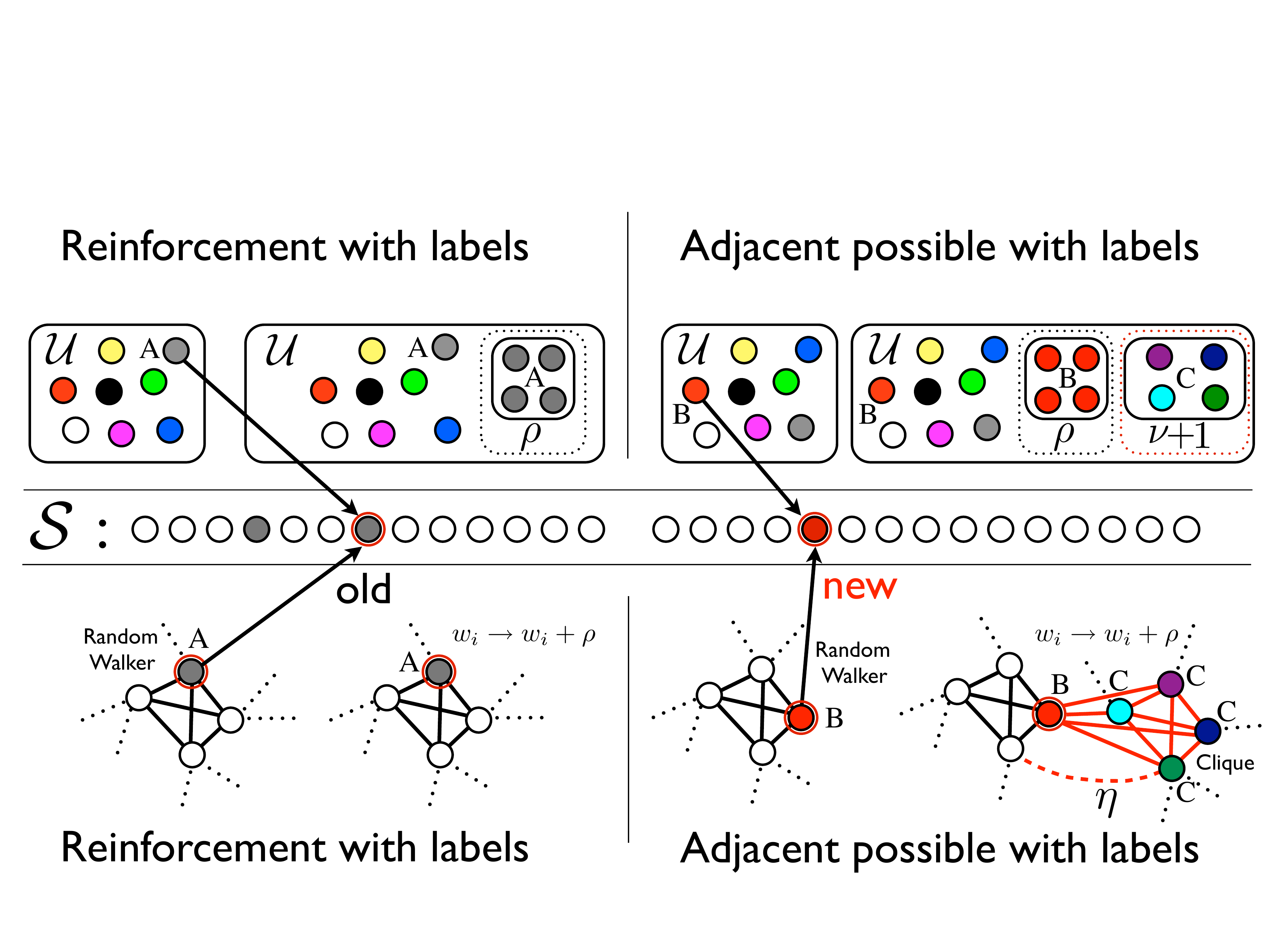}}
\caption{{\bf Models} {\em Top}: scheme of the \emph{urn model with
    semantic triggering}. On the left panel we describe a generic
  reinforcement step of the dynamics, where one element already drawn
  earlier on time is drawn from the urn $\mathcal{U}$ (the gray ball).
  In this case one adds this element to $\mathcal{S}$ (depicted at the
  center of the figure) and, at the same time, put $\rho$ additional
  gray balls to $\mathcal{U}$, all with the same label $A$ of the
  parent gray ball. On the right panel we illustrate a generic
  adjacent possible step of the dynamics. Here, upon drawing a new
  ball (red) from $\mathcal{U}$, $\nu+1$ brand new balls are added to
  $\mathcal{U}$, all sharing a brand new label $C$, along as the
  $\rho$ red balls of the reinforcement step that takes place at each
  time step. {\em Bottom}: scheme of the random walk (RW) based model
  for the dynamics of novelties. Whenever a RW visits an already
  visited node (gray node on the left panel) one adds a gray element
  to $\mathcal{S}$ and reinforce the node's weight according to the
  formula $w_i \rightarrow w_i + \rho$. Whenever the RW visits for the
  first time a node $i$ (red node in the right panel), a new clique
  (representing the newly created adjacent possible) with $\nu+1$
  nodes is added to the graph, all the nodes sharing a brand new label
  $C$. Each node of the clique is connected to the red node, and with
  a probability $\eta$ to the other already existing nodes. At the
  same time one adds the red element to $\mathcal{S}$, always
  reinforcing the node's weight according to the formula $w_i
  \rightarrow w_i + \rho$.\label{fig:RW_cartoon}}
\end{figure}
Our urn
model with triggering, both with and without semantics, can
be mapped in the framework of the exploration of an evolving graph
$\mathcal{G}$ through a random walker (RW). In particular, the RW
dynamics can be constructed as follows (see also
figure~\ref{fig:RW_cartoon}).

\begin{figure}[htp]
\centerline{%
  \includegraphics[width=0.5\columnwidth]{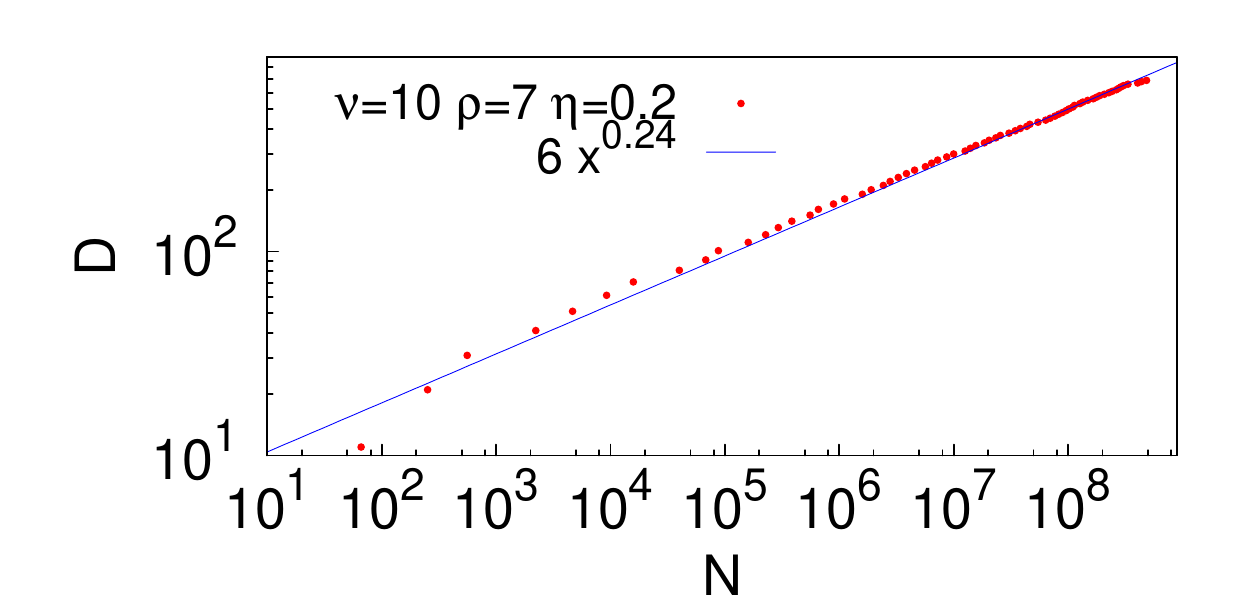}%
  \includegraphics[width=0.5\columnwidth]{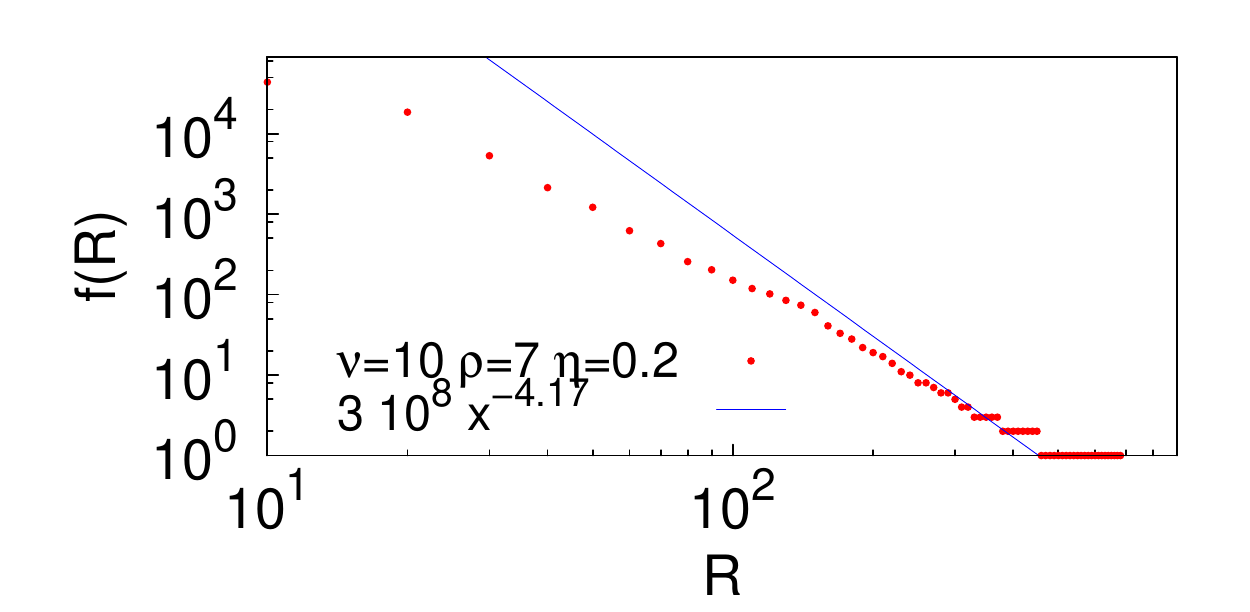}}
\centerline{%
  \includegraphics[width=0.5\columnwidth]{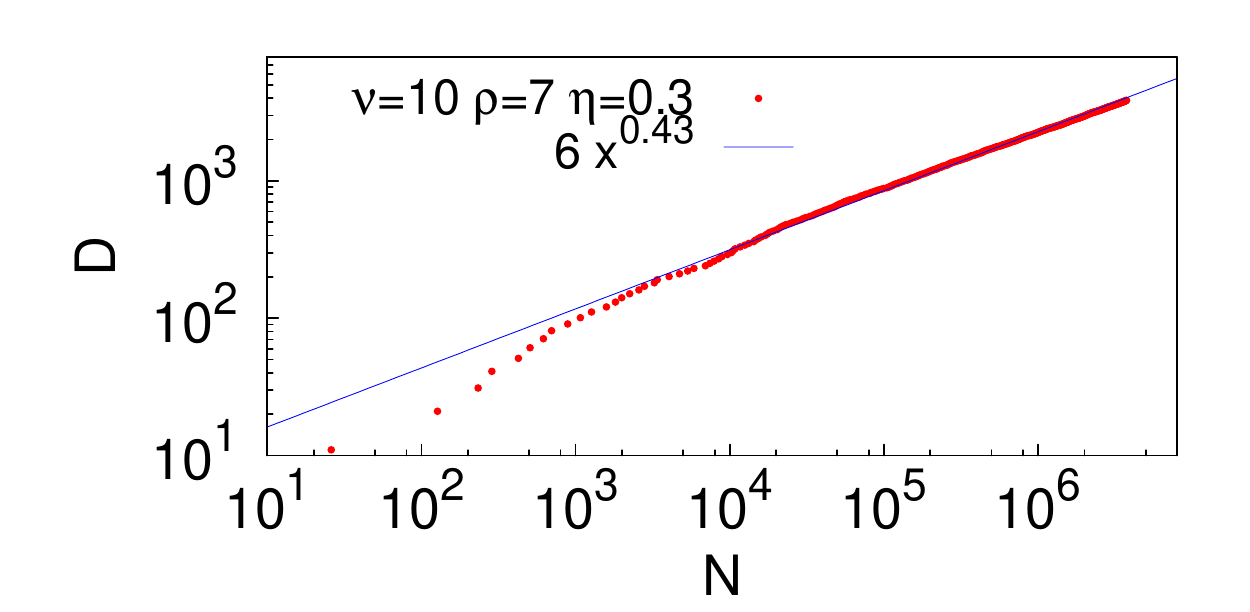}%
  \includegraphics[width=0.5\columnwidth]{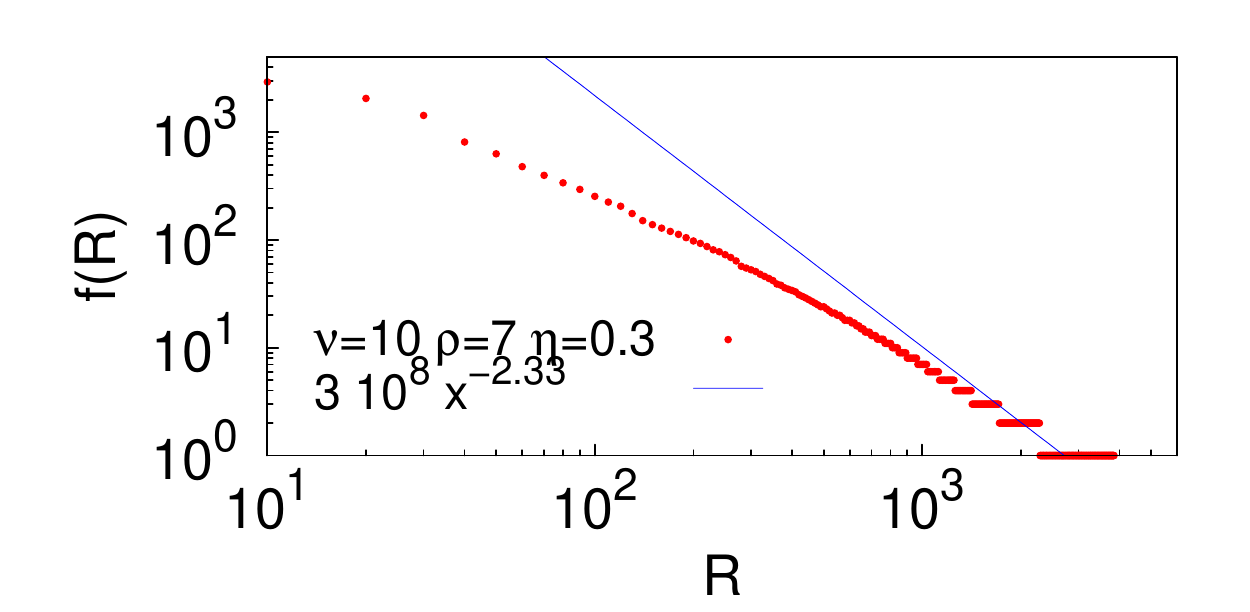}}
\centerline{%
  \includegraphics[width=0.5\columnwidth]{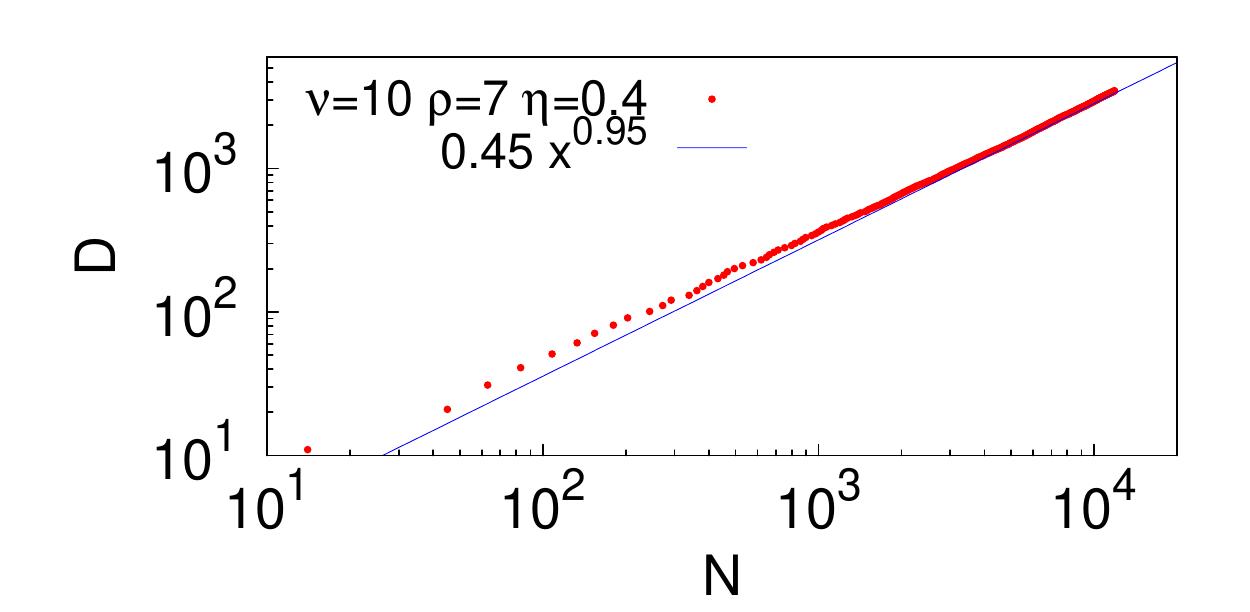}%
  \includegraphics[width=0.5\columnwidth]{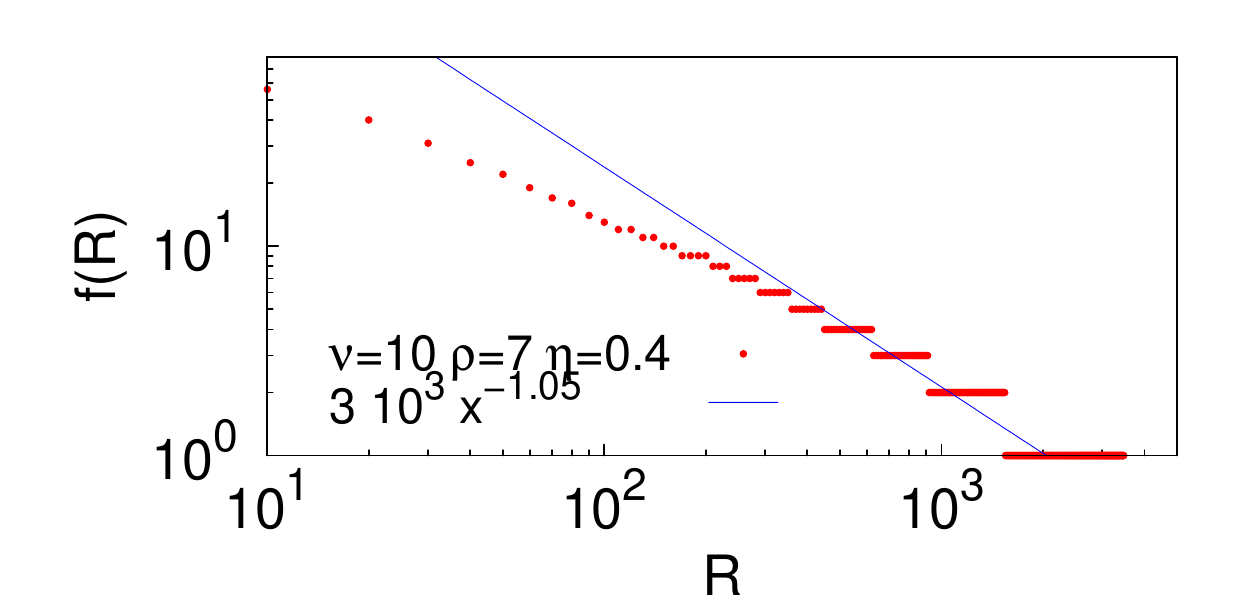}}
\caption{{\bf Growth of the number of distinct elements (Heaps' law)
    and frequency-rank distribution (Zipf's law) for the RW model.}
  Left: Heaps' law for several values of the parameters of the random
  walk model mapping the urn model with semantic triggering. Straight
  lines show functions of the form $a x^{\beta}$, where $a$ is a
  constant. Right: Zipf's law for the
  corresponding values of the parameters of the random walk model. The
  exponent $\alpha$ of the tail of the distributions is compatible
  with the exponent $\beta$ of the Heaps' law. Straight lines show
  functions of the form $a x^{-1/\beta}$, where $a$ is a constant. In
  all the simulations $N_0=\nu+1$.}\label{fig:RW_heaps_and_Zipf}
\end{figure}

We start with a graph $\mathcal{G}$ of $N_0$ nodes, divided in
$N_0/(\nu+1)$ cliques, each node in the same clique sharing a common
label. We then draw a link between each pair of nodes belonging to
different cliques with probability $\eta \leq 1$. Starting with the RW
in a random position, and with a weight $w_j=1$ for each node $j$, at
each time step:
\begin{itemize}
\item[(i)] move the RW to a neighbour node or keep it on the present
  node (self-loops allowed) with a weight-dependent probability;
\item[(ii)] reinforce the selected node weight $w_i \rightarrow w_i+ \rho$; 
\item[(iii)] if the node visited is new (i.e., it is visited for the
  {\em first time}) add a clique with $\nu+1$ new nodes connected to
  the just visited node, each node in the new clique sharing a common
  label, different from all the preexisting ones. In addition draw a
  link between each node in the newly added clique and all the
  preexisting nodes of the network with probability $\eta$.
\end{itemize}
\noindent If $\eta=1$ this model maps one-to-one to the urn model with
triggering introduced in the main text. When $\eta <1$ the
correspondence with the {\em urn model with semantic triggering} is
not one-to-one: in the case of the graph the connections between two
nodes are fixed (or {\em quenched}), i.e. either they are there or
they are not, whether the possibility of going from one element to
each of the others in the urn model is always probabilistic (one can
imagine that this corresponds to an {\em annealed} version of the
graph model, where links are continuously re-drawn according to a
fixed probability). Despite this difference, the statistical
properties of the two models turn out to be equivalent from a
qualitative point of view also in the case $\eta <1$. In
figure~\ref{fig:RW_heaps_and_Zipf} we report some examples of the
Heaps' and Zipf's laws for the RW model, for different values of the
parameters $\nu$, $\rho$ and $\eta$, while in
figure~\ref{fig:RW_triggering} we give an example of the triggering
events as measured by the entropy $S$ associated to the labels and the
distribution $f(l)$ of triggering time intervals between two
successive appearance in the sequence $\mathcal{S}$ of the same label
(see Section~\ref{sec:triggering}).

As a final remark, we note that the RW modeling scheme allows one to more
naturally extend the structure of the semantic relations between the
different elements. The semantic relations are in fact encoded
in the growing graph topology, and one can imagine different ways
of linking the new nodes, corresponding to more complex and realistic
semantic structures.

%\include{SM_triggering_events}

%%%%%%%%%%%%%%%%%%%%%%%%%%%%%%%%%%%%%%%%%%%%%%%%%%%%%%
\section{Details of the datasets used}

\subsection{Gutenberg Corpus}
The corpus of English texts used in the analysis was collected by a
crawl of the material available at the Gutenberg Project ebook
collection~\cite{gutenberg}. The crawl was carried on February 2007
and resulted in a set of about \ 7500 non-copyrighted ebooks in plain
ASCII format. After a filtering procedure used to remove from the
analysis all non-English text we came up with ca.\ 4600 texts, dealing
with diverse subjects and including both prose and poetry. In total,
the corpus consisted of about \ $2.8\times 10^8$ words, with about \
$5.5\times 10^5$ different words. In the analysis we ignored
capitalization. Words sharing the same lexical root were considered as
different, i.e.,\ the word \emph{tree} was considered different from
\emph{trees}. Homonyms, as for example the verbal past perfect
\emph{saw} and the substantive \emph{saw}, were treated as the same
word.
%% Moreover, words with the same root but with different spellings
%% (eg.\ singular and plural form of substantives) were considered as
%% different, while synonyms as the same word (eg.\ \emph{saw}, either
%% intended as the pp.\ of \emph{to see} or as the carpenter tool).
The aggregated analysis is performed by putting all the books in a
random order one after the other in a single text. The texts used in
the non aggregated analysis are listed in Table~\ref{tab:texts}.
\begin{table}
	\centering
	\footnotesize
	\begin{tabular}{|c|c|c|c|c|c|}
		\hline 
		Author & Work & Total nr of words & Nr of distinct words & $\alpha$ & $\beta$\\ 
		\hline
		C.~Dickens		& Hard Times				& 124109	& 8747 & 1.17 & 0.58\\
		C.~Dickens		& David Copperfield	& 426904	& 14026 & 1.43 & 0.53\\
		C.~Dickens		& Oliver Twist				& 191395	& 10177 & 1.30 & 0.55\\
		H.~Melville		& Moby-Dick 					& 252571	& 17136 & 1.22 & 0.60\\
		S.~Butler		& Odyssey (prose)		& 131444	& 6363 & 1.51 & 0.50\\
		A.~Pope			& Odyssey (verse)		& 132461	& 8292 & 1.37 & 0.50\\
		Homer				& Odyssey						& 86868		& 17506 & 1.03 & 0.70\\
		Homer				& Iliad 								& 112082	& 21853 & 1.05 & 0.68\\
		\hline
	\end{tabular}
	\caption{\textbf{Texts from the Gutenberg site used in the
      non-aggregated analysis.} For each text we report the total
    number of words, total number of \emph{distinct} words and the
    estimated values of the (minus) the Zipf's exponent and Heaps'
    exponent. Note that $1/\alpha>\beta$ since the single texts are
    not sufficiently long to allow the asymptotic regime to be
    visible, and the  frequency-rank distribution curve	has not yet
    gone through the crossover visible around $10^4\sim 10^5$ in the
    analogous curve of the whole Gutenberg dataset, showed in the main
    article. 
    % dictionary growth is able to sample well the tail of
    % the % frequency-rank distribution, while the frequency-rank
    % distribution curve has not yet gone through the crossover
    % visible around $10^4\sim 10^5$ in the analogous curve of
    % the % whole Gutenberg dataset, showed in the main article.
		\label{tab:texts}
 	}		
\end{table}

\subsection{Delicious}
Delicious~\cite{delicious} is an online social annotation platform of
bookmarking where users associate keywords (tags) to web resources
(URLs) in a post, in order to ease the process of their retrieval. The
dataset used for the present analysis \cite{cattuto2007heaps} consists
of approximately $5\times 10^6$ posts, comprising about 650,000 users,
$1.9\times 10^6$ resources and $2.5\times 10^6$ distinct tags (for a
total of about \ $1.4\times 10^8$ tags), and covering almost 3 years
of user activity, from early 2004 up to November 2006.
%In processing the data, we discarded all posts containing no tags (about 7\% of the total). 
Since \textit{Delicious} is case-preserving but not case sensitive, we
ignored capitalization in tag comparison, and counted all different
capitalization of a given tag as instances of the same lower-case
tag. The time stamp of each post was used to establish post ordering
and determine the temporal evolution of the system.

In the non-aggregated analysis we extracted from the Delicious dataset
the posts of the three most active users (RangerRick, hidekii,
PeterPeter) and two random ones (Vitelot, AndreaB).

\subsection{Last.fm}
Last.fm \cite{last.fm} is a music website equipped with a music
recommender system. Last.fm builds a detailed profile of each user's
musical taste by recording details of the songs the user listens to,
either from Internet radio stations, or the user's computer or many
portable music devices. The data set we used~\cite{lastfm_data,Celma}
contains the whole listening habits of 1000 users till May, 5th 2009,
recorded in plain text form.
% In its compressed version it occupies only 650\,MB of disk space,
% while in its full uncompressed form about \ 2.5\,GB.
It contains about \ $1.9\times 10^7$ listened tracks with information
on user, time stamp, artist, track-id and track name.

For the non-aggregated analysis we consider only the data of the five
most active listeners.

\subsection{English Wikipedia}
The English Wikipedia database we analyzed consists of 323 compressed
files summing up to a total of 48\,GB of disk space. The uncompressed
overall size is around 20\,TB. The Wikipedia database we
collected~\cite{wikipedia.en},
dates back to March 7th, 2012. \\
Due to the database huge dimension, we had to develop a special
procedure to extract the information we needed. The computer we used
to process the database is a multi-core machine mounting 8 Intel(R)
Xeon(R) X3470 CPU, with a 2.93\,GHz working clock frequency, with
a RAM of 16\,GB.\\
The database contains a copy of all pages with all their edits in
plain text by using the XML structure.
%Entries are arranged without a particular order, whereas
%for a given page its edits are sorted chronologically.\\

In order to perform the analysis related to the detection of
triggering events, we extracted from the database the following
information. First of all, we identified for each new born page, say
$B$, the page, say $A$, that internally linked the new born page for
the first time. We call the page $A$ the {\em mother page} of $B$ and
we identify for each edit its mother page as its label (note that
several edits can have the same mother page, i.e., the same label). We
then follow the steps below:
\begin{itemize}
\item[(1)] To each edit event we associate: (i) the wikipedia page
  exclusive identification number (ID), (ii) the user (wikipedia
  contributor) ID (UID), (iii) the edit ID (EID), (iv) its time stamp
  (TS), (v) the PID of its mother page;
\item[(2)] from the list of all edits endowed with the information
  discussed in (1), we removed the multiple edits of the same page
  done by the same user, retaining his/her first edit;
\item[(3)] we sorted the list (2) according to increasing time stamp.

\end{itemize}

For the non-aggregated analysis we focused on seven randomly chosen
editors. Special care was needed to understand whether a selected user
was human. In fact, the most active editors of Wikipedia are robots
performing minor changes routinely.

\begin{table}
	\centering
	\footnotesize
	\begin{tabular}{|c|c|c|c|}
		\hline 
		User ID & Total Nr of edits & Nr of distinct edits & $\alpha  $ \\ 
		\hline
		1188594		& 14613	& 8619 	& 0.45 \\
		1638938		& 6776	& 3094	& 0.56 \\
		23958			& 19226	& 7295 	& 0.70 \\
		281454		& 1480	& 974 	& 0.41 \\
		2829979		& 11642	& 4622 	& 0.50 \\
		356300		& 10415	& 3738 	& 0.83 \\
		62662			& 6118	& 975 	& 1.06 \\
		82835			& 937852	& 716418	& 0.41 \\
		99037			& 128802	& 78961  & 0.57  \\
		\hline
	\end{tabular}
	\caption{\textbf{Editors of Wikipedia used in the
      non-aggregated analysis.}  For each editor we report: the
    total number of edited articles; the total number of
    \emph{distinct} edited articles; the observed values
    $\alpha$ of the  (minus) the Zipf's exponent.
    The  values of the Heaps' exponent for all the considered users turn
    out to be $\beta \simeq 1$, in agreement with the alpha values $\alpha
    \leq 1$ as predicted by the model.}
		\label{tab:editors}		
\end{table}

\section{Results for non aggregated data}

The analysis performed in the main text, involving the previously
described datasets as a whole, is here repeated for some of their
selected records. In case of the Gutenberg dataset, we chose texts; in
Wikipedia, Last.fm and Delicious, we chose editors, listeners and
tagging users respectively.

\begin{figure}[htp]
\centerline{%
	\includegraphics[width=0.5\textwidth]{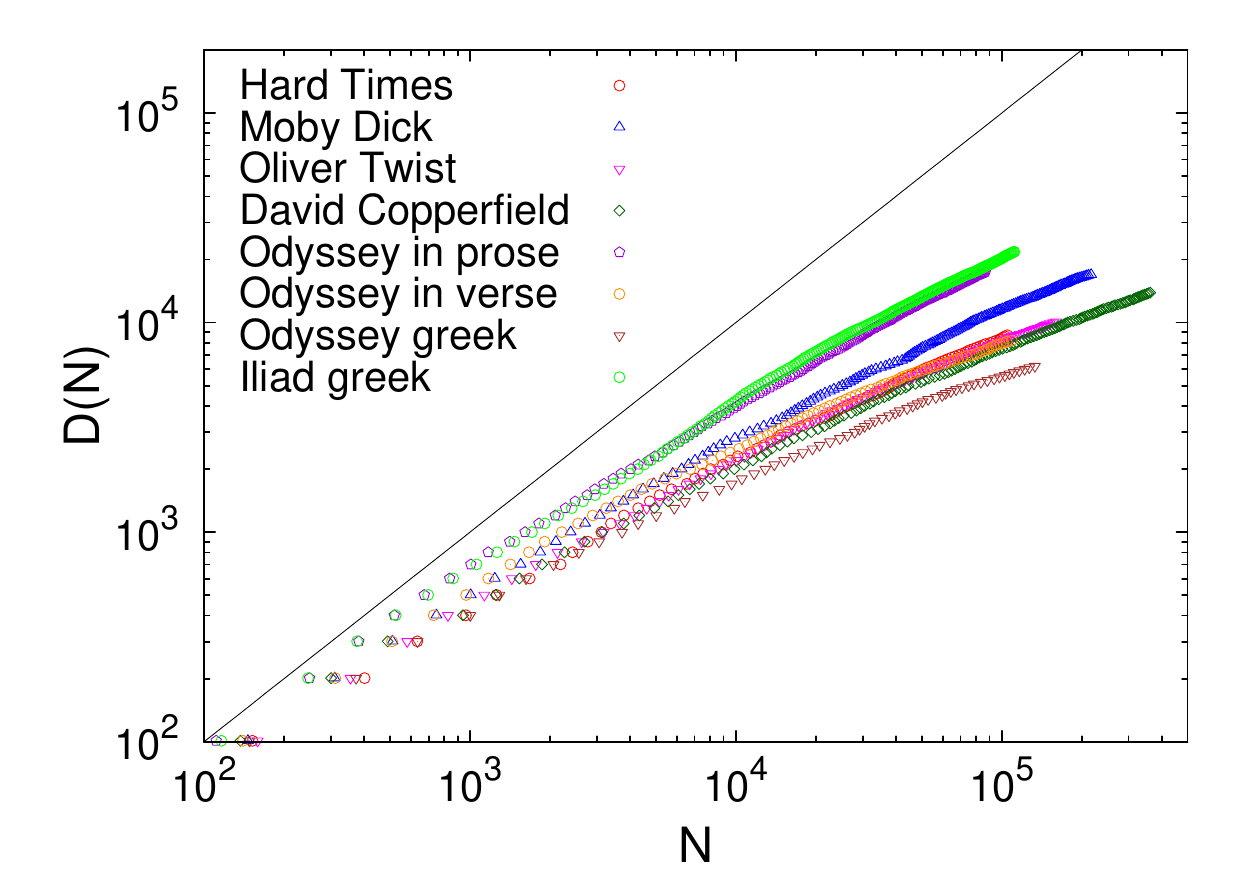}%
	\includegraphics[width=0.5\textwidth]{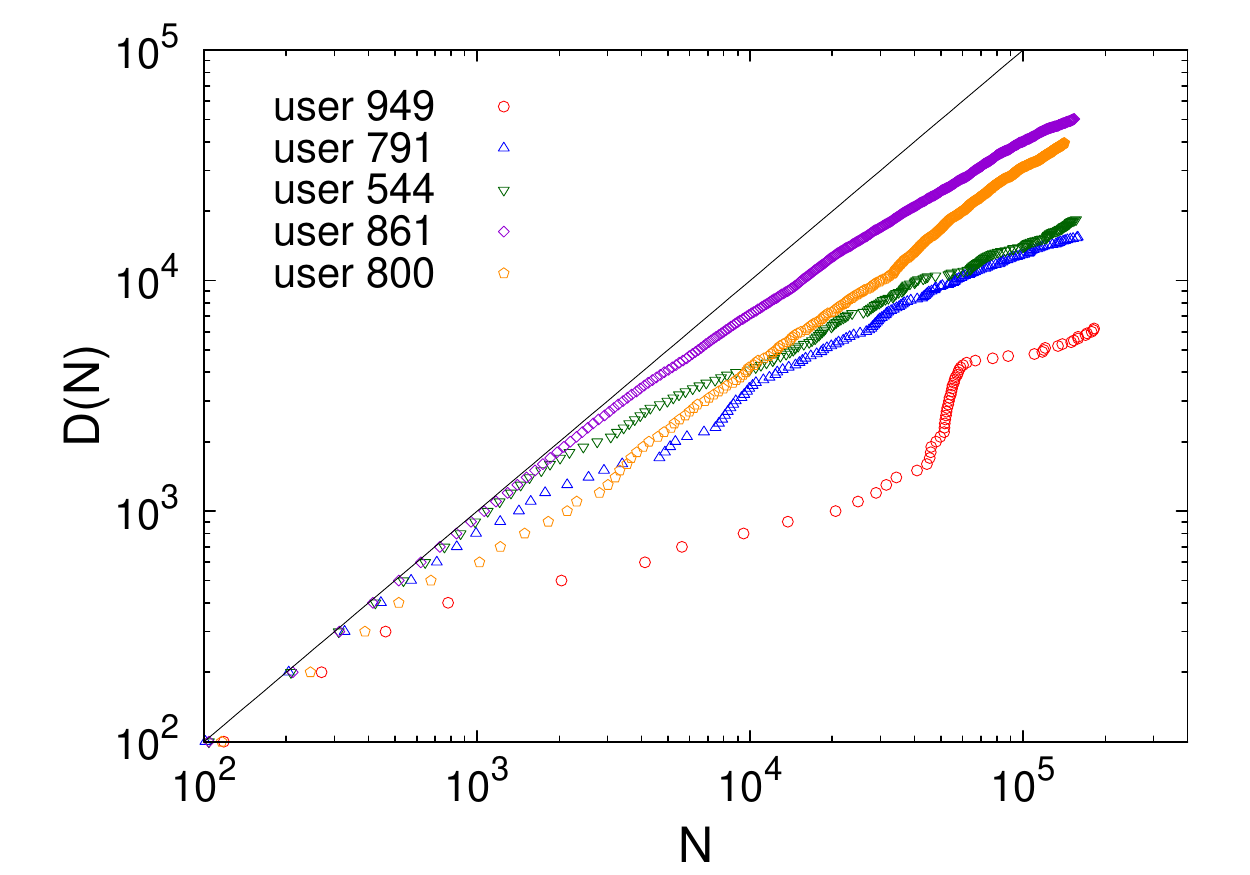}}
\centerline{%
	\includegraphics[width=0.5\textwidth]{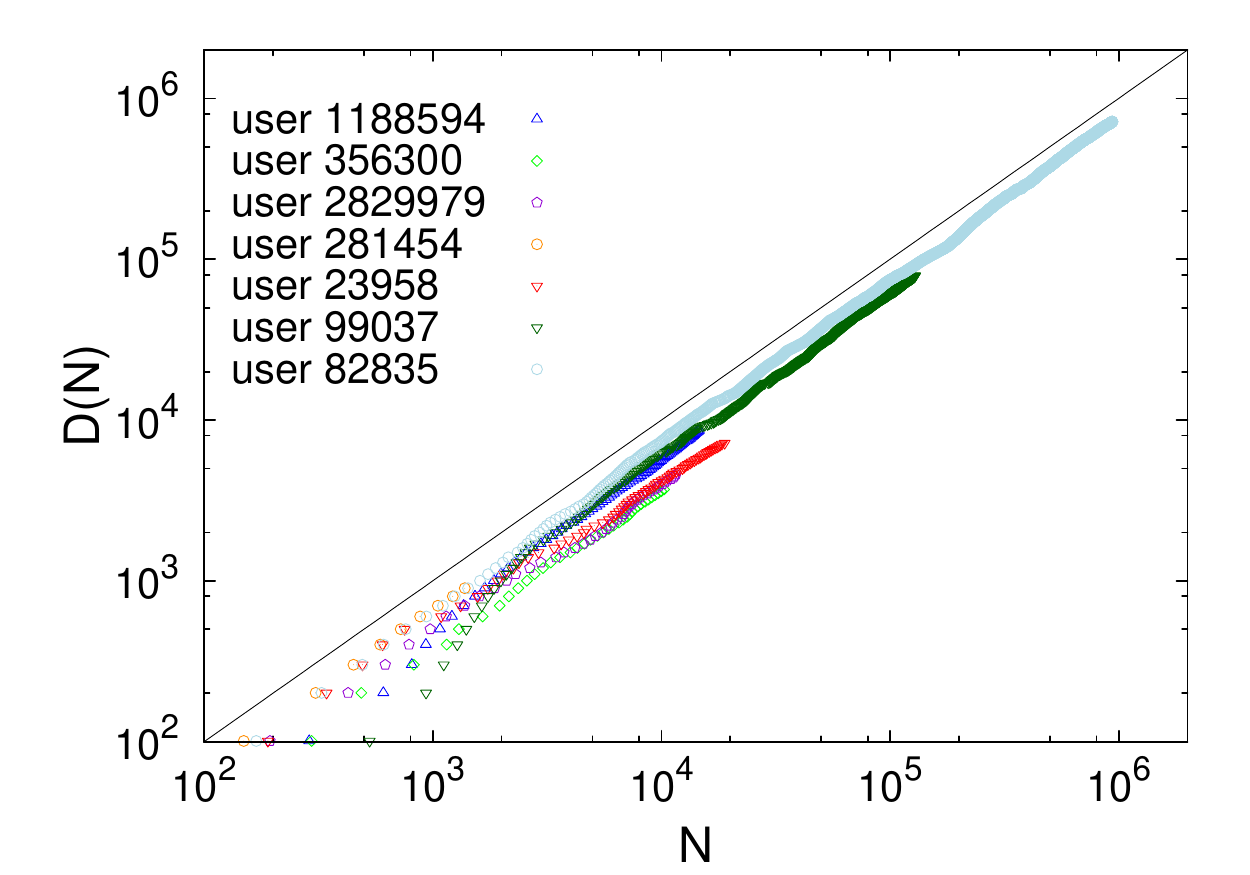}%
	\includegraphics[width=0.5\textwidth]{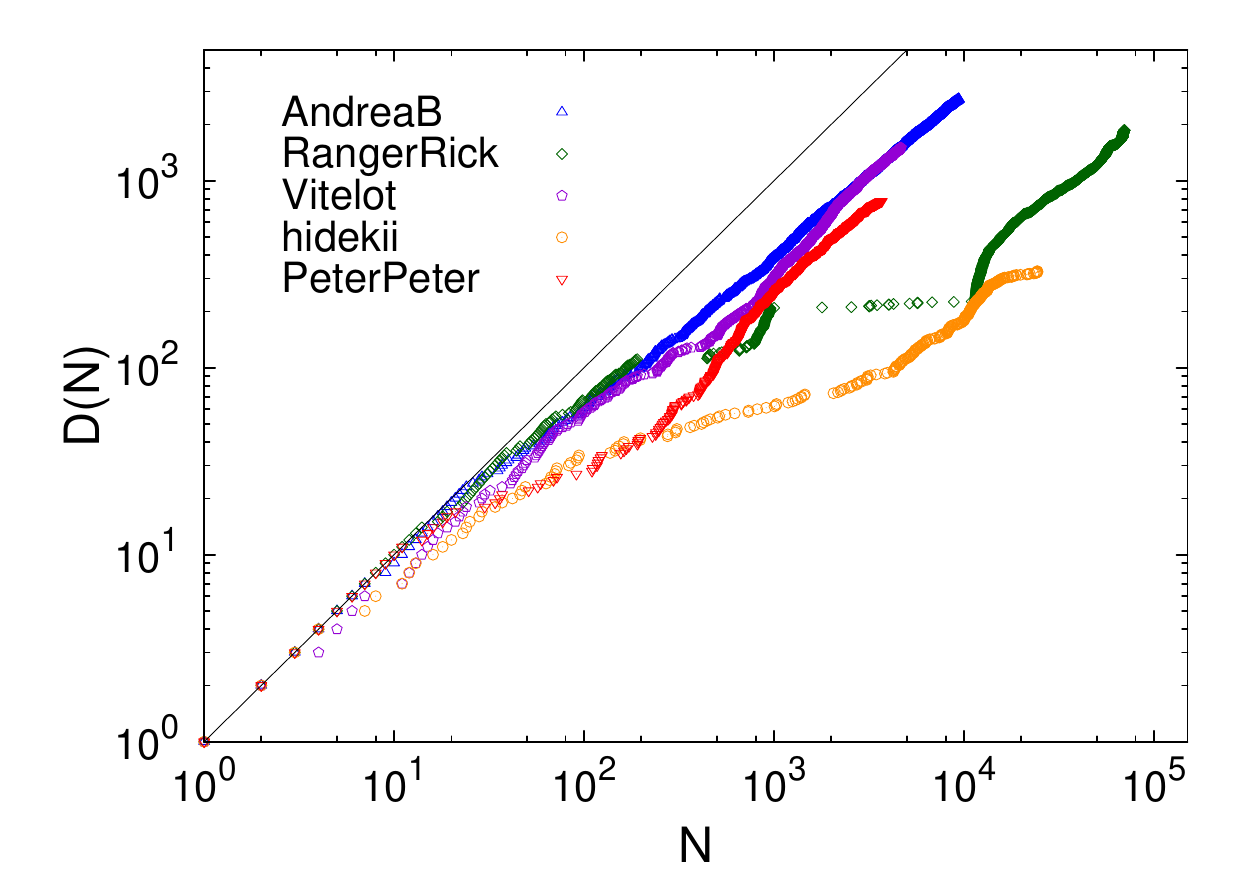}}
\caption{
	\textbf{Growth of the number of distinct elements (Heaps' law).}
	Top-left: Selected masterpieces from the Gutenberg dataset (words as elements);
	Top-right: most active users in Last.fm (lyrics as elements);
	Bottom-left: selected (human) random editors of Wikipedia with appreciable activity (wiki-articles as elements);
%% 	\vito{se fosse a crescita lineare, verificare che $\alpha<1$ e consistente con la teoria, ma ad occhio pare che sia cos\`i. CHE FIGO!!!}
	%
	Bottom-right: Selected users of Delicious (tags as elements). The
  linear growth is indicated by the straight line. The discontinuities
  in both right panels can be ascribed to a data import from other
  sources (local playlists to Last.fm, local bookmarks to Delicious).
\label{fig:heaps_non_aggregated}
}
\end{figure}

\begin{figure}[htp]
\centerline{%
	\includegraphics[width=0.5\textwidth]{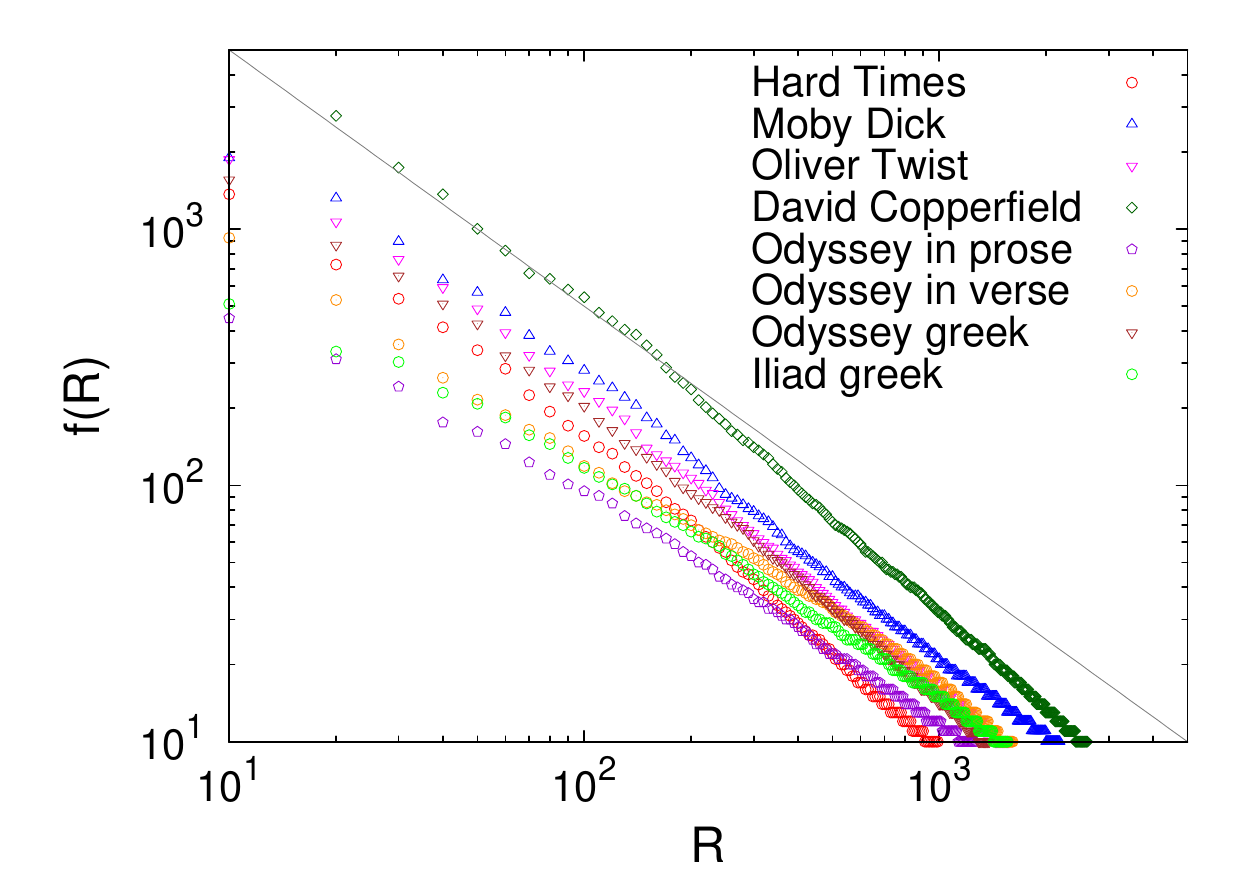}%
	\includegraphics[width=0.5\textwidth]{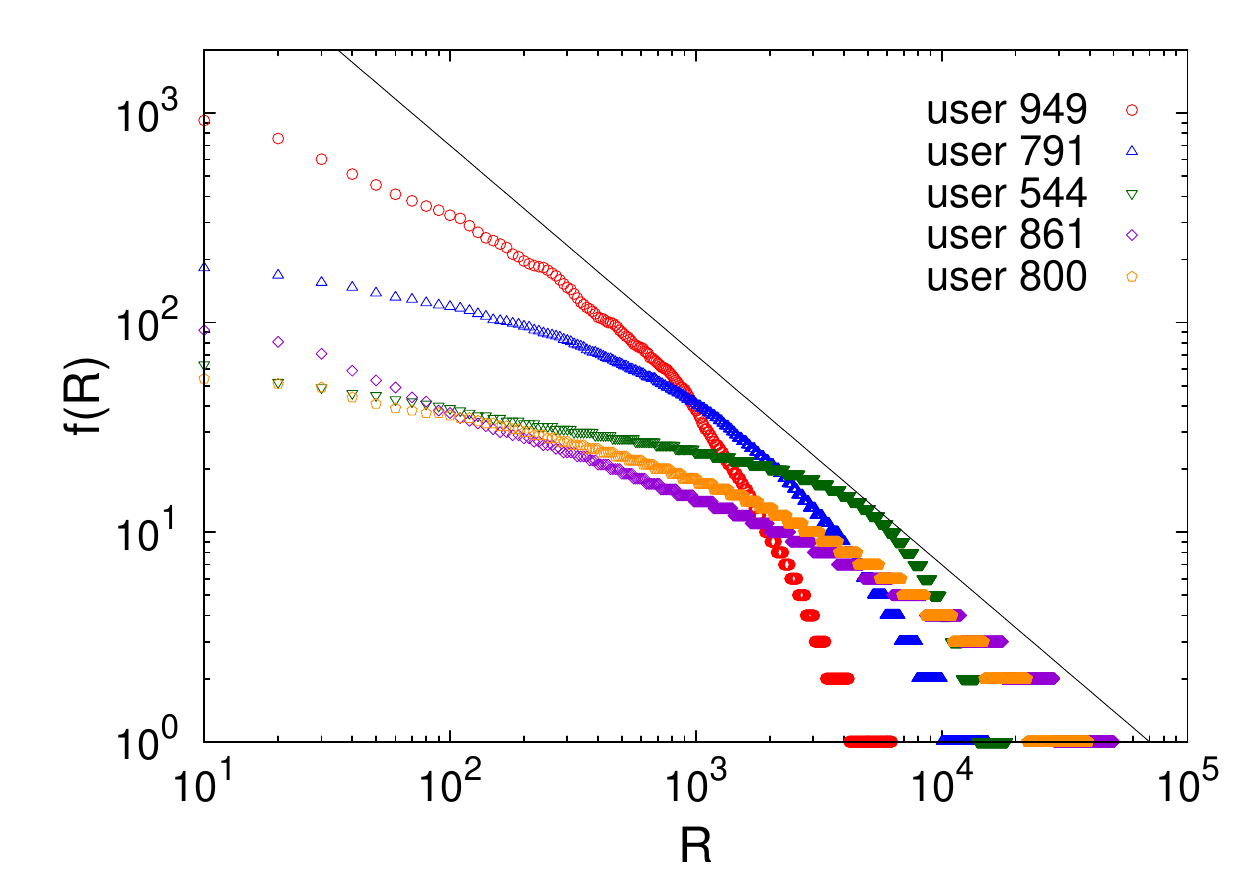}}
\centerline{%
	\includegraphics[width=0.5\textwidth]{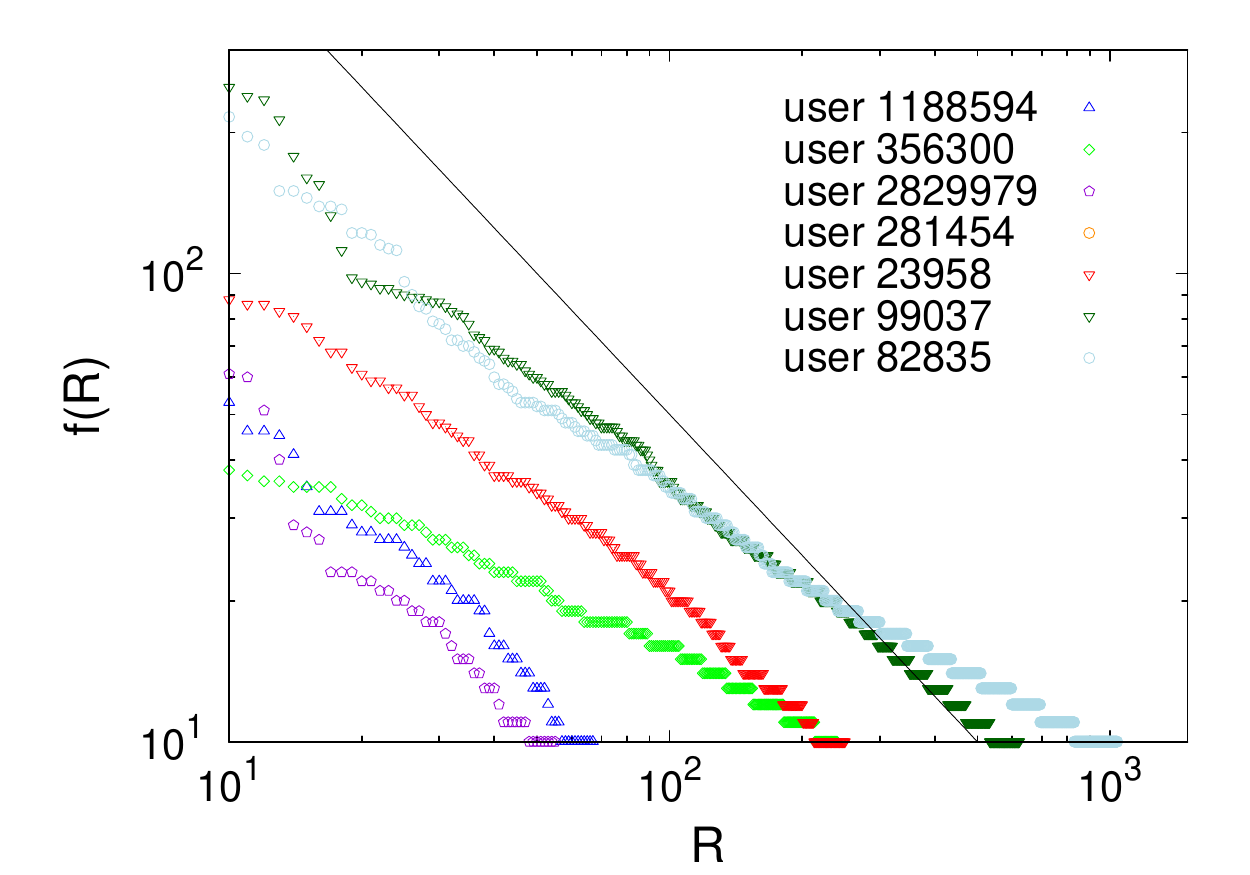}%
	\includegraphics[width=0.5\textwidth]{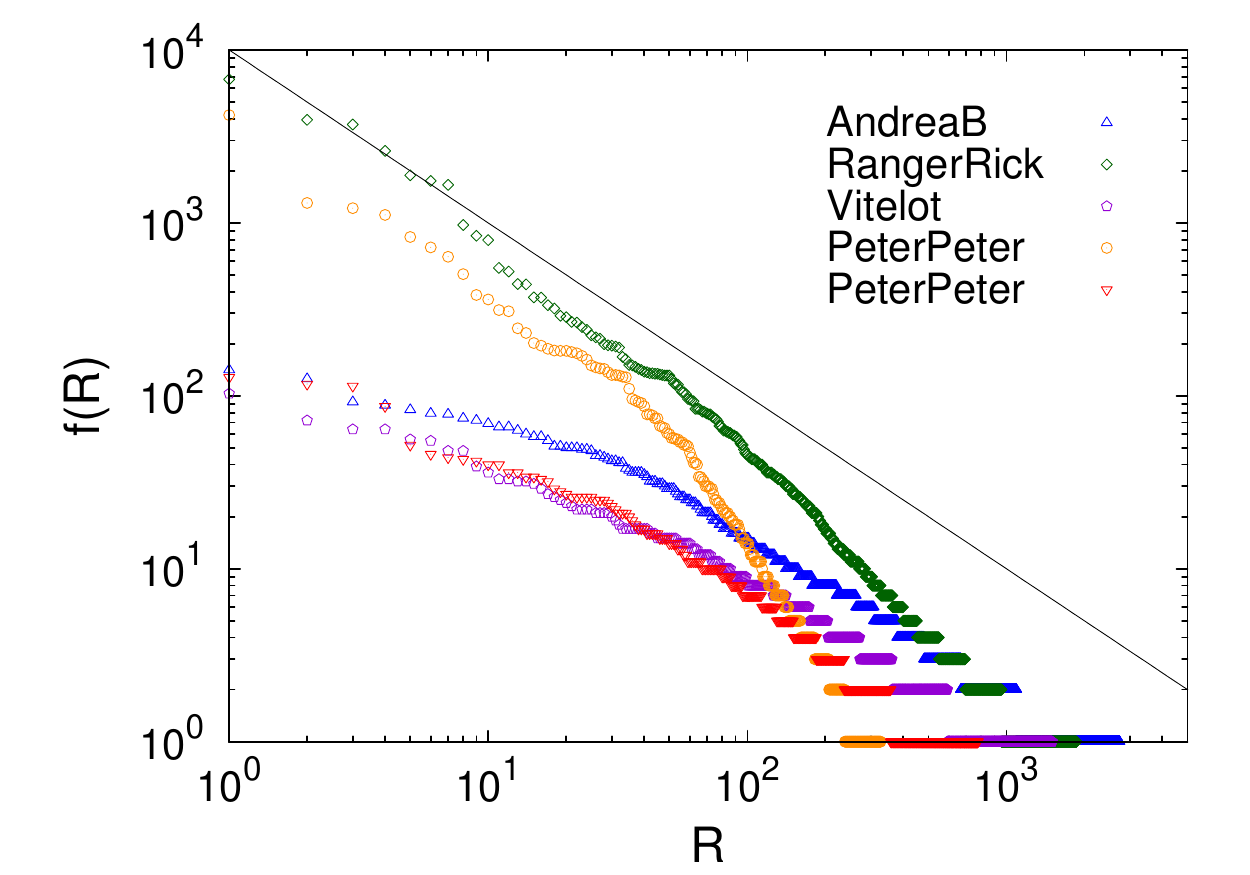}}
\caption{{\bf Frequency-rank distribution (Zipf's law).} 
	Top-left: Selected masterpieces from the Gutenberg dataset (words as elements);
	Top-right: most active users in Last.fm (lyrics as elements);
	Bottom-left: selected (human) random editors of Wikipedia with appreciable activity (wiki-articles as elements);
	Bottom-right: Selected users of Delicious (tags as elements).
	The straight line shows the strict Zipf's law with $\alpha=1$ as a guide for the eye.
\label{fig:zipf_non_aggregated}
}
\end{figure}

\paragraph{Heaps' and Zipf's law\\}
The analysis of Heaps' law is displayed in
Fig.~\ref{fig:heaps_non_aggregated} and shows an asymptotic sublinear
power-law behaviour in the case of texts (see Table~\ref{tab:texts})
and a possible linear behavior for Wikipedia editors (see
Table~\ref{tab:editors}). In the case of Last.fm and Delicious, the
sublinear behavior can still be spotted but the dictionary curves are
less smooth than those of Wikipedia and Gutenberg. The reason is that
in both Last.fm and Delicious, users may import large blocks of music
tracks and web-site bookmarks from their local storage, thus
introducing a sort of discontinuity in time.
This discontinuity is obviously less appreciable in
figure~\ref{fig:zipf_non_aggregated}, were we show the frequency-rank
distribution of words in selected texts, lyrics in selected listeners
using Last.fm, wiki-articles for selected editors in Wikipedia and
tags for selected users of Delicious. In fact, the frequency-rank is
insensible to the temporal ordering of the elements, being a global
statistical property of the sample. Note how the more inflected
ancient Greek language results in a smaller Zipf's exponent than that
of English texts and correspondingly in a larger Heaps' exponent (see
Table~\ref{tab:texts}). It is also worth noting that the measured
exponent $\beta$ of the Heaps' law in the selected texts does not
happen to be the reciprocal of the measured Zipf's exponent $\alpha$.
In the main text we have shown that the frequency-rank curve of the
whole Gutenberg corpus displayed two main behaviors with different
exponents (an analogous observation was shown in
Ref.~\cite{Montemurro_2001}) so that, when inferring $\alpha$ from
texts containing $10^4\sim 10^5$ distinct words, one tends to
underestimate it. The Heaps' law, instead, is already sufficiently
sensible to sample the tail of the distribution so that the measured
$\alpha$ and $\beta$ are such that $1/\alpha>\beta$.

By looking at Fig.~\ref{fig:heaps_non_aggregated} we find that the
growth of the number of distinct article edited in Wikipedia by users
is linear. Our Polya's urn model accounts for this possibility as
well, by predicting a connection between the Zipf's exponent and the
slope of the linear dictionary growth.

%% By exploiting this connection we can assess which are the values of the model parameters
%% $\nu$ and $\rho$ that asymptotically reproduce the measured curves. A look to the last two columns of Table~\ref{tab:editors} reveals plausible values, interestingly different from
%% each editor.

%

\paragraph{Triggering events\\}
To detect whether in a sequence there is a triggering mechanism in
play, we make use of the definition of entropy~(\ref{eq:entropy}) and
look at the distribution of time intervals between elements of the
same class (see Section~\ref{sec:triggering}).

For example, when listening to a certain lyric of a given artist, we
could be tempted to listen to other of her lyrics. In that case, the
occurrences of the lyrics' artist will be clusterized in the sequence
more than an uncorrelated poissonian process. At the same time, we
expect that the distribution of time intervals between the lyrics of
the same artist will be more biased toward small time intervals than a
poissonian process. In the case of lyrics, the class of elements is
given by their artist, in Wikipedia by the wiki-article (\emph{mother
  page}) that first linked to a new wiki-page, while in texts we
considered each word as bearing its own class, lacking of a
satisfactory classification of words in semantic areas.

\begin{figure}[t!]
\centerline{%
	\includegraphics[width=0.33\textwidth]{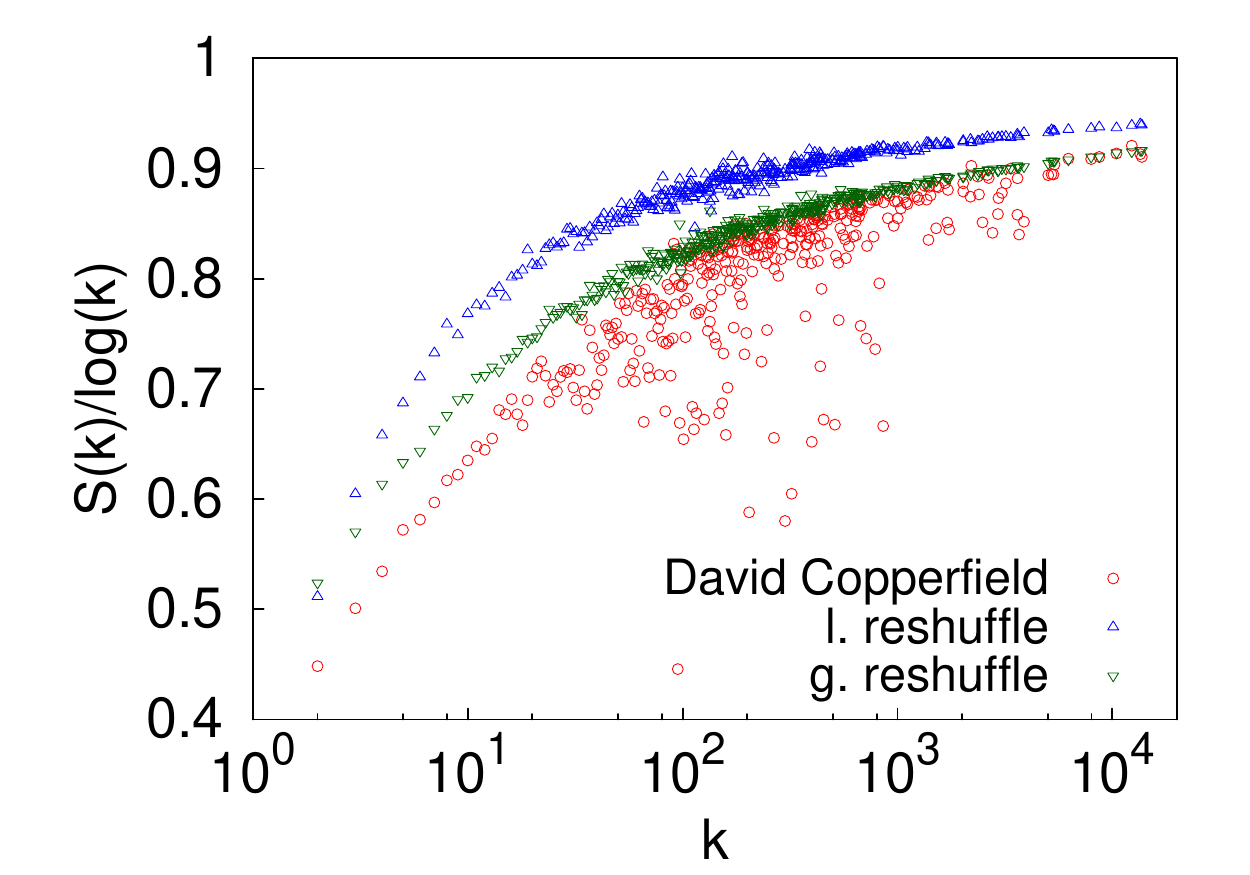}%
	\includegraphics[width=0.33\textwidth]{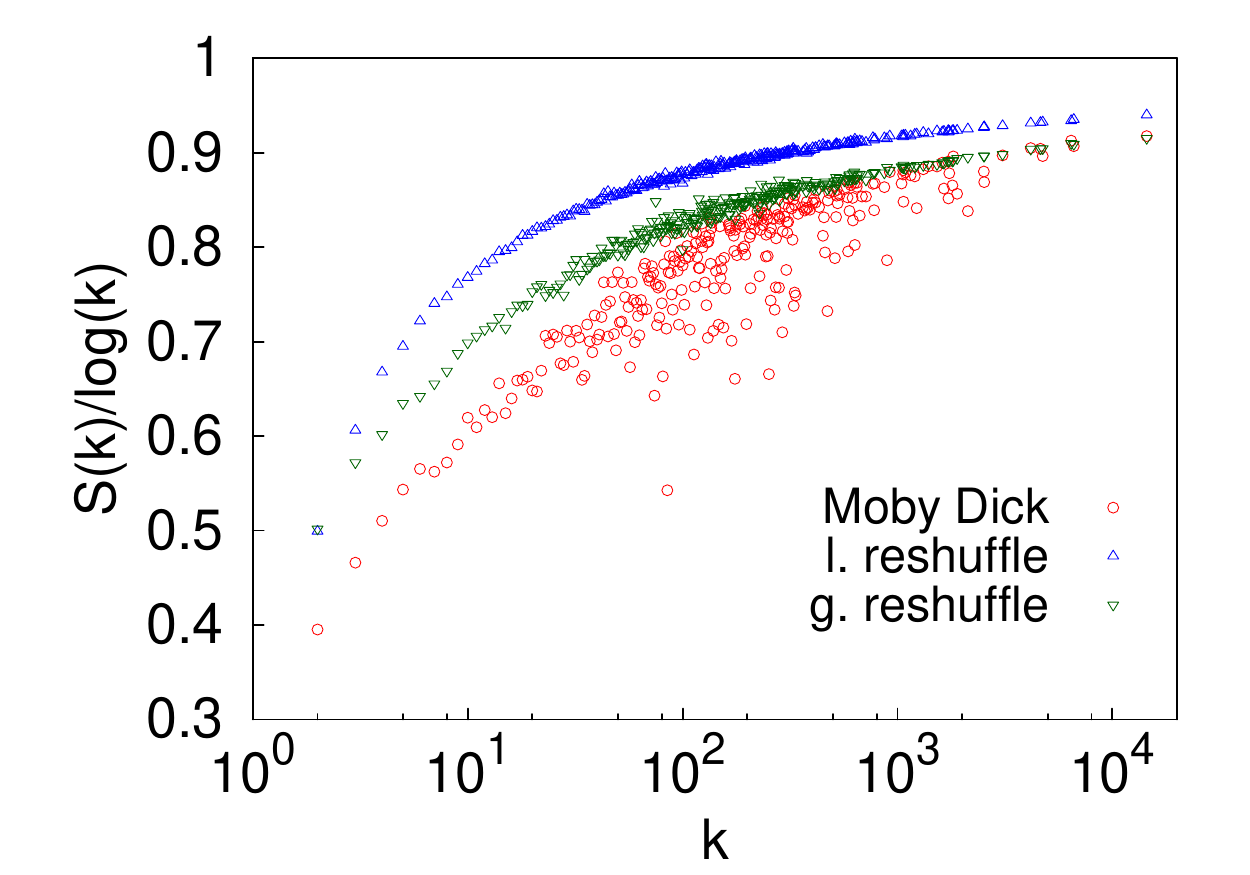}%
	\includegraphics[width=0.33\textwidth]{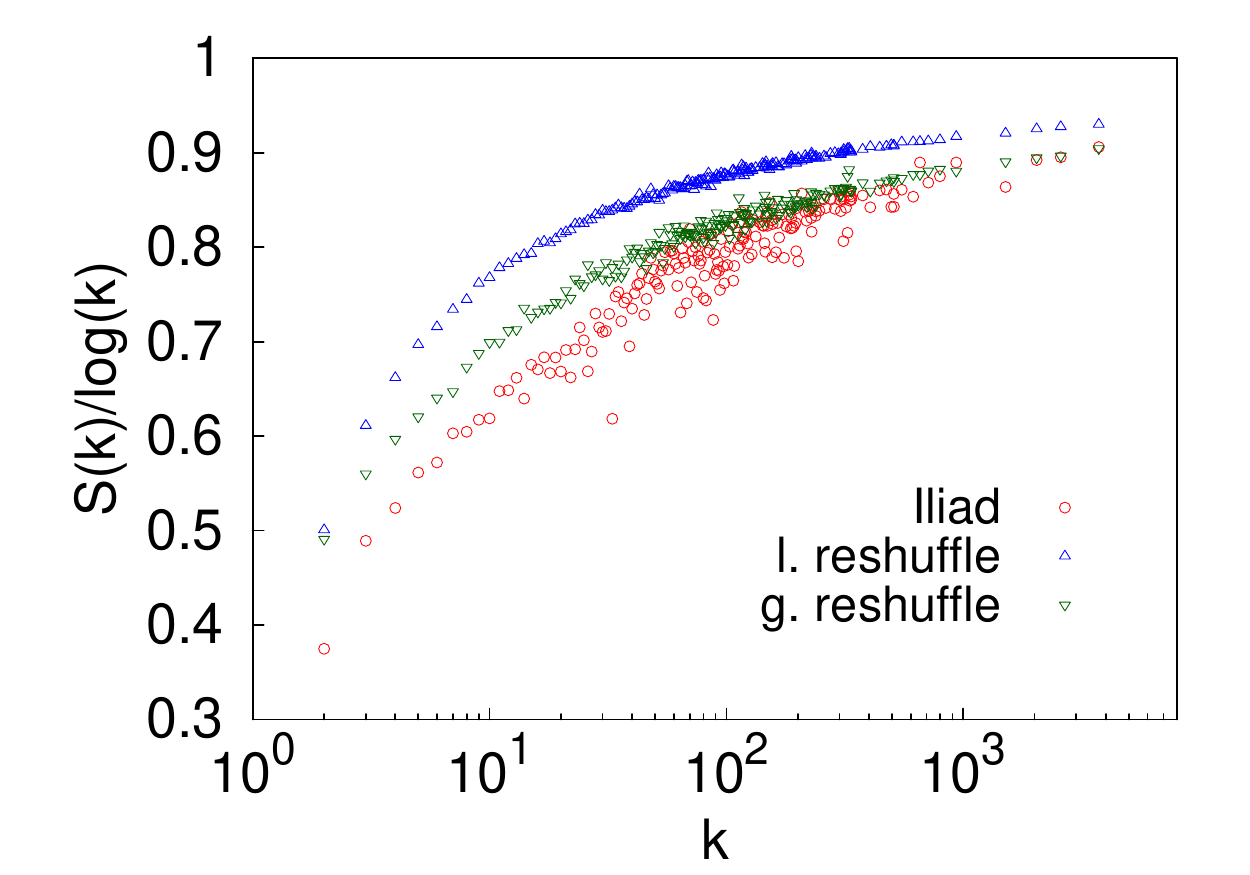}}
\centerline{%
	\includegraphics[width=0.33\textwidth]{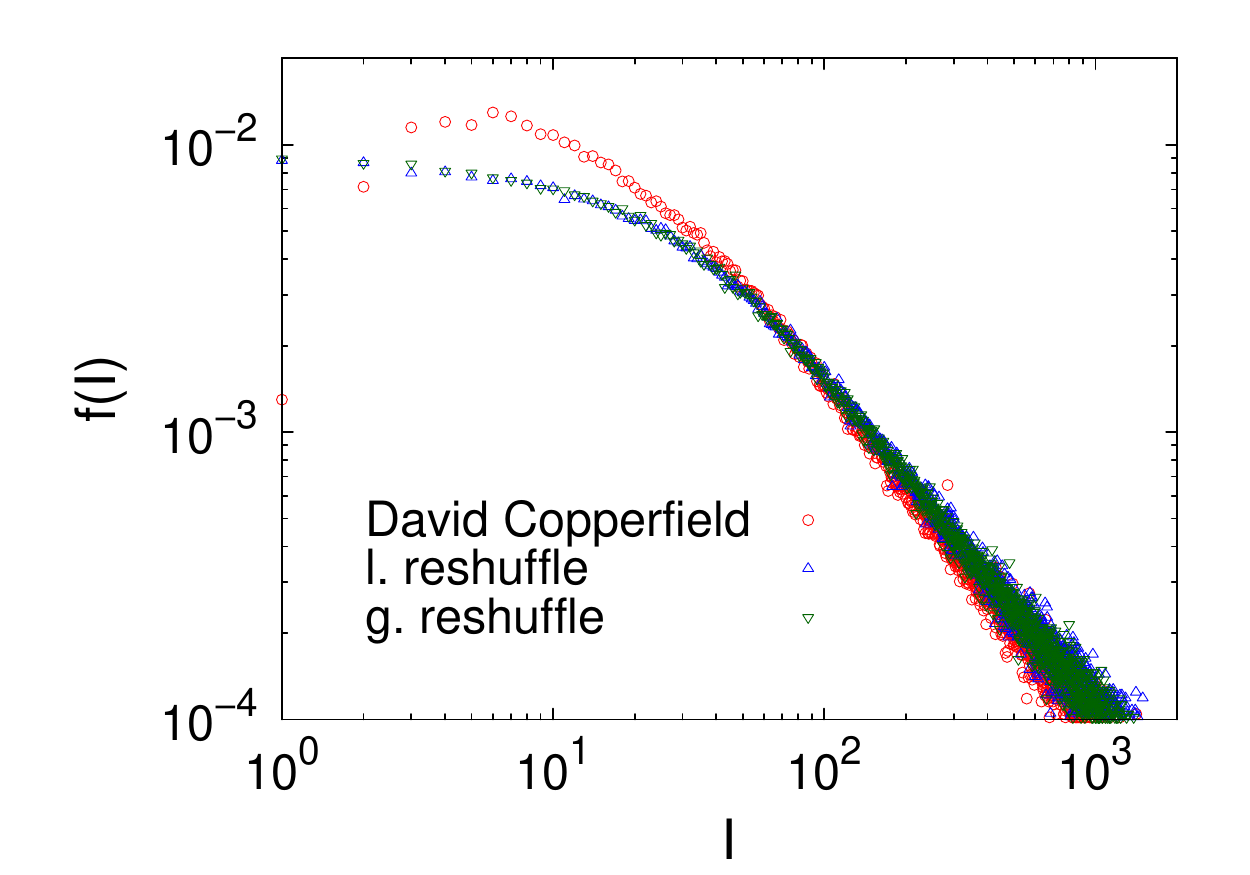}%
	\includegraphics[width=0.33\textwidth]{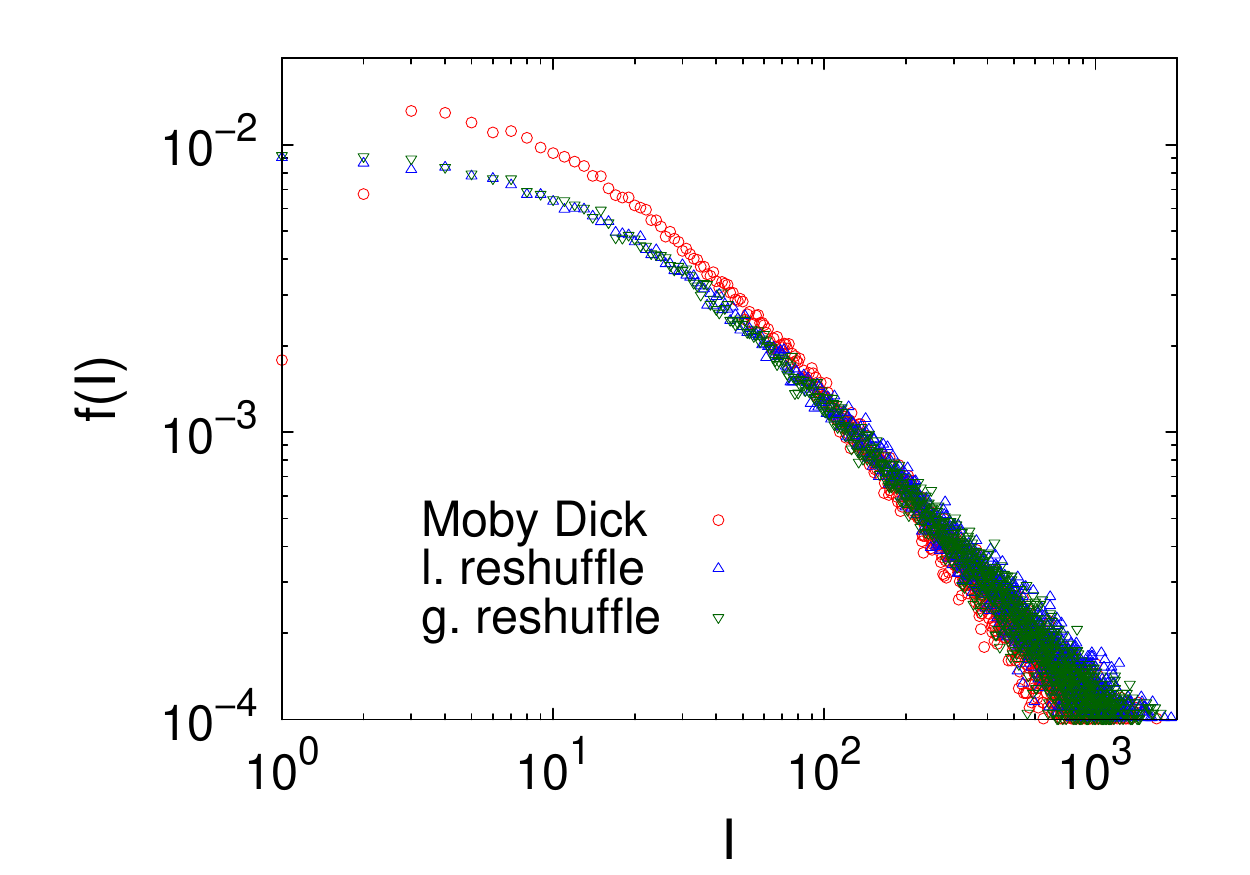}%
	\includegraphics[width=0.33\textwidth]{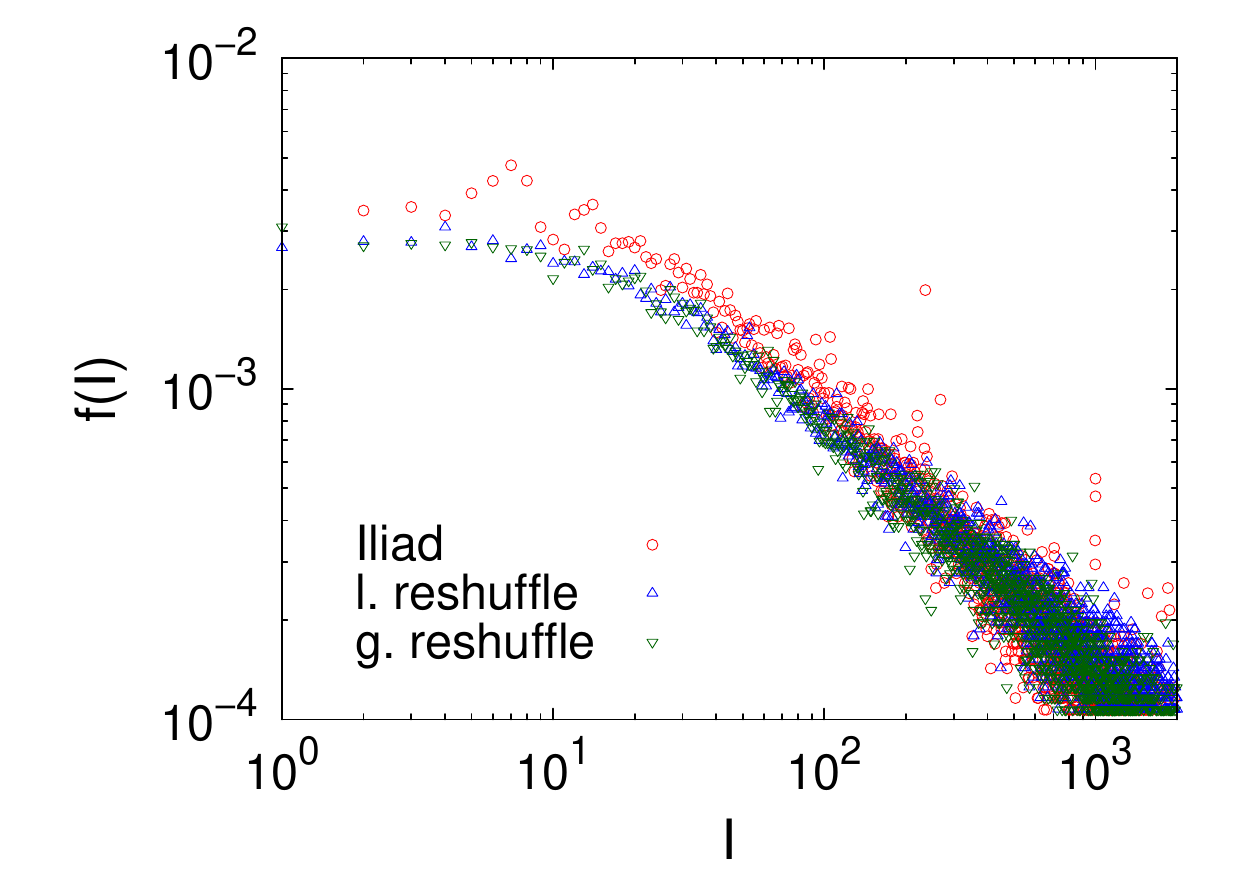}}
\caption{\textbf{Triggering events in single books from the Gutenberg dataset.} 
	Top: normalized average entropy in selected texts (red dot) and in the locally  (blue dots) an globally (green dots)
	reshuffled texts. Lower values of the entropy correspond to higher clusterized occurrences of elements. 
	Bottom: Time intervals distribution. More clusterized data result in higher values of the 
	distribution at low interval lengths.
\label{fig:triggering_gutenberg}	
	}
\end{figure}

\begin{figure}[t!]
\centerline{%
	\includegraphics[width=0.33\textwidth]{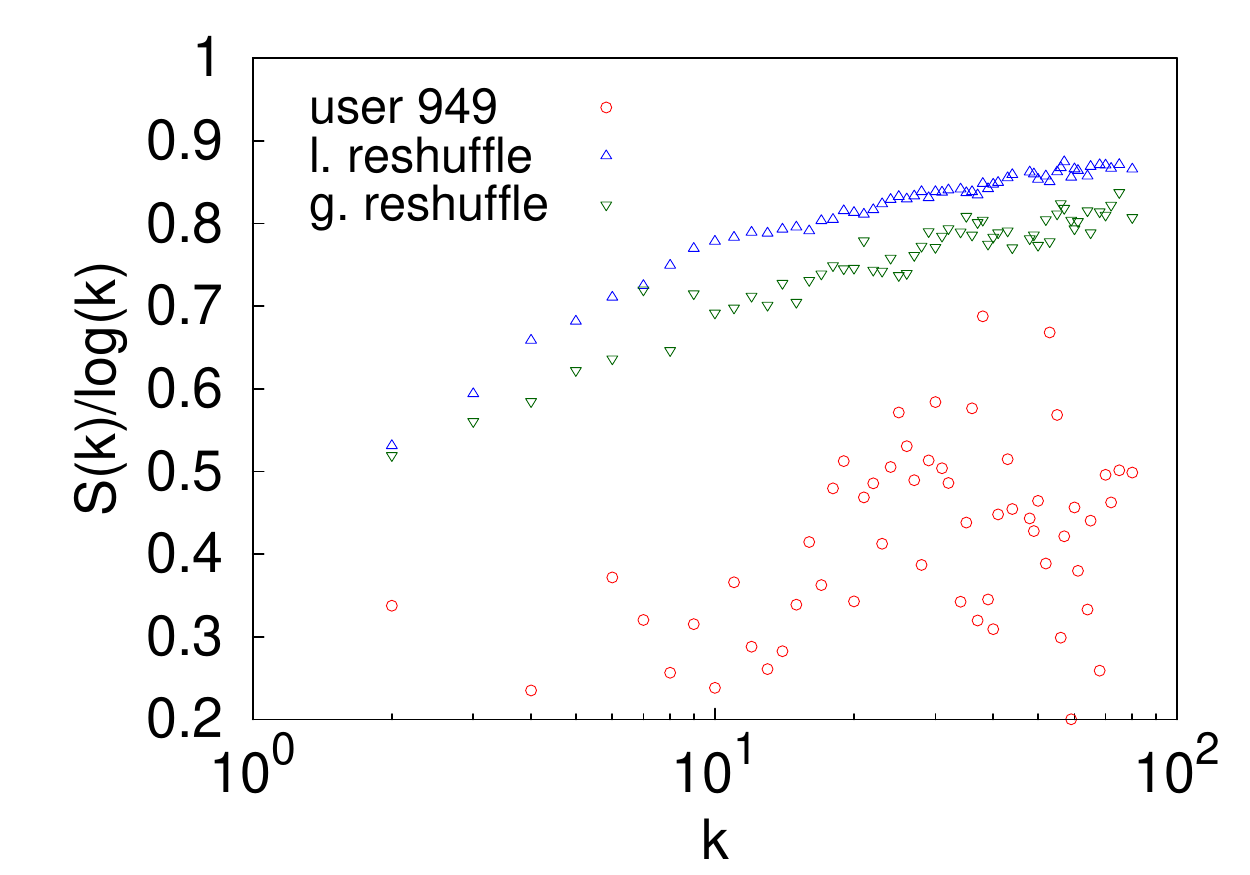}%
	\includegraphics[width=0.33\textwidth]{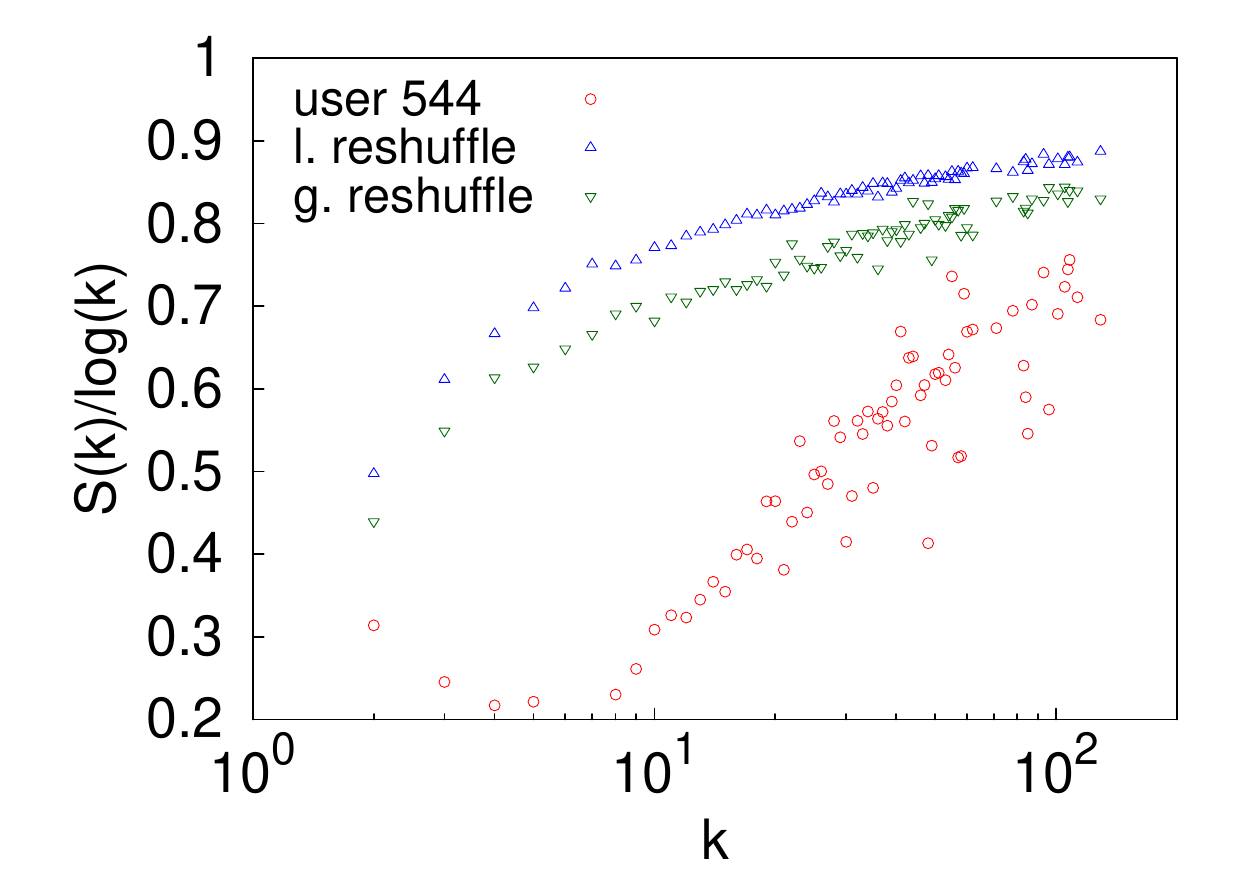}%
	\includegraphics[width=0.33\textwidth]{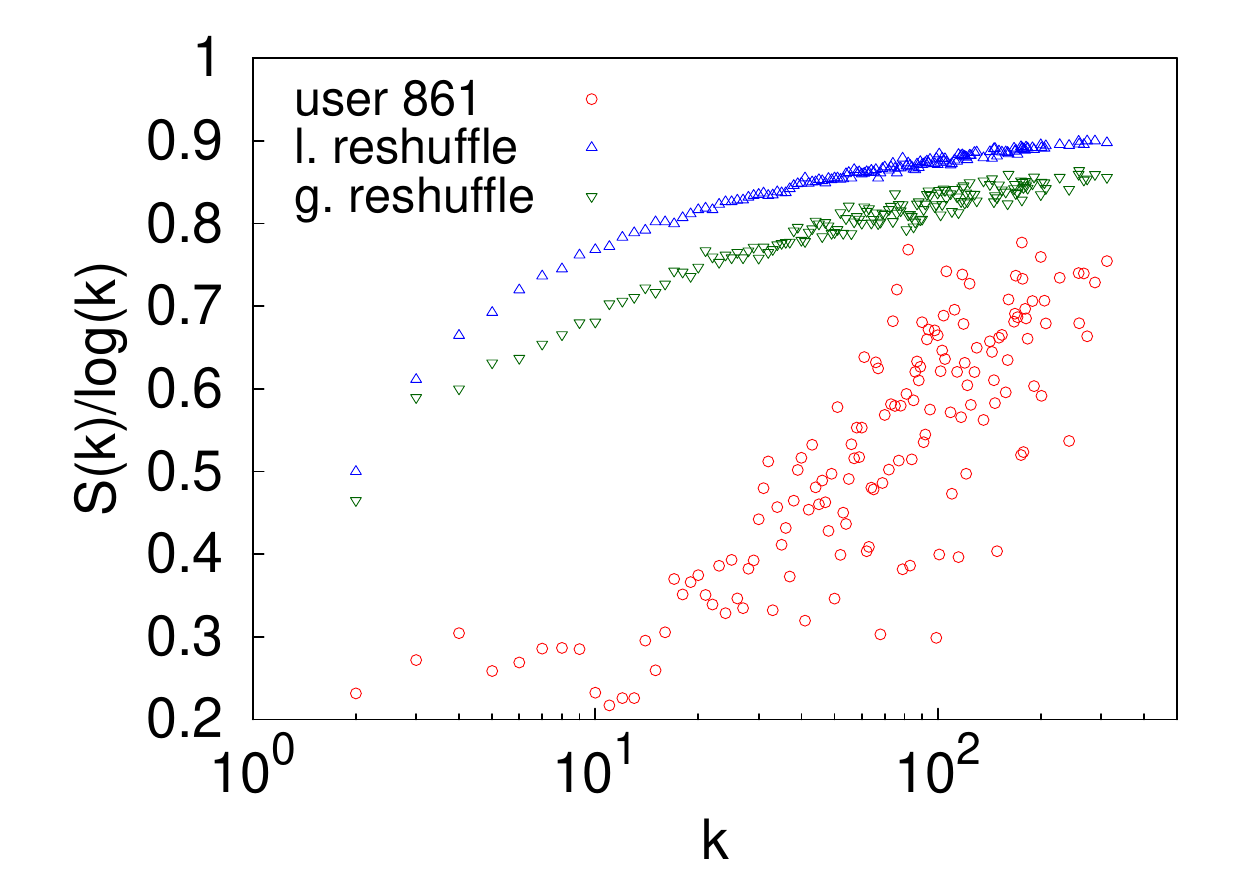}}
\centerline{%
	\includegraphics[width=0.33\textwidth]{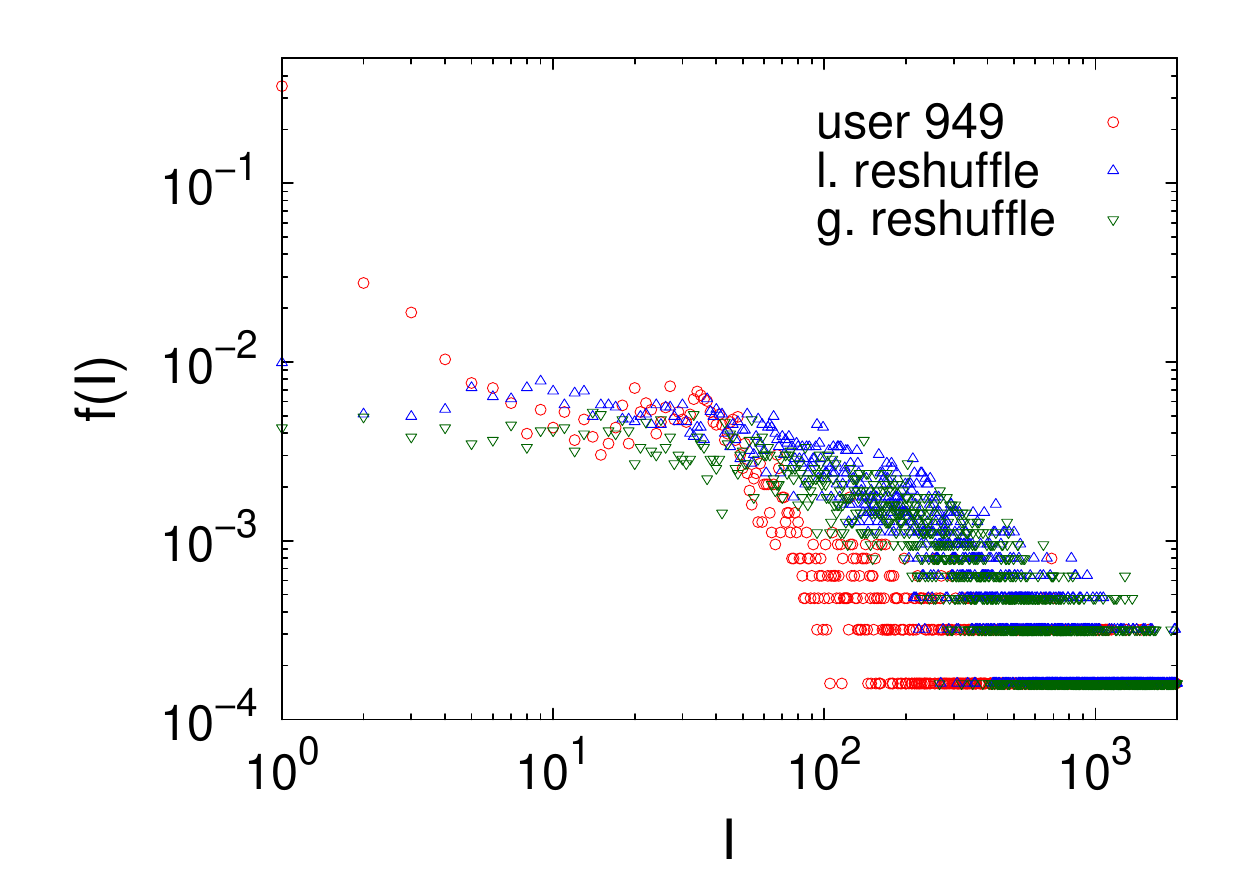}%
	\includegraphics[width=0.33\textwidth]{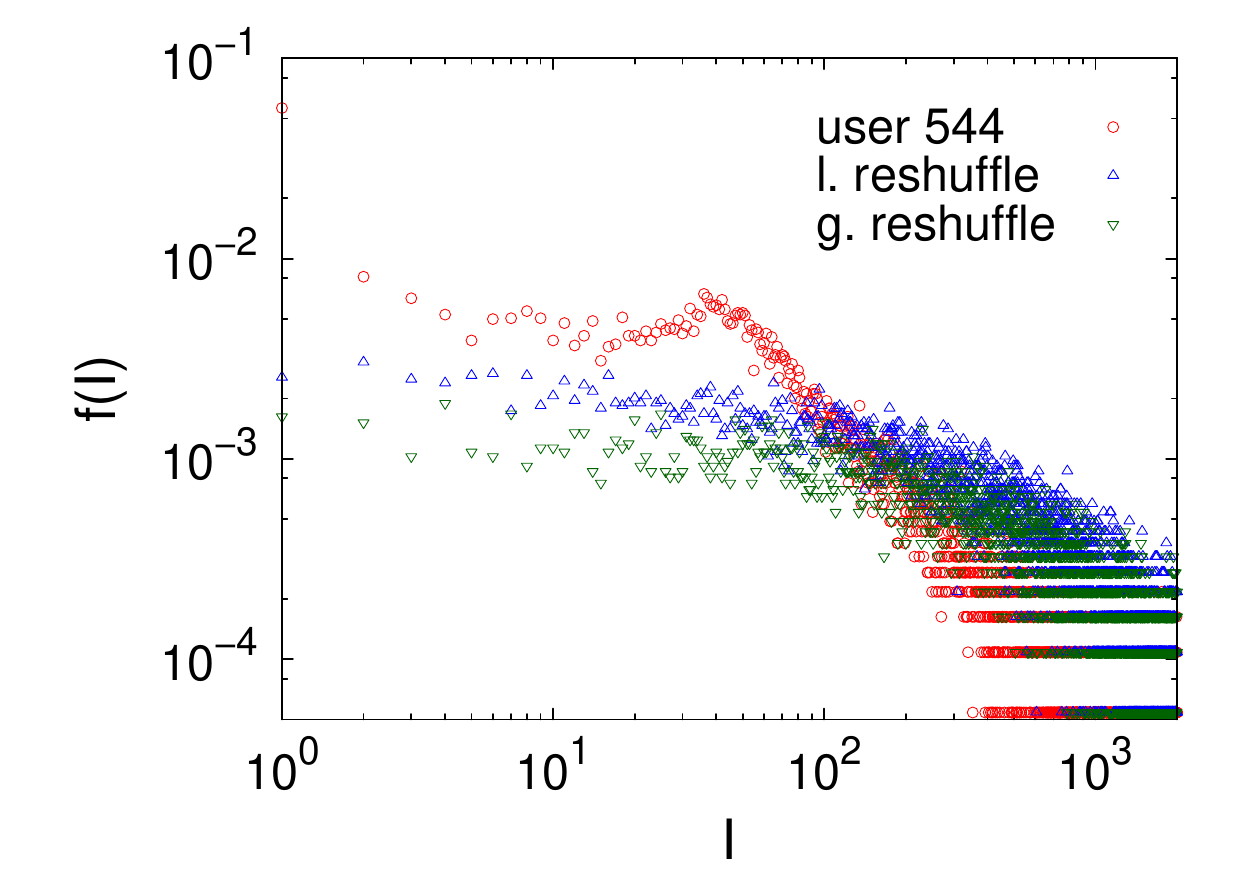}%
	\includegraphics[width=0.33\textwidth]{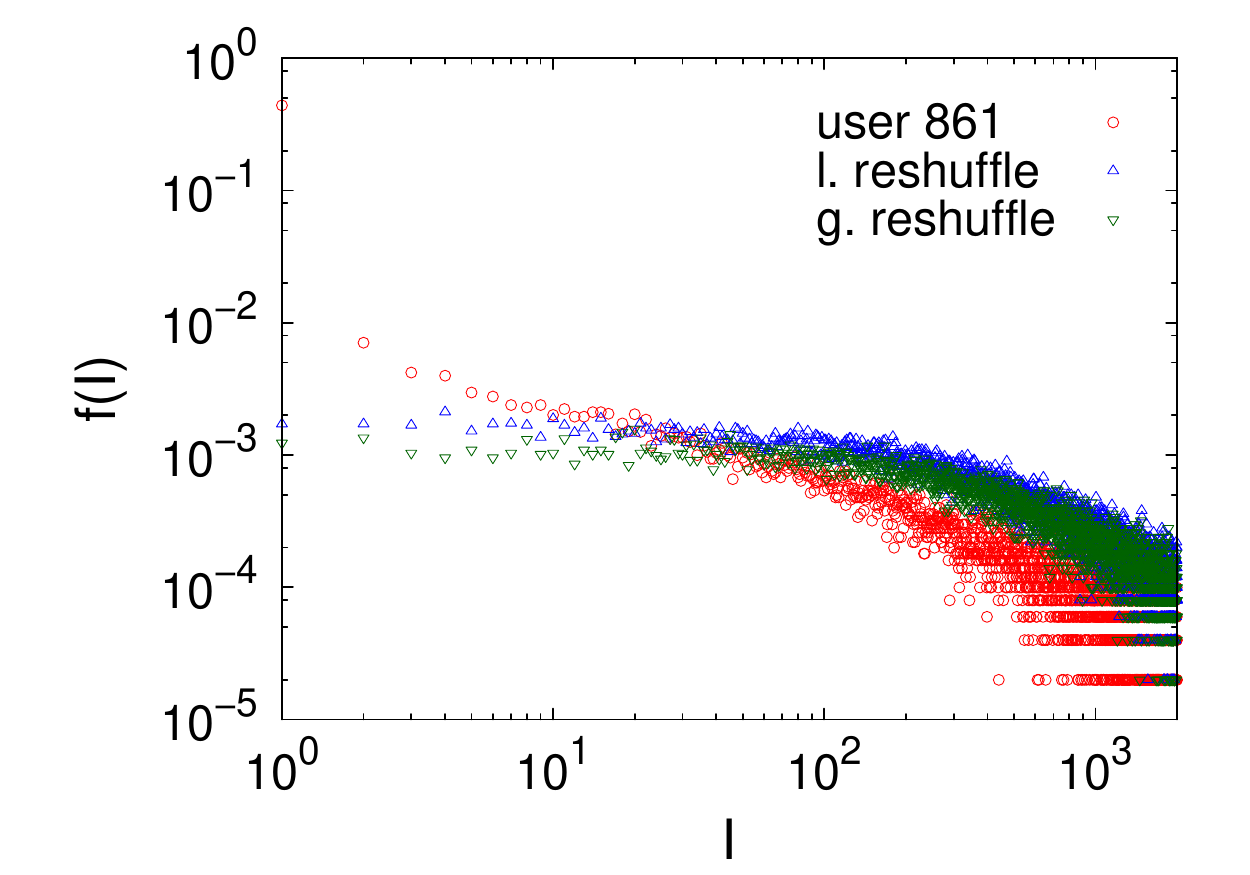}}
\caption{\textbf{Triggering events for single users in the Last.fm dataset.} 
	Top: normalized average entropy in selected listeners (red dot) and in the locally  (blue dots) an globally (green dots)
	reshuffled playlists. Lower values of the entropy correspond to higher clusterized occurrences of elements. 
	Bottom: Time intervals distribution. More clusterized data result in higher values of the 
	distribution at low interval lengths.
\label{fig:triggering_lastfm}	
}
\end{figure}

\begin{figure}[t!]
\centerline{%
	\includegraphics[width=0.33\textwidth]{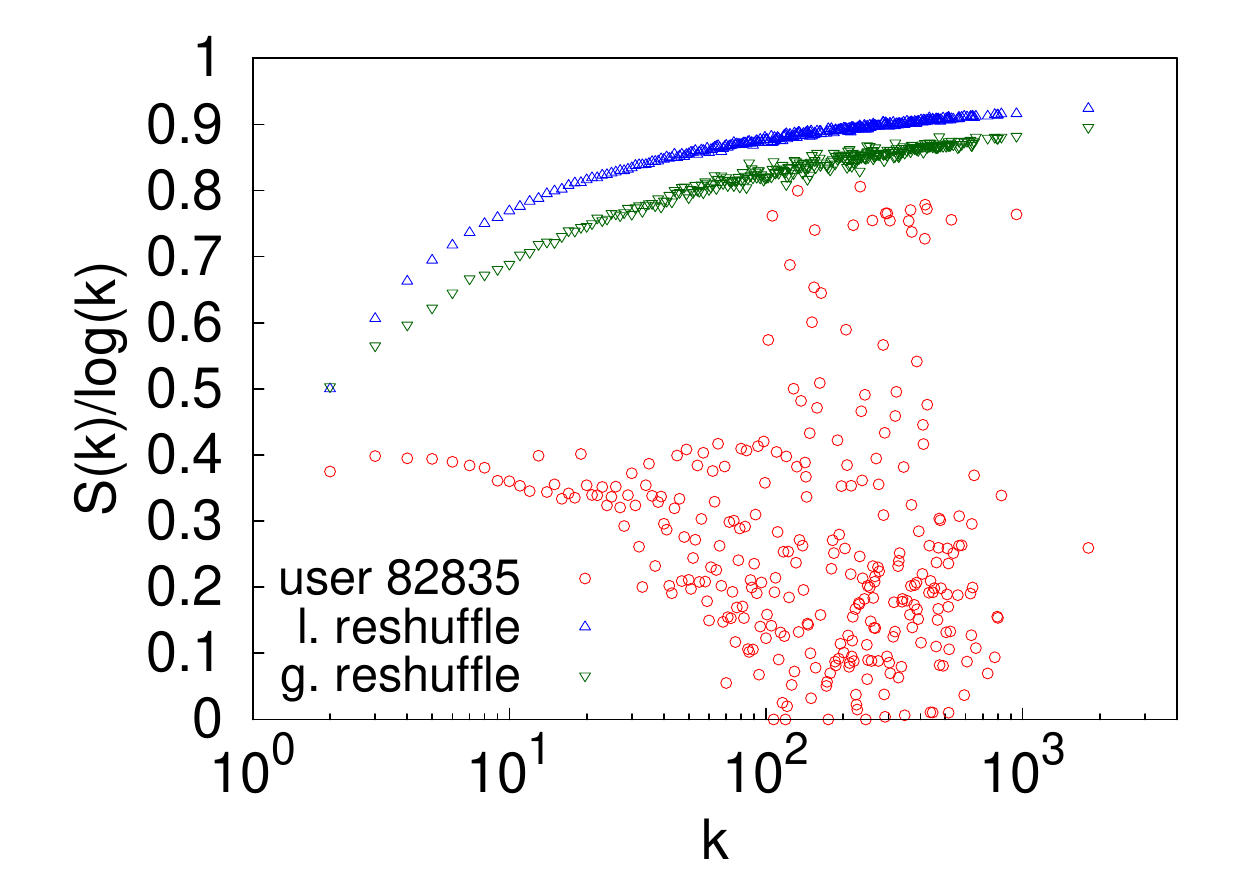}%
	\includegraphics[width=0.33\textwidth]{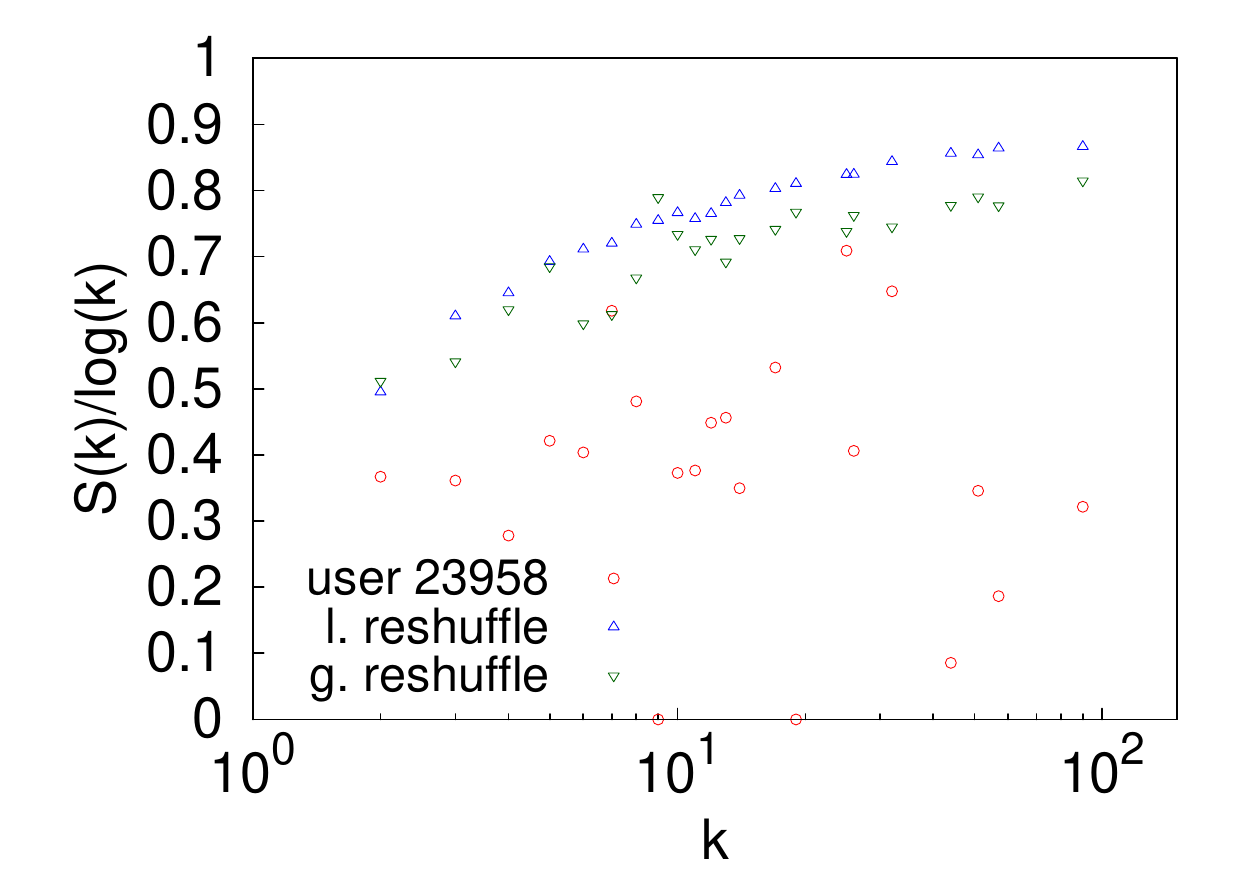}%
	\includegraphics[width=0.33\textwidth]{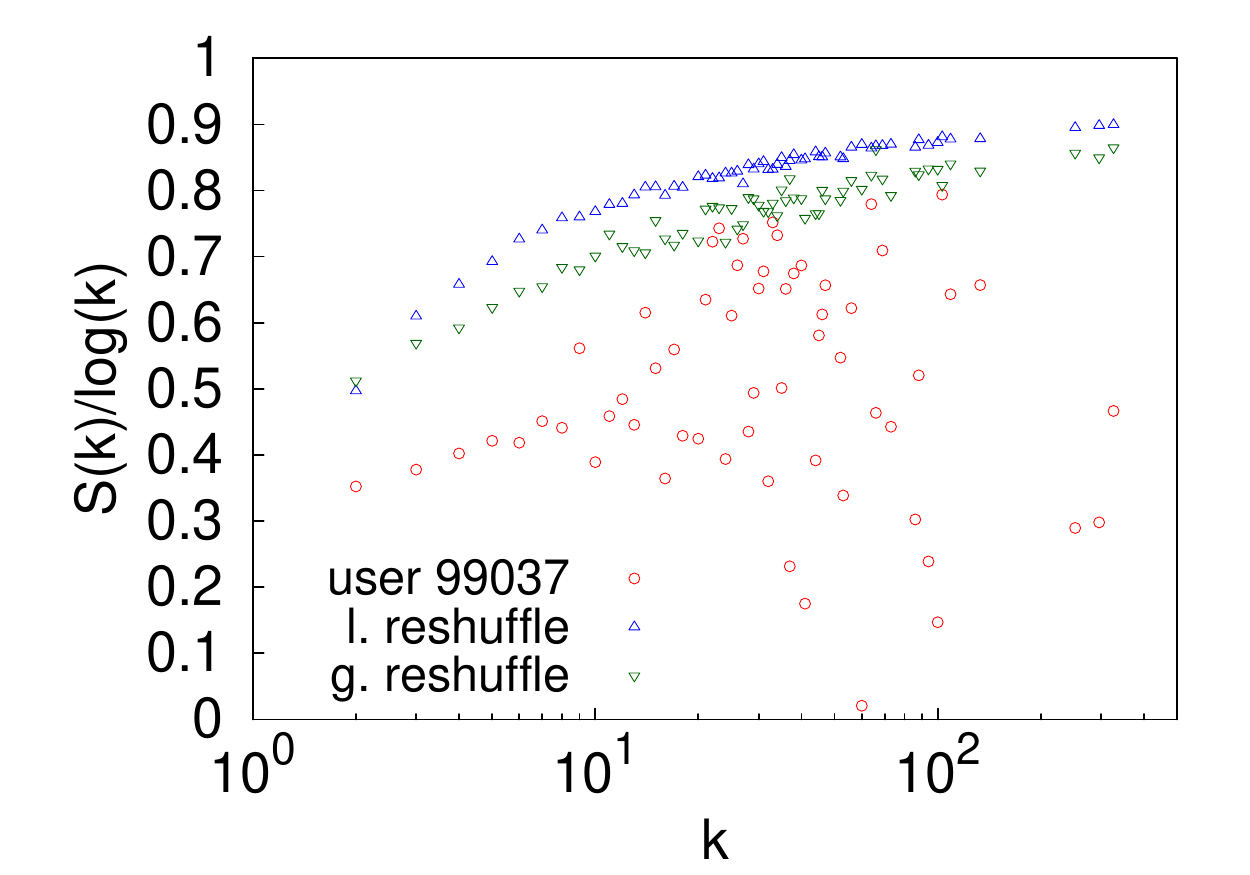}}
\centerline{%
	\includegraphics[width=0.33\textwidth]{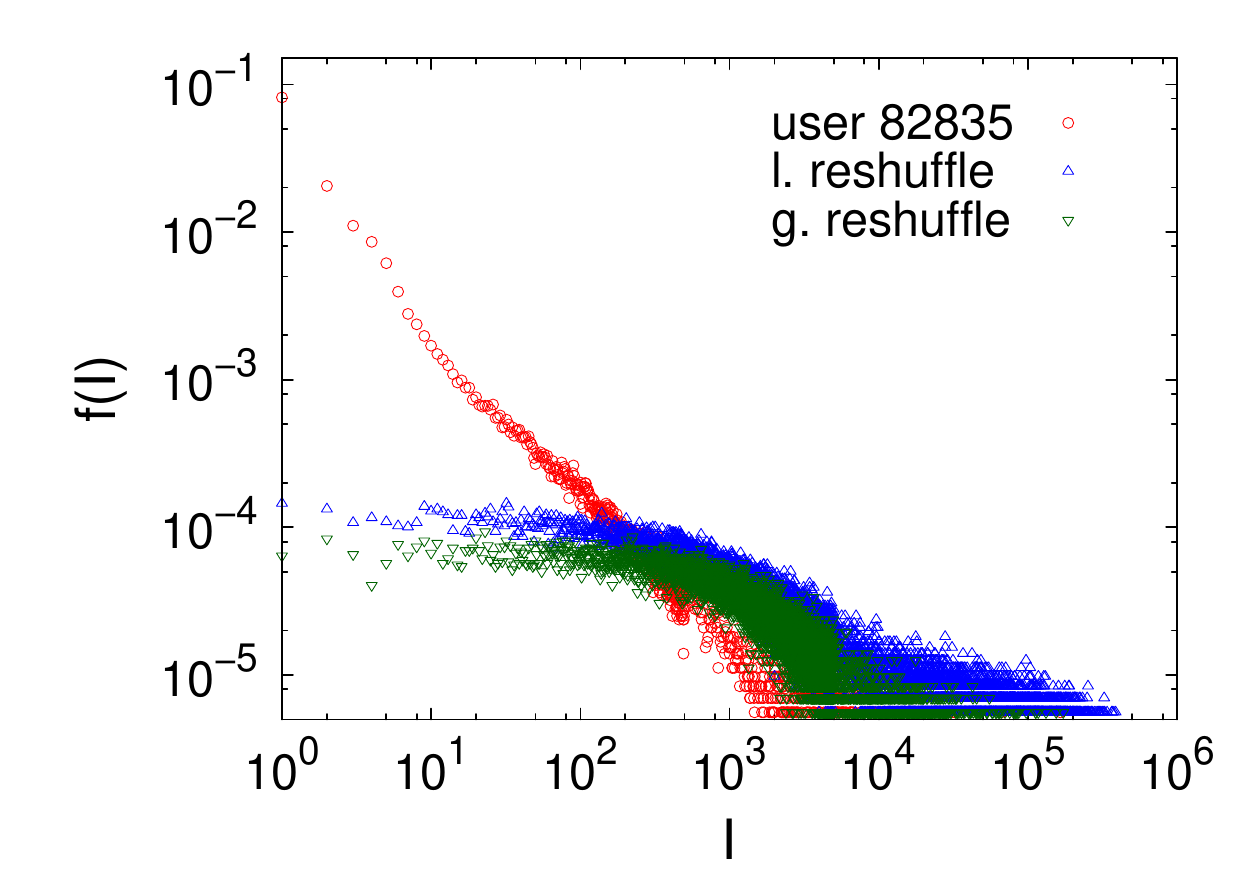}%
	\includegraphics[width=0.33\textwidth]{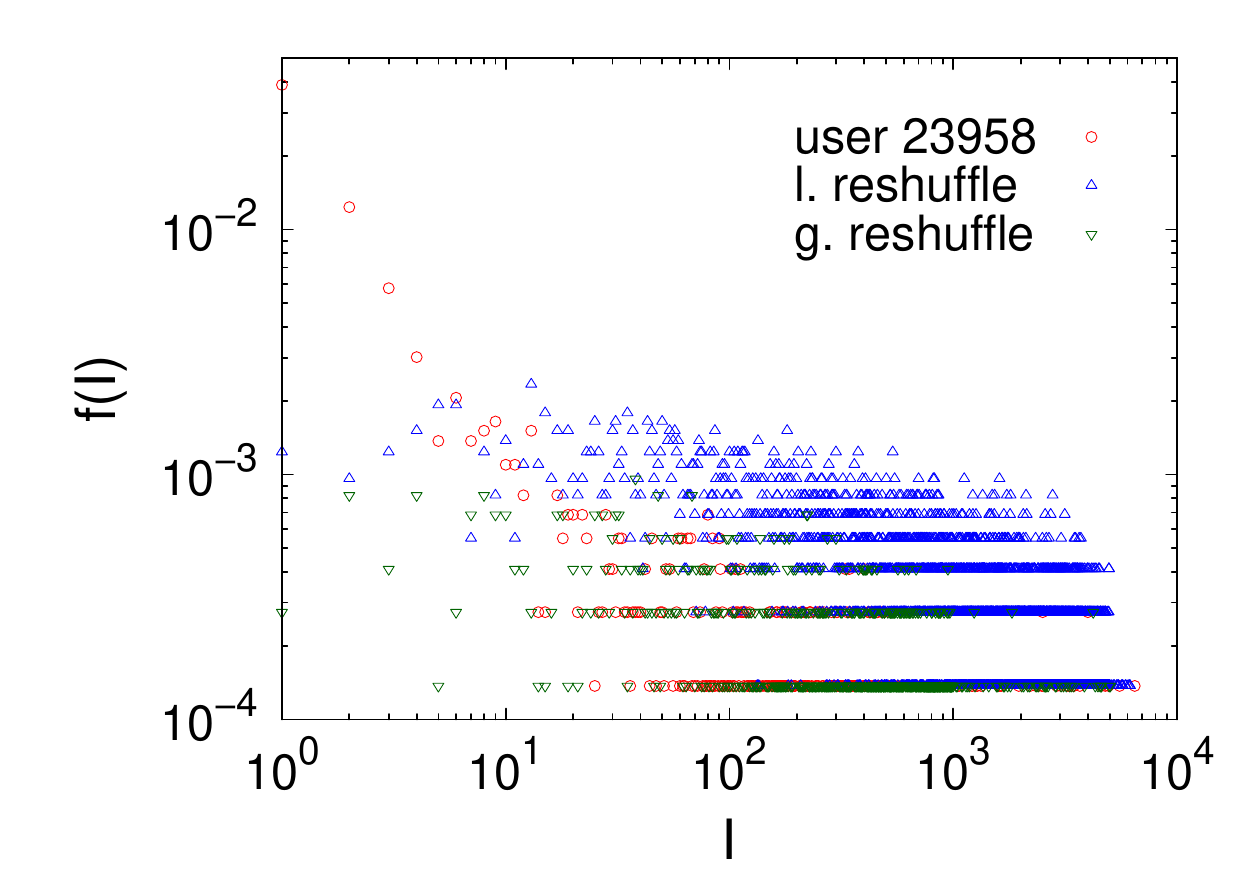}%
	\includegraphics[width=0.33\textwidth]{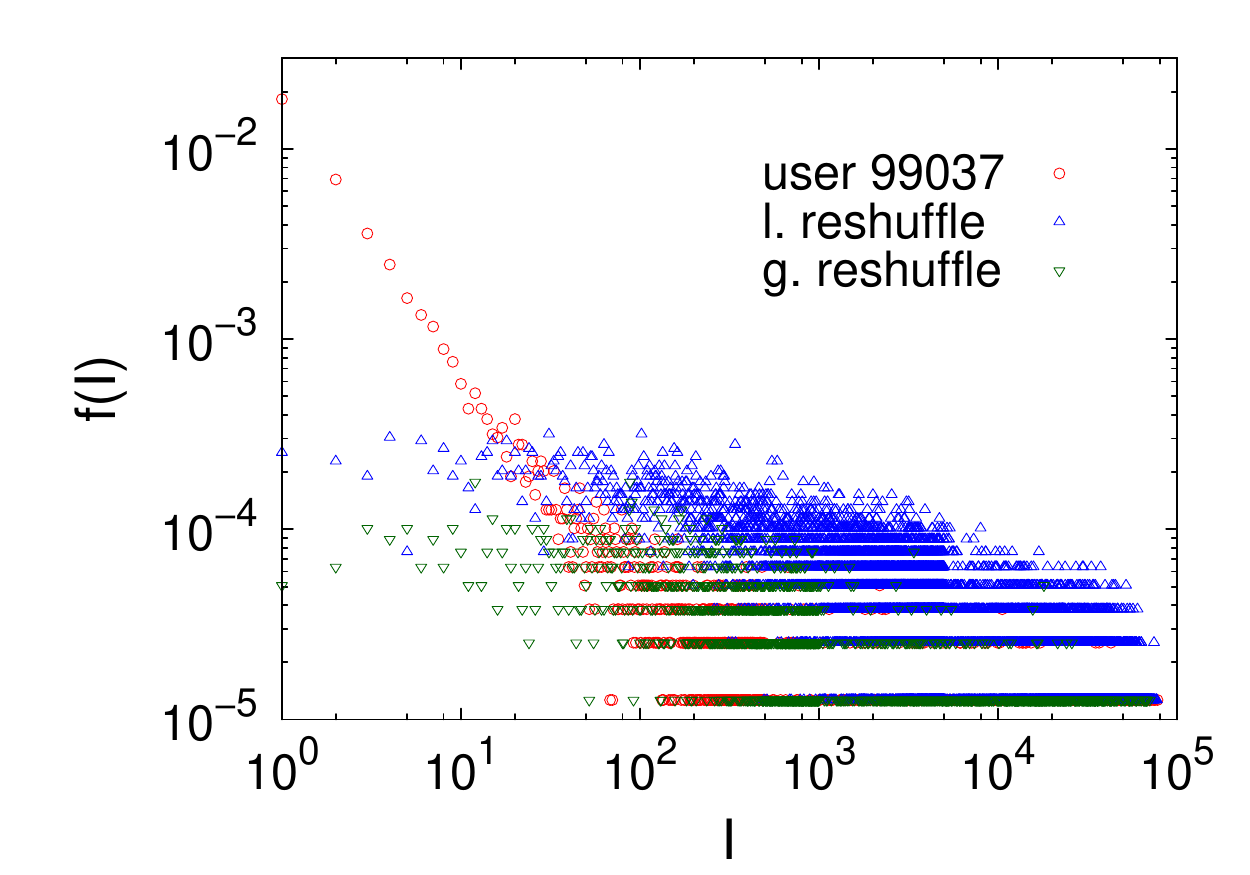}}
\caption{\textbf{Triggering events for single users in the Wikipedia dataset.} 
	Top: normalized average entropy in selected editors (red dot) and in the locally  (blue dots) an globally (green dots)
	reshuffled wiki-articles. Lower values of the entropy correspond to higher clusterized occurrences of elements. 
	Bottom: Time intervals distribution. More clusterized data result in higher values of the 
	distribution at low interval lengths.
\label{fig:triggering_wikipedia}	
}
\end{figure}

In order to distinguish between sequences ruled by a random poissonian
process from sequences featuring triggering events, we show in
figures~\ref{fig:triggering_gutenberg}, \ref{fig:triggering_lastfm}
and \ref{fig:triggering_wikipedia} the entropy and interval
distribution curves of selected texts, Last.fm listeners and wiki
editors (red dots), together with the correspondingly randomly
shuffled sequences (blue dots) and the \emph{locally} shuffled
sequences (green dots). The latter are achieved by shuffling the
subsequence that goes from the element following the first occurrence
of a given element, to the end. These figures confirm that also at the
user level one obtains the same results of the whole datasets. In
particular, the drop of the entropy around the value of $10$ in the
three selected Last.fm listeners can be a consequence of the typical
number of songs in a song album: who listens one song of an album,
tends to browse all of it, so that a dozen of songs with the same
artist appear heavily clusterized at short times, thus dropping the
associated entropy value.

The interest of looking at triggering events on single books, or
considering a single contributor of Wikipedia or a single Last.fm user
is to investigate the nature of the correlations observed in the whole
databases. In particular, the question is whether the statistical
signatures we detected emerge as an effect of a collective process or
are present also at the single user level. The results reported in
figures~\ref{fig:triggering_gutenberg}, \ref{fig:triggering_lastfm}
and \ref{fig:triggering_wikipedia} show that the adjacent possible
mechanism plays a role also on the individual level, and its effect is
enhanced in collective processes.

\end{document}